\documentclass[aps,prd,tightenlines,showpacs,groupedaddress,superscriptaddress,10pt,nofootinbib,floatfix,
preprintnumbers,letterpaper,twocolumn,longbibliography]{revtex4-1}
\usepackage{hyperref,amssymb,amsmath,graphicx,xcolor,cancel,comment}
\usepackage{bm,bbold}

\newcommand{\lsim}{\raisebox{-0.7ex}{$\stackrel{\textstyle <}{\sim}$ }}

\DeclareSymbolFont{symbolstx}{OMS}{txsy}{m}{n}
\SetSymbolFont{symbolstx}{bold}{OMS}{txsy}{bx}{n}
\DeclareSymbolFontAlphabet{\mathcal}{symbolstx}

\def\cO{\mathcal{O}}
\def\cQ{\mathcal{Q}}

\begin{document}

\begin{figure}[!t]
  \vskip -1.1cm \leftline{
    \includegraphics[width=3.0 cm]{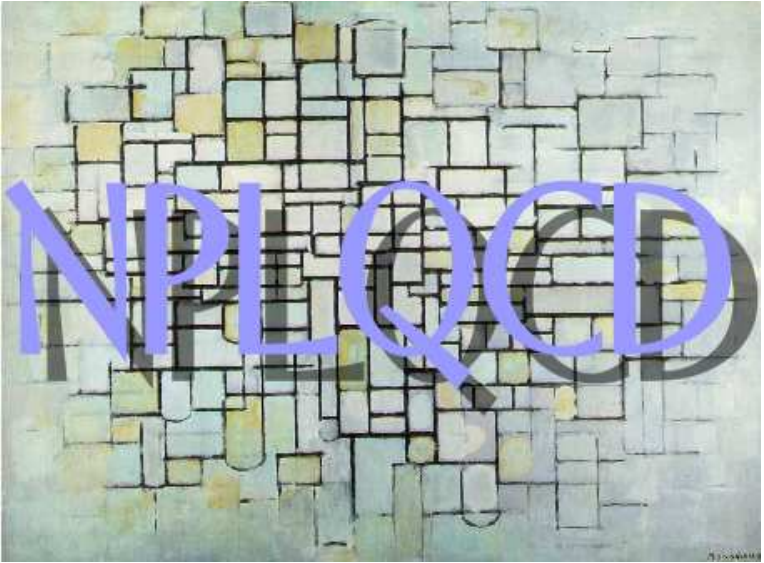}} \vskip
  -0.5cm
\end{figure}

\title{Octet Baryon Magnetic Moments from Lattice QCD:
Approaching Experiment from a Three-Flavor Symmetric Point}

\author{Assumpta~Parre\~no} 
\email{assum@fqa.ub.edu}
\affiliation{Dept.~d'Estructura i
  Constituents de la Mat\`eria.~Institut de Ci\`encies del Cosmos
  (ICC), Universitat de Barcelona, Mart\'{\i} Franqu\`es 1,
  E08028-Spain}

\author{Martin J. Savage}
\email{mjs5@uw.edu}
\affiliation{Institute for Nuclear Theory, University of Washington, Seattle, WA 98195-1560, USA}
\affiliation{Kavli Institute for Theoretical Physics, University of California, Santa Barbara, CA 93106, USA}

\author{Brian C. Tiburzi} 
\email{bctiburz@gmail.com}
\affiliation{Institute for Nuclear Theory, University of Washington, Seattle, WA 98195-1560, USA}
 \affiliation{Kavli Institute for Theoretical Physics, University of California, Santa Barbara, CA 93106, USA}
\affiliation{ Department of Physics, The City
  College of New York, New York, NY 10031, USA} 
\affiliation{Graduate
  School and University Center, The City University of New York, New
  York, NY 10016, USA} 
\affiliation{RIKEN BNL Research Center,
  Brookhaven National Laboratory, Upton, NY 11973, USA}

\author{Jonas Wilhelm} 
\email{jonas.wilhelm@physik.uni-giessen.de}
\affiliation{Justus-Liebig-Universit\"at Gie\ss en, Ludwigstra\ss e 23,  Gie\ss en
35390, Germany}
\affiliation{Department of Physics, University of Washington, Box 351560, Seattle, WA 98195, USA}

\author{Emmanuel~Chang} 
\email{changezy@uw.edu}
\affiliation{Institute for Nuclear Theory, University of Washington, Seattle, WA 98195-1560, USA}

\author{William Detmold} 
\email{wdetmold@mit.edu}
\affiliation{Kavli Institute for Theoretical Physics, University of California, Santa Barbara, CA 93106, USA}
\affiliation{ Center for Theoretical Physics,
  Massachusetts Institute of Technology, Cambridge, MA 02139, USA}

\author{Kostas~Orginos} 
\email{kostas@jlab.org}
\affiliation{Department of Physics, College of
  William and Mary, Williamsburg, VA 23187-8795, USA}
\affiliation{Jefferson Laboratory, 12000 Jefferson Avenue, Newport
  News, VA 23606, USA}

\collaboration{NPLQCD Collaboration}
\noaffiliation

\date{\today}

\preprint{NSF-KITP-16-140}  
\preprint{INT-PUB-16-028}  
\preprint{MIT-CTP-4833}


\begin{abstract}
Lattice QCD calculations with background magnetic fields are used to determine the magnetic moments of the octet baryons. 
Computations are performed at the physical value of the strange quark mass,  and two values of the light quark mass,
one corresponding to the  $SU(3)_F$-symmetric point, 
where the pion mass is $m_\pi\sim 800 \, \texttt{MeV}$, 
and the other corresponding to a pion mass of $m_\pi\sim 450 \, \texttt{MeV}$. 
The  moments are found to exhibit only mild pion-mass dependence 
when expressed in terms of appropriately chosen magneton units---the natural baryon magneton. 
A curious pattern is revealed among the anomalous baryon magnetic moments which  is linked to the constituent quark model, 
however, careful scrutiny exposes additional features. 
Relations expected to hold in the large-$N_c$ limit of QCD are studied;
and, 
in one case, a clear preference for the 
quark model over the large-$N_c$ prediction is found.
The magnetically coupled 
$\Lambda$--$\Sigma^0$ system is treated in detail at the 
$SU(3)_F$  
point,  
with the lattice QCD results comparing favorably with predictions based on $SU(3)_F$ symmetry.
This analysis enables the first extraction of the isovector transition  magnetic polarizability.
The possibility that large magnetic fields  stabilize strange matter is explored, but such a scenario is found to be unlikely. 
\end{abstract}

\maketitle

\section{Introduction}
\label{sec:intro}

The precisely measured values of magnetic moments of the lowest-lying octet of 
$J^\pi=\frac{1}{2}^+$ 
baryons, 
along with the rate of the radiative transition 
$\Sigma^0\rightarrow\Lambda + \gamma$, 
have been essential in elucidating important aspects of the structure of hadrons.
One of the major early  successes in the phenomenological modeling of hadrons was recovering the pattern of these magnetic moments 
from the na\"ive nonrelativistic quark model (NRQM)~\cite{GellMann:1964nj,Zweig:1981pd,Zweig:1964jf}.
In this model, 
baryons are comprised of three nonrelativistic constituent quarks with Dirac magnetic moments.  
When the three quark masses are fit to best reproduce the masses of the octet baryons, 
the magnetic moments predicted by this 
simple model compare with those of nature surprisingly well.
The predicted NRQM rate of 
$\Sigma^0\rightarrow\Lambda+\gamma$ 
and of the radiative transitions from the 
lowest-lying decuplet of 
$J^\pi=\frac{3}{2}^+$ 
baryons to the octet baryons are also in impressive agreement with experiment.

A closer connection can be made to the underlying theory of the strong interactions.
The global flavor symmetries of Quantum Chromodynamics (QCD), 
the two-flavor isospin 
$SU(2)$ 
and 
chiral 
$SU(2)_L \times SU(2)_R$, 
as well as the three-flavor analogues, 
$SU(3)_F$~\cite{GellMann:1962xb,Ne'eman:1961cd}
and 
$SU(3)_L \times SU(3)_R$, 
have been used to explore the
magnetic moments~\cite{Coleman:1961jn,Okubo:1963zza,Beg:1964nm,Sakita:1964qr}, 
and provide frameworks with which to systematically refine theoretical predictions in terms of small expansion 
parameters~\cite{Caldi:1974ta}.
In the case of two flavors, 
the proton and neutron magnetic moments 
(and additionally those of the 
$\Lambda$, 
$\Sigma$ 
and 
$\Xi$ 
baryons)
are described by isoscalar and isovector contributions, 
and have chiral expansions in terms of a small parameter determined by the 
mass of the pion, 
$m_\pi$.
In the three-flavor case, 
the expansion of the magnetic moments is determined by the kaon mass, 
$m_K$;
and, 
as with all quantities in 
$SU(3)_L \times SU(3)_R$ 
chiral perturbation theory 
($\chi$PT), 
is found to only slowly converge~\cite{Jenkins:1992pi}.
A connection between the constraints imposed upon the form of the magnetic moments from the flavor symmetries of QCD and results obtained 
from the NRQM has been made by considering the large-$N_c$ limit of QCD%
~\cite{Dashen:1994qi,Ahuatzin:2010ef,Jenkins:2011dr} 
in which the number of colors becomes large%
~\cite{'tHooft:1973jz}.
In this limit,  
the relations between low-energy predictions of the NRQM, 
the Skyrme model and $\chi$PT 
coincide as the emergent, 
approximate spin-flavor symmetries of QCD become manifest.

Starting with the pioneering works of  
Bernard {\it et al}.~\cite{Bernard:1982yu}  and Martinelli {\it et al}.~\cite{Martinelli:1982cb}, 
there have been a number of lattice QCD (LQCD) 
calculations of the magnetic moments of the octet baryons~\cite{Lee:2005ds,Detmold:2010ts,Primer:2013pva},
the decuplet baryons~\cite{Aubin:2008qp}, 
and the vector mesons~\cite{Luschevskaya:2016epp}.
Additionally,  
there have been LQCD calculations of the electric and magnetic polarizabilities of the 
lowest-lying hadrons~\cite{Fiebig:1988en,Christensen:2004ca,Lee:2005dq,Detmold:2006vu,Detmold:2009dx,Lujan:2014kia,Freeman:2014kka,Luschevskaya:2015bea,Luschevskaya:2015cko}, 
and of baryons in strongly interacting models of dark matter%
~\cite{Appelquist:2015zfa}.
Techniques developed in these works have also been used to determine the magnetic moments and polarizabilities of light 
nuclei~\cite{Beane:2014ora,Chang:2015qxa} with LQCD.
Further,  the rate for the radiative capture process
$n+ p\rightarrow d+\gamma$~\cite{Beane:2015yha},
which is dominated by the magnetic dipole amplitude at low energies,
has been calculated by making use of formal developments that combined the background-field technique 
with L\"uscher's method~\cite{Luscher:1986pf,Luscher:1990ux} 
to describe a two-particle coupled-channel system in a finite volume~\cite{Detmold:2004qn,Meyer:2012wk}.
While the very earliest calculations were quenched 
(i.e.~the calculations did not include quark-antiquark vacuum fluctuations), 
they produced ratios of proton to neutron magnetic moments of 
$\mu_p/\mu_n=-1.6\pm 0.2$~\cite{Bernard:1982yu} 
and 
$-1.6\pm 0.15$~\cite{Martinelli:1982cb},
which are consistent with the precisely known experimental value of 
$\mu_p^{\rm (expt)}/\mu_n^{\rm (expt)}=-1.459898075(12)$.
In later studies, 
it was found that, while the magnetic moments and polarizabilities of the electrically neutral baryons are 
 straightforward to 
extract from LQCD correlation functions generated in uniform background magnetic fields, 
those of the charged baryons are somewhat more challenging due to their eigenfunctions being spatial Landau levels
as opposed to momentum eigenstates.
For light nuclei, the source and sink structures used in the calculations of the magnetic structure of light nuclei~\cite{Beane:2014ora,Chang:2015qxa} have permitted determinations of both the magnetic moments and polarizabilities, 
as the effective mass of each spin state reaches a plateau after a  modest (but usually different)
number of time slices.
The magnetic moments of the baryons correspond to just one kinematic point of the magnetic form factor, and the more general
 behavior of the form factor provides further insight into the distribution of charged currents within the baryon.
There are  extensive studies of the baryon electromagnetic 
form factors, for example 
Refs.~\cite{Lin:2008mr,Bhattacharya:2013ehc,Shanahan:2014uka},
from which the magnetic moments and associated radii can be extracted.
General nonuniform background fields have been recently proposed to extract higher electromagnetic moments, 
as well as charge radii from LQCD~\cite{Davoudi:2015cba,Davoudi:2015zda}.
Further developments include 
accessing the hadronic vacuum polarization from magnetic susceptibilities%
~\cite{Bali:2015msa},
and an approach to hadron structure based on the Feynman-Hellman theorem%
~\cite{Detmold:2004kw,Freeman:2012ry,Horsley:2012pz,Chambers:2014qaa,Chambers:2015kuw,Savage:2016kon,Bouchard:2016heu}.

Previous  LQCD calculations of the magnetic moments of the proton, neutron and light 
nuclei~\cite{Beane:2014ora,Chang:2015qxa}
have found that  nearly all of their light-quark mass dependence 
is captured by the nucleon mass defining the unit of nuclear magnetons.
\footnote{Such natural magneton units have been used earlier by others, 
for example, 
to calculate nucleon magnetic moments in the Skyrme model
in which the nucleon mass is characteristically too large%
~\cite{Meissner:1986js}.
}  
In other words, 
$M_N(m_\pi) |{\bm\mu}_i(m_\pi)|$ 
is found to be 
approximately constant over a wide range of pion masses extending up to  
$\sim 1 \, \texttt{GeV}$, 
and possibly beyond,
where 
$M_N(m_\pi)$ 
is the mass of the nucleon at a given pion mass, and 
${\bm\mu}_i(m_\pi)$ 
is the magnetic moment of the nucleon or nucleus at that same pion mass.
This behavior is quite intriguing for a number of reasons.  
Empirically, 
it is found that the nucleon mass is essentially linearly dependent on the pion mass for 
$m_\pi \gtrsim 250 \, \texttt{MeV}$ 
with a coefficient very close to unity%
~\cite{WalkerLoud:2008bp,Walker-Loud:2014iea},
but expected to tend towards the chiral behavior for smaller pion masses,
see Ref.~\cite{BMWLatticeConf} for recent progress.  
At even larger pion masses, 
this behavior is expected to evolve toward 
$M_N \sim \frac{3}{2} \, m_\pi$.
In the chiral expansion of the nucleon magnetic moments, 
the leading correction to the 
$SU(3)_F$-symmetric predictions depends linearly on the pion mass, 
which is consistent with the observed behavior, but only for a nucleon mass that depends linearly on the pion mass.
In the context of the NRQM, 
$M_N  |{\bm\mu}_i |$ 
is required to be approximately independent of the pion mass as the nucleon magnetic moments result from 
combinations of quark spins, each normalized by the constituent quark mass which is  
$\sim \frac{1}{3} M_N$.

In this work, 
we extend our studies of the magnetic moments of the nucleons and light nuclei to baryons in the lowest-lying octet.
Our calculations of the magnetic moments of baryons are accomplished by modifying the LQCD gauge link variables to include a background electromagnetic gauge potential in calculations of the quark propagators. 
The magnetic moment of a baryon is extracted from the component of the energy splitting between its two spin states that 
depends linearly on the magnetic field. 
Essentially, the Zeeman effect for each baryon is determined. 
In particular, 
calculations of the magnetic moments and of the 
$\Sigma^0\rightarrow\Lambda+\gamma$  
radiative decay matrix element are performed at the 
$SU(3)_F$-symmetric point at two lattice spacings with the physical strange-quark mass for which
$m_\pi\sim 800 \, \texttt{MeV}$, 
and further, 
calculations at a single lattice spacing are performed at a pion mass of 
$m_\pi\sim 450 \, \texttt{MeV}$  
with the physical strange quark mass.
Contributions to the magnetic moments from quark-disconnected diagrams 
are not included in these calculations, 
which impacts the magnetic moments obtained at 
$m_\pi\sim 450 \, \texttt{MeV}$,
but these contributions are estimated to be small.
Details of our computational approach are given in Sec.~\ref{s:details}. 
The main results of this work are summarized as follows. 
\begin{itemize}
\item
Natural baryon magnetons 
$\texttt{[nBM]}$, 
where the mass of each baryon is used to define its magnetic moment, 
capture the majority of the quark-mass dependence of magnetic moments for the entire multiplet, 
even away from the limit of 
$SU(3)_F$ 
symmetry, 
see 
Sec.~\ref{s:magmom}.

\item
In 
$\texttt{[nBM]}$  units,  the anomalous moments of the proton and 
$\Sigma^+$  are  $\delta\mu_B \sim +2$,  of the neutron and $\Xi^0$ are  $\delta\mu_B \sim -2$, 
and those of the  $\Sigma^-$  and  $\Xi^-$  are   $\delta\mu_B \sim 0$.
Such values are consistent with the 
$SU(3)_F$-symmetric moments, 
$\mu_D$
and 
$\mu_F$~\cite{Coleman:1961jn,Okubo:1963zza,Beg:1964nm,Sakita:1964qr},  
assuming the values  
$\mu_D \sim +3$ and  $\mu_F \sim +2$, 
see 
Sec.~\ref{s:CG}.

\item
These values for $SU(3)_F$-symmetric moments, and the mild quark-mass dependence they exhibit, are suggestive of the NRQM.
The magnetic moment relations predicted by the NRQM are scrutinized, 
and interesting features are found in comparing the LQCD results and experiment, 
see
Sec.~\ref{s:NRQM}.

\item
Large-$N_c$ relations between magnetic moments, 
and the parametric scaling of their corrections
are compared with the results of the LQCD calculations.
In general, relations independent of 
$SU(3)_F$ 
breaking are found to be compatible, 
however, 
those dependent on the level of  
$SU(3)_F$ 
breaking are less clear. 
In particular, 
one relation is violated at the $\sim 40\%$ level; but, 
importantly, is considerably more consistent with 
the NRQM prediction, 
see
Sec.~\ref{s:LargeN}.

\item
Extracting the matrix element of the radiative transition 
$\Sigma^0\rightarrow\Lambda+\gamma$ 
is found to be more challenging because of the small mass splitting between  
$\Sigma^0$ 
and 
$\Lambda$. 
At the 
$SU(3)_F$-symmetric point, 
a matrix of correlation functions is diagonalized to reveal the closely spaced energy eigenstates, 
from which this matrix element is determined, 
see 
Sec.~\ref{s:LamSig}.

\end{itemize}

In addition to the magnetic moments, higher-order magnetic interactions, such as the magnetic polarizabilities, 
can also be determined from the LQCD calculations~\cite{Chang:2015qxa}. 
The energy dependence of each spin state, 
moreover, 
can be calculated over a range of magnetic fields, 
and allows for an exploration of the possibility that a large magnetic field could stabilize strange matter in dense astrophysical objects. 
Our results in Sec.~\ref{sec:LargeB} indicate that considerably larger baryon densities than are conceivably achieved in neutron stars are needed to stabilize strange matter. 
Various details related to the analysis of baryon correlation functions computed with LQCD appear in 
Appendix~\ref{s:A}, 
and further technical details concerning the transition correlation functions are discussed in 
Appendix~\ref{s:B}. 
Our presentation ends in Sec.~\ref{s:end}  with a summary of the main results.

\section{Computational Overview}
\label{s:details}

In the present study, 
lattice calculations are performed using three ensembles of 
QCD gauge configurations. 
Each ensemble was generated using the 
L\"uscher-Weisz gauge action%
~\cite{Luscher:1984xn}
with a tadpole-improved%
~\cite{Lepage:1992xa}, 
clover-fermion action%
~\cite{Sheikholeslami:1985ij}.
Configurations used in this work were taken at intervals of ten hybrid Monte Carlo trajectories.  
A summary of these gauge configurations is provided in 
Table~\ref{t:configs}.

Two of the gauge-field ensembles,  which we label Ensembles I and II, 
feature $N_f = 3$
degenerate, 
dynamical quark flavors with mass close to that of the physical strange quark.  
The resulting mass of non-singlet pseudoscalar mesons for these ensembles is found to be
$\sim 800 \, \texttt{MeV}$ (Ensemble I) and $\sim 760 \, \texttt{MeV}$ (Ensemble II). 
Ensemble III has been generated with 
$N_f = 2+1$
dynamical light-quark flavors, 
where the strange quark mass is taken at its physical value. 
The isospin-degenerate light quark mass on this ensemble corresponds to a pion mass of 
$\sim 450 \, \texttt{MeV}$. 
Ensemble I has been extensively used to study properties of single- and multi-baryon 
systems~\cite{Beane:2012vq,Beane:2013br,Beane:2014ora,Detmold:2015daa,Chang:2015qxa,Beane:2015yha}, 
while
Ensemble III has been recently detailed in Refs.~\cite{Orginos:2015aya,Beane:2015yha}.

The lattice spacing on each of the ensembles has been determined using quarkonium hyperfine splittings.
\footnote{We thank Stefan Meinel for these determinations.}
The limited statistics on the finer lattice ensemble, 
Ensemble II, which has a  pseudoscalar mass similar to that of Ensemble I,
has been included for an initial investigation of the continuum limit of magnetic moments. 
Further investigation of single- and multi-baryon systems with higher statistics on Ensemble II is left to future work.

%
\begin{table}
\caption{%
Summary of the three ensembles of QCD gauge-field configurations used in this work. 
Further details regarding Ensemble I can be found in Ref.~\cite{Beane:2012vq,Beane:2013br}, 
while Ensemble III has been detailed in Ref.~\cite{Orginos:2015aya}. 
}
\begin{center}
\resizebox{\linewidth}{!}{
\begin{tabular}{|c|cccccccc|}
\hline
\hline
  &  $L / a$ &  $T/ a$ & $\beta$ & $ a m_l$ & $ a m_s $ & $a  [\texttt{fm}]$ & $m_\pi [\texttt{MeV}]$ & $ N_{\text{cfg}}$ 
\tabularnewline
\hline
\hline
I & $32$ & $48$ & $6.1$ & $-0.2450$ & $-0.2450$ & $0.1453(16)$ &  $806.9(8.9) $ & $1006$
\tabularnewline
II & $48$ & $64$ & $6.3$ & $-0.2050$ & $-0.2050$ & $0.1036(11)$ & $766.9(8.1)$ & $94$
\tabularnewline 
III & $32$ & $96$ & $6.1$ & $-0.2800$ & $-0.2450$ & $0.1167(16)$ &  $449.9(4.6)$ & $544$
\tabularnewline
\hline
\hline
\end{tabular}
}
\end{center}
\label{t:configs}
\end{table}

The background magnetic fields are implemented by post-multiplication of the dynamical
$SU(3)$
color gauge links by fixed
$U(1)$
electromagnetic links, 
$U^{(Q)}_\mu(x)$,
having the form
\begin{eqnarray}
U^{(Q)}_1(x)
&=&
\begin{cases}
1 & \text{for } x_1 \neq L - a \\
\exp \left( - i Q \, n_\Phi \frac{2 \pi x_2}{L} \right)
& \text{for } x_1 =  L - a
\end{cases},
\notag \\
U^{(Q)}_2(x) 
&=&
\exp \left( i Q \, n_\Phi \frac{2 \pi a x_1}{L^2} \right)
,\notag \\
U^{(Q)}_3 (x) &=& U^{(Q)}_4(x) = 1
\label{eq:U1}
,\end{eqnarray}
where the integer $n_\Phi$ is the magnetic flux quantum of the torus which satisfies
$| n_\Phi | \leq \frac{1}{4} L^2 / a^2$,  see Ref.~\cite{'tHooft:1979uj}.
Typically, 
this multiplication is carried out individually for each quark flavor due to flavor-symmetry breaking introduced by 
quark mass differences and quark electric charges,   
$Q$,
which appear above in units of the magnitude of the electron's charge, $e >0$. 
Using Eq.~\eqref{eq:U1},  the $U(1)$ flux through each elementary plaquette in the ($\mu$-$\nu$)-plane is identically equal to
$\exp ( i Q e F_{\mu \nu} )$, 
where
\begin{eqnarray}
Q e B_z = \frac{2 \pi}{L^2} n_\Phi 
,\end{eqnarray}  
with
$B_z$ 
 the 
$z$-component of the magnetic field, 
$B_z = F_{12} = - F_{21}$, 
and all other components of the electromagnetic field-strength tensor 
vanish.~\footnote{The electromagnetic gauge links in 
Eq.~\eqref{eq:U1} 
also give rise to two non-trivial holonomies, which
however, 
 are only relevant for quarks propagating around the torus. 
Due to confinement, 
such long-distance effects scale as 
$\sim \exp ( - m_\pi L )$
and are negligible in the present study of magnetic moments;
see Ref.~\cite{Chang:2015qxa,Tiburzi:2014zva} for further details. 
} 
Throughout this work,   the flux quanta
$n_\Phi = 3$, $-6$,  and $12$ are employed. 
The factors of three result from the fractional nature of quark charges in units of $e$, 
and the doubling of flux is employed to economize on the computation of quark propagators. 
For example, the up-quark propagator with  $n_\Phi = 3$ is the same as the down-quark propagator with 
$n_\Phi = -6$.  On Ensembles I and II, 
the latter is identical to the strange-quark propagator due to mass degeneracy. 
This equality of down and strange propagators preserves an 
$SU(2)$ symmetry, commonly called $U$-spin,  
that can be thought of as a rotation in three-dimensional flavor space about the 
up-quark axis
\begin{equation}
\begin{pmatrix}
u \\ d \\ s
\end{pmatrix}
\longrightarrow
\left(
\begin{array}{cc}
  1 & 0 \phantom{s} 0  \\
  0 & \raisebox{-7.5pt}{{\Large\mbox{{$\mathcal{U}$}}}} \\[-4ex]
\\  0 &
\end{array}
\right)
\begin{pmatrix}
u \\ d \\ s
\end{pmatrix}
\label{eq:Uspin}
,\end{equation}  
where 
$\mathcal{U} \in SU(2)$. 

Post-multiplication of the $U(1)$ 
gauge links onto QCD gauge links is an approximation that ignores effects of the electromagnetic field on the sea quarks and, indirectly, 
the gluonic sector. 
In a complete calculation, 
the background electromagnetic field  couples to sea-quark degrees of freedom through the fermionic determinant. 
The present computations should thus be thought of as partially quenched (PQ) due to the omission of such contributions.%
~\footnote{Our computations are not otherwise PQ, because the valence and sea quark masses employed are degenerate.}
Because magnetic moments arise from a response that is linear in the external field, 
however, 
there are cases for which sea-quark contributions vanish. 
Computations of magnetic moments at an 
$SU(3)_F$-symmetric point, 
for example, 
are complete. 
In this case, 
the sea-quark contributions arising from expanding the fermionic determinant to linear order in the external field are necessarily proportional to the sum
$\sum\limits_f Q_f = Q_u + Q_d + Q_s$, 
which vanishes.  
With  
$SU(3)_F$ 
breaking, 
the sum of sea-quark current effects no longer vanishes because contributions from each flavor are no longer identical. 
Decomposing the electromagnetic current into isoscalar and isovector contributions
allows for a separation of these matrix elements into sea-quark charge dependent and sea-quark charge independent terms, 
respectively. 
Thus computations of magnetic moments on Ensembles I and II are complete, 
while only the isovector magnetic moments computed on Ensemble III are complete. 
Omitted contributions to the current from sea quarks on Ensemble III are nonetheless expected to be small,
see, 
e.g.,  
Ref.~\cite{Abdel-Rehim:2015lha}.

To determine QCD energy eigenstates in the presence of external magnetic fields, interpolating
operators are chosen which have the quantum numbers of the octet baryons. 
In particular, 
interpolating operators that have been tuned to produce strong overlap with the ground-state baryons in vanishing magnetic fields are employed. 
The imposition of sufficiently weak magnetic fields should not alter the operator overlaps substantially. 
While field-strength dependent overlaps are observed in practice,
diminished overlaps have not impeded the ground-state saturation of correlations functions. 
For a complete discussion of these points
and further details concerning the smeared-smeared (SS) and smeared-point (SP) correlation functions computed in this work,
see Ref.~\cite{Chang:2015qxa}.~\footnote{A further ingredient is the projection of the correlation functions onto vanishing three momentum. 
For electrically neutral baryons, 
the energy eigenstates remain momentum eigenstates in non-vanishing magnetic fields. 
While the same is not true for charged baryons, 
effects from the tower of Landau levels cancel to a high degree in the ratio of spin-projected correlation functions that are used to determine magnetic moments~\cite{Lee:2014iha}.
}

Consider an octet baryon, denoted by $B$,  that is subject to a constant and  
uniform magnetic field oriented along the $z$-direction, $\bm{B} = B_z \hat{\bm{z}}$. 
The energy eigenvalues of this baryon with its spin polarized in the $z$-direction,
magnetic quantum number $s = \pm \frac{1}{2}$, and zero longitudinal momentum, $p_z = 0$, have the form
\begin{eqnarray}
  E^{(s)}_{B}({B_z}) 
  &=& 
  M_B 
  + 
  \frac{| Q_B e { B_z}|}{M_B} 
  \left( n_L + \frac{1}{2} \right)
   - 2 \mu_B s  B_z
  + \ldots,
  \notag \\
  \label{eq:Eshift}
\end{eqnarray}
where $M_B$ is its mass,  $Q_B$ its charge in units of  $e$, and  $n_L$ 
is the quantum number of the Landau level that it occupies.  
For a spin-$\frac{1}{2}$ baryon, 
there is a structure-dependent contribution from the magnetic moment, 
$\mu_B$, 
that is linear in the magnetic field. 
The ellipses denote contributions that involve two or more powers of the
magnetic field, 
such as that from the magnetic polarizability.
The moments are determined from LQCD computations of Zeeman splittings,  $\Delta E$. 
These are defined to be energy differences between eigenstates of differing spin polarizations
\begin{eqnarray}
\Delta E
&\equiv&
E_{B}^{(+\frac{1}{2})}(\bm{B})
-
E_{B}^{(-\frac{1}{2})}(\bm{B})
\label{eq:Zee} 
,\end{eqnarray}
where the baryon label and magnetic-field dependence of $\Delta E$ are suppressed for notational ease. 
Using the expected magnetic-field dependence of the energy eigenvalues in Eq.~\eqref{eq:Eshift}, 
this reduces to 
\begin{equation}
\Delta E 
=
- 2 \mu_B \, B_z + \ldots
\label{eq:Linear}
,\end{equation}
where the ellipsis represents contributions that are higher order in the magnetic field. 
The procedure used to determine the  magnetic moments relies on the precise determination of the Zeeman splittings in 
Eq.~\eqref{eq:Zee}
from ratios of spin-projected correlation functions, 
and subsequent extrapolation to vanishing magnetic field using the expectation in 
Eq.~\eqref{eq:Linear}. 
A detailed description of the analysis is relegated to  Appendix~\ref{s:A}.

\section{Baryon Magnetic Moments}
\label{s:magmom}

The magnetic moments of the octet baryons are determined from LQCD calculations performed in background 
magnetic fields, using the procedures 
detailed in Appendix~\ref{s:A}.  As with any LQCD calculation, the results are  dimensionless quantities, 
made so by compensating powers of the lattice spacing.
In what follows,  
conversions of these results into units that can be compared with experiment are discussed, 
and various features, including the 
values of anomalous magnetic moments, 
pion-mass dependence, 
lattice-spacing dependence, 
are discussed.

\subsection{Units for Magnetic Moments}
\label{s:units}

%
%
%
\begin{figure}[t!]
\resizebox{0.9\linewidth}{!}{
\includegraphics{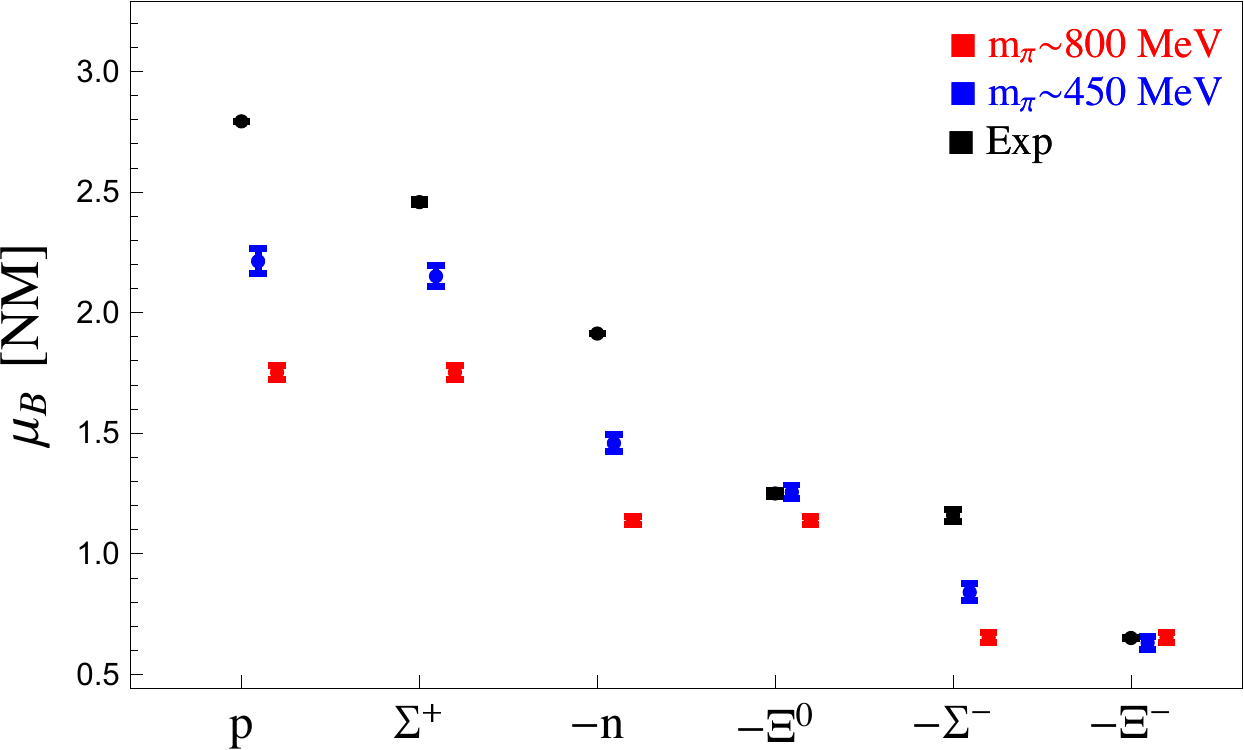}
}
\vskip 0.2in
\resizebox{0.9\linewidth}{!}{
\includegraphics{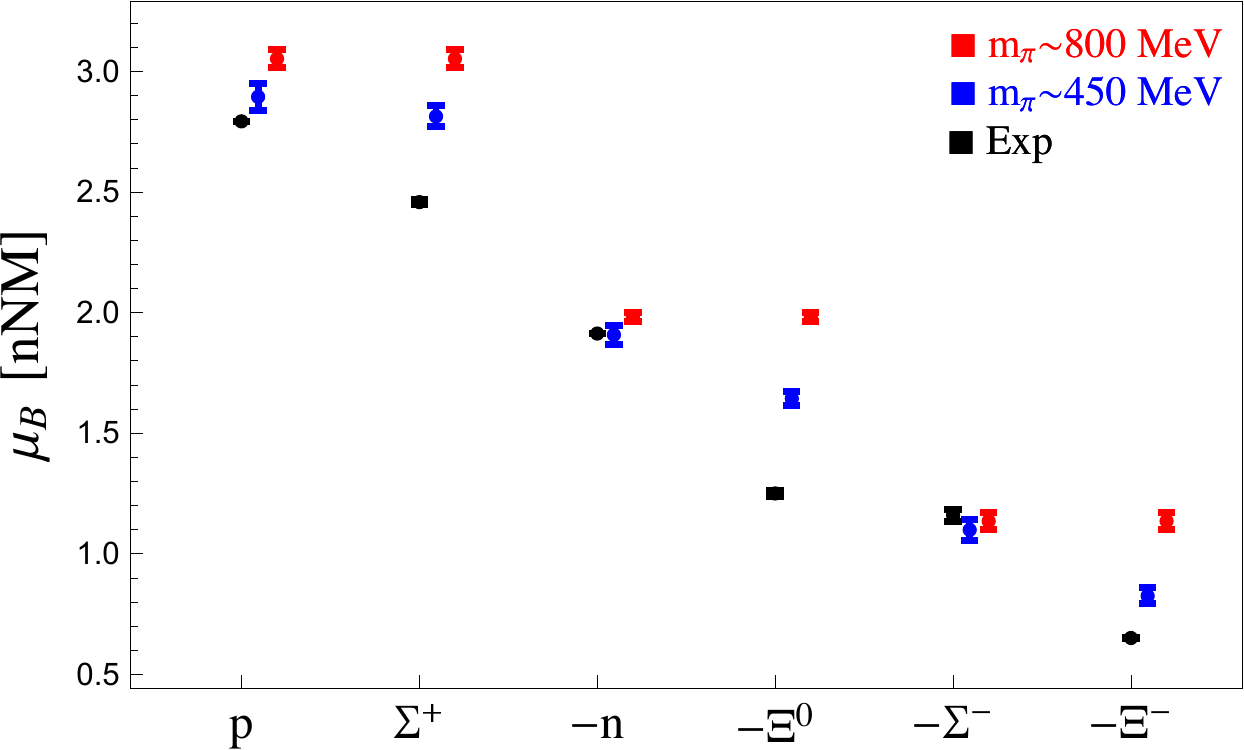}
}
\vskip 0.2in
\resizebox{0.9\linewidth}{!}{
\includegraphics{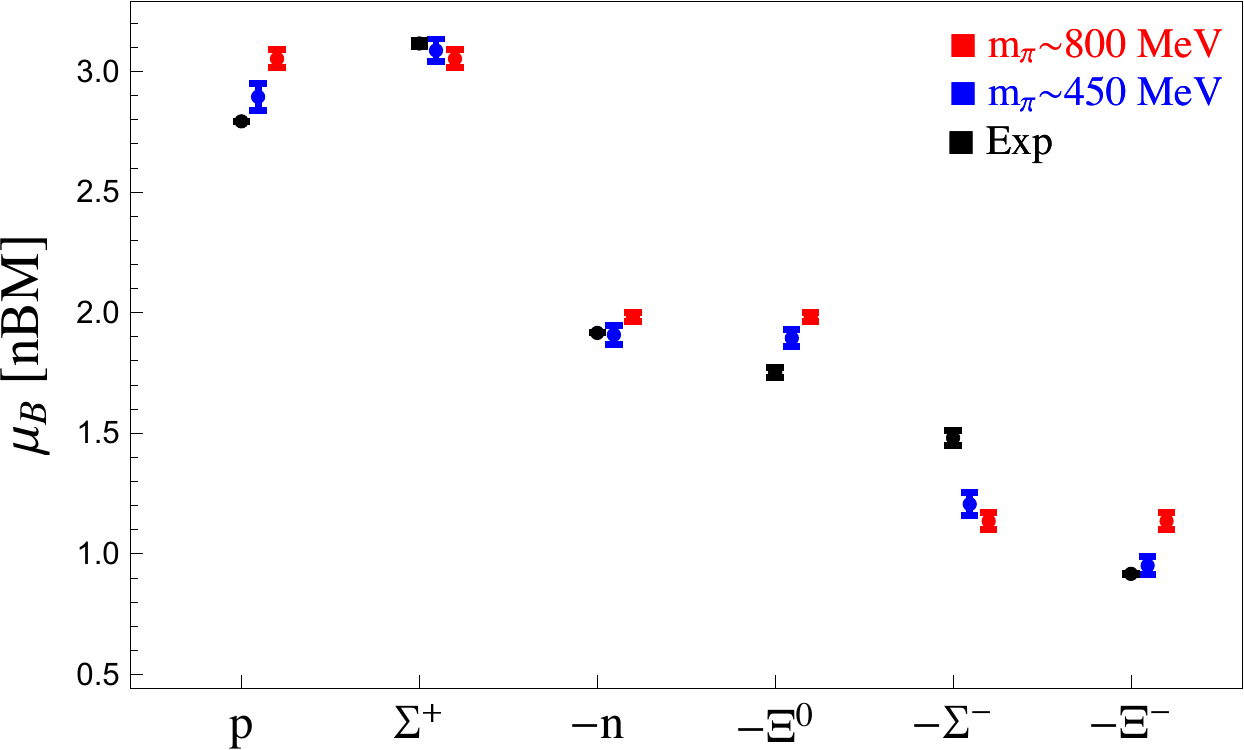}
}
\caption{
Magnetic moments of the octet baryons determined from LQCD calculations at 
$m_\pi \sim 800 \, \texttt{MeV}$ (Ensemble I) and
$m_\pi \sim 450 \, \texttt{MeV}$ (Ensemble III), along with their experimental values. 
The quark-disconnected contributions to the magnetic moments 
at  $m_\pi \sim 450 \, \texttt{MeV}$ are not included,
and they vanish by $SU(3)_F$ symmetry 
at 
$m_\pi \sim 800 \, \texttt{MeV}$.
Comparisons with the experimental values are made in units of 
$\texttt{[NM]}$ in the upper panel, 
$\texttt{[nNM]}$ in the middle panel, and 
$\texttt{[nBM]}$ in the lower panel.
The notation ``$-B$'' indicates the negative value of the moment, 
i.e. $\mu_{-B}  \equiv - \mu_B$, 
so that all displayed quantities are positive.  
The  uncertainties of the  LQCD results reflect quadrature-combined statistical and systematic uncertainties.
The values of the  moments in
$\texttt{[NM]}$ 
and 
$\texttt{[nNM]}$
are given in  Table~\ref{t:bigtable}  in Appendix~\ref{s:A}, 
while those in 
$\texttt{[nBM]}$ follow  from applying Eq.~\eqref{eq:anomalous} 
to the results appearing in Table~\ref{t:nBM}. 
}
\label{f:Mplot}
\end{figure}
%
%
%

The magnetic moment of an octet baryon, $B$,  can be described using units of 
{\it baryon magnetons} or {\it natural baryon magnetons},  which are defined by
\begin{eqnarray}
\texttt{[BM]}
&=&
\frac{e}{2 \mathcal{M}_B}, 
\quad
\texttt{[nBM]} 
=
\frac{e}{2 M_{B} (m_\pi)}
\label{eq:BM}
,\end{eqnarray}
respectively,
where $\mathcal{M}_B$
is the experimentally-measured mass of the baryon,  and 
$M_B (m_\pi)$ is the mass of the baryon computed with LQCD
(which depends on the input light quark mass through the lattice-determined value of the pion mass,  $m_\pi$). 
From a phenomenological point of view, 
it is conventional to use 
nuclear magnetons, 
$\texttt{[NM]}$,
for all baryons, 
for which we also define the corresponding 
\emph{natural nuclear magnetons},
$\texttt{[nNM]}$. 
These are simply the special cases with 
$B = N$ 
of the above units, 
namely
\begin{eqnarray}
\texttt{[NM]}
&=&
\frac{e}{2 \mathcal{M}_N}, 
\quad
\texttt{[nNM]} 
=
\frac{e}{2 M_{N} (m_\pi)}
\label{eq:nNM}
.
\end{eqnarray}
Such units proved advantageous in our studies of magnetic moments of light nuclei%
~\cite{Beane:2014ora,Chang:2015qxa,Beane:2015yha}.

%
\begin{table}
\caption{%
Baryon anomalous magnetic moments,  $\delta \mu_B$, 
in $\texttt{[nBM]}$, Eq.~\eqref{eq:BM}.  
The first uncertainty  is statistical, 
while the second is the fitting systematic including that from the choice of fit functions. 
Ensemble I necessarily maintains exact 
$U$-spin symmetry, leading to repeated entries. 
Experimental values derived from Ref.~\cite{Agashe:2014kda}
are given in $\texttt{[BM]}$. 
}
\begin{center}
\begin{tabular}{|c|cc|c|}
\hline
\hline
& 
\multicolumn{2}{c|}{$\delta \mu _B\, \texttt{[nBM]}$}
&
$\delta \mu_B \, \texttt{[BM]}$
\\
$B$
& 
I 
& 
III
& 
Experiment~\cite{Agashe:2014kda}
\tabularnewline
\hline
\hline
$p$
&
$\phantom{-} 2.052(14)(34)$
&
$\quad
\phantom{-} 1.895(22)(51)
\quad$
&
$\phantom{-} 1.7929(0)$
\tabularnewline
$\Sigma^+$
&
$\phantom{-} 2.052(14)(34)$
&
$\phantom{-} 2.087(18)(44)$
&
$\phantom{-} 2.116(13)$
\tabularnewline
$n$
&
$-1.982(03)(19)$
&
$-1.908(08)(37)$
&
$-1.9157(0)$
\tabularnewline
$\Xi^0$
&
$-1.982(03)(19)$
&
$-1.894(10)(33)$
&
$-1.752(20)$
\tabularnewline
$\Sigma^-$
&
$-0.136(14)(32)$
&
$-0.206(21)(43)$
&
$-0.480(32)$
\tabularnewline
$\Xi^-$
&
$-0.136(14)(32)$
&
$\phantom{-} 0.049(16)(34)$
&
$\phantom{-5}  0.0834(35)$
\tabularnewline
\hline
\hline
\end{tabular}
\end{center}
\label{t:nBM}
\end{table}

%
%
%
\begin{figure}
\resizebox{0.9\linewidth}{!}{
\includegraphics{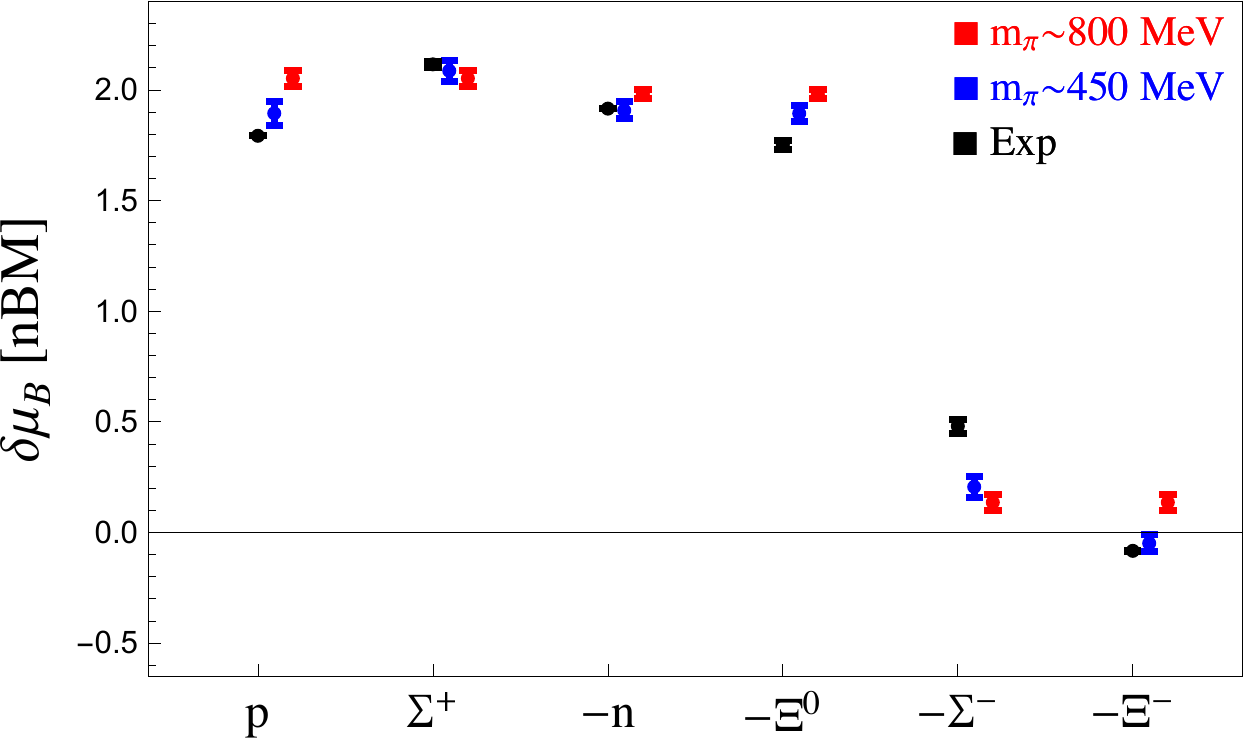}
}
\caption{
The  anomalous magnetic moments 
of the octet baryons
in $\texttt{[nBM]}$ compared with experiment in $\texttt{[BM]}$.
The shorthand notation $\mu_{-B} \equiv - \mu_{B}$ is used for display purposes.
The  $\Sigma^-$ and $\Xi^-$ baryons  magnetically behave close to point-like Dirac particles. 
The non point-like structure of the remaining baryons is approximately the same
(up to clockwise versus counter-clockwise circulation of current). 
}
\label{f:BMplot}
\end{figure}
%
%
%

To convert magnetic moments from lattice magneton units to $\texttt{[NM]}$, 
 they are multiplied by $a \mathcal{M}_N$, requiring knowledge of the lattice spacing.
 The results of our LQCD calculations of the octet baryon magnetic moments in 
 $\texttt{[NM]}$ are given in Table~\ref{t:bigtable} in Appendix~\ref{s:A}, 
and are shown in Fig.~\ref{f:Mplot}, along with their experimental values.
In these units, considerable pion-mass dependence is  generally observed but, curiously, 
the magnetic moments of the $\Xi$ baryons appear relatively insensitive. 
This situation changes somewhat when the moments are instead converted into 
$\texttt{[nNM]}$.
Results in these units are also shown in Fig.~\ref{f:Mplot}, 
and are obtained by multiplying the lattice magneton values by 
$a M_N (m_\pi)$. 
Note that this does not introduce scale-setting uncertainties.  
The situation clarifies even further when using
$\texttt{[nBM]}$, 
for which the lattice magneton values are multiplied by $a M_B(m_\pi)$. 
These values are also shown in Fig.~\ref{f:Mplot}, 
and can be obtained from Table~\ref{t:nBM}.
Magnetic moments expressed in $\texttt{[nBM]}$ show the mildest pion-mass dependence, 
moreover, 
they are close to the experimental values, even at large quark masses.

Another salient feature of  the $\texttt{[nBM]}$ units is that 
 the Dirac contribution to the magnetic moment can be readily subtracted,
 leaving the anomalous magnetic moment,
\begin{equation}
\delta \mu_B
\, \texttt{[nBM]}
=
\mu_B \, \texttt{[nBM]} - Q_B
\label{eq:anomalous} 
,
\end{equation}
which vanishes for a point-like particle. 
The Dirac moment is a short-distance contribution to the magnetic moment, and in our LQCD calculations 
it is fixed through the implementation of the external field through link variables. 
This emerges from the lattice Ward-Takahashi identity,  
because the corresponding electromagnetic current is the conserved point-split current.  
Thus non-vanishing anomalous magnetic moments provide a more direct probe of bound-state structure.
The values are given in 
Table~\ref{t:nBM}, 
and are shown graphically in 
Fig.~\ref{f:BMplot}.
On the scale of fractions of 
$\texttt{[nBM]}$, 
we strikingly see  anomalous magnetic moments only having values
$\delta \mu_B \sim \pm 2$
and 
$\delta \mu_B \sim 0$, 
for all six baryons with $I_3\ne0$ (the $\Lambda$ and $\Sigma^0$ will be discussed later).  
The latter value is approximately attained for both the
$\Sigma^-$ 
and 
$\Xi^-$
baryons, 
and suggests that their magnetic structure deviates very little from point-like particles. 
These striking features in  Fig.~\ref{f:BMplot}  will be subsequently linked to the NRQM, 
and discussed in the context of the large-$N_c$ limit of QCD.

\subsection{Extrapolations}
\label{s:extrap}


\subsubsection{Pion-Mass Extrapolation}

The LQCD results reveal rather mild pion-mass dependence of the baryon magnetic moments when 
given in $\texttt{[nBM]}$. 
As a result, rudimentary extrapolations of the LQCD results to the physical pion mass are attempted.
\footnote{%
Note that with the limited data sets available a combined pion mass and continuum extrapolation is not practicable. 
} 
Due to missing sea-quark contributions on Ensemble III, 
the isovector and isoscalar magnetic moments are extrapolated separately;
and, 
as the Dirac contribution is free from pion-mass dependence in $\texttt{[nBM]}$, 
only the anomalous parts of the isovector and isoscalar magnetic moments are extrapolated.  
We assume the dependence on the pion mass is quadratic, 
and perform extrapolations using 
\begin{eqnarray}
\delta \mu^I_B  (m_\pi^2)
&=&   
\delta \mu^I_B  (0)
+
A^I_B \, \, m^2_\pi 
\label{eq:Xtrap}
,\end{eqnarray}
%

\begin{widetext}

%
\begin{table}
\caption{%
Rudimentary extrapolation of anomalous magnetic moments, 
$\delta \mu_B$,  
to the physical pion mass. 
The first uncertainties are statistical, 
while the second uncertainties are from systematics.
The extrapolated values are obtained from assuming quadratic dependence on the pion mass, 
and compared with the experimentally determined values.
For each moment, 
we extrapolate LQCD results from Ensembles I and III to the physical pion mass, 
as well as results from Ensembles II and III, 
which have quite similar lattice spacings.}
\begin{center}
\resizebox{0.66\linewidth}{!}{
\begin{tabular}{|c|ccc|c|c|}
\hline
\hline
$\delta \mu _B\, \texttt{[nBM]}$
&   I  &  II & III 
&
Extrapolated
&
Experiment $\texttt{[BM]}$
\\
\hline
\hline
$p-n$
&
$\phantom{-} 4.034(15)(40)$
&
&
$\phantom{-} 3.802(25)(67)$
&
$\phantom{-} 3.71(12) \phantom{8}$
&
$\phantom{-} 3.71(00) \phantom{8}$
\tabularnewline
&
&
$\phantom{-} 3.70(07)(15) \phantom{8}$
&
$\phantom{-} 3.802(25)(67)$
&
$\phantom{-} 3.85(18) \phantom{8}$
&
$\phantom{-} 3.71(00) \phantom{8}$
\tabularnewline
\hline
$\Sigma^+ -\Sigma^-$
&
$\phantom{-} 2.188(21)(48)$
&
&
$\phantom{-} 2.293(28)(63)$
&
$\phantom{-} 2.34(12) \phantom{8}$
&
$\phantom{-} 2.60(05) \phantom{8}$
\tabularnewline
&
&
$\phantom{-} 1.91(08)(17) \phantom{8}$
&
$\phantom{-} 2.293(28)(63)$
&
$\phantom{-} 2.47(20) \phantom{8}$
&
$\phantom{-} 2.60(05) \phantom{8}$
\tabularnewline
\hline
$\Xi^0 - \Xi^-$
&
$- 1.846(14)(35)$
&
&
$- 1.943(18)(46)$
&
$- 1.983(88)$
&
$- 1.835(23)$
\tabularnewline
&
&
$- 1.784(30)(74)$
&
$- 1.943(18)(46)$
&
$- 2.02(11) \phantom{8}$
&
$- 1.835(23)$
\tabularnewline
\hline
\hline
$p+n$
&
$\phantom{-} 0.071(14)(38)$
&
&
$- 0.013(22)(59)$
&
$- 0.05(11) \phantom{8}$
&
$-0.12(00) \phantom{8}$
\tabularnewline
&
&
$\phantom{-} 0.02(07)(14) \phantom{8}$
&
$- 0.013(22)(59)$
&
$- 0.03(17) \phantom{8}$
&
$-0.12(00) \phantom{8}$
\tabularnewline
\hline
$\Sigma^+ + \Sigma^-$
&
$\phantom{-}1.917(19)(44)$
&
&
$\phantom{-} 1.881(27)(60)$
&
$\phantom{-} 1.87(12) \phantom{8}$
&
$\phantom{-} 1.64(05) \phantom{8}$
\tabularnewline
&
&
$\phantom{-}1.80(07)(15) \phantom{8}$
&
$\phantom{-} 1.881(27)(60)$
&
$\phantom{-} 1.92(17) \phantom{8}$
&
$\phantom{-} 1.64(05) \phantom{8}$
\tabularnewline
\hline
$\Xi^0 +\Xi^-$
&
$-2.117(15)(39)$
&
&
$-1.845(19)(50)$
&
$-1.73(10) \phantom{8}$
&
$-1.67(02) \phantom{8}$
\tabularnewline
&
&
$-1.896(30)(77)$
&
$-1.845(19)(50)$
&
$-1.82(12) \phantom{8}$
&
$-1.67(02) \phantom{8}$
\tabularnewline
\hline
\hline
\end{tabular}
}
\end{center}
\label{t:nBM_mpi}
\end{table}

\end{widetext}

\noindent
for each of the baryon isospin multiplets, 
$B = N$, 
$\Sigma$,
and
$\Xi$. 
The superscript  $I$ is used to reflect that separate extrapolations are performed for the anomalous 
part of isovector and isoscalar moments.~\footnote{%
In the case of isovector magnetic moments, 
linear pion-mass dependence emerges when one considers 
$SU(2)$ 
chiral corrections to the baryon magnetic moments about the 
$m_\pi = 0$  limit. 
While the isoscalar magnetic moments receive  
$m_\pi^2$ 
corrections about the  
$SU(2)$  
chiral limit, see, e.g., 
Ref.~\cite{Bernard:1992qa},
the relevant symmetry group for our  LQCD calculation at 
$m_\pi\sim 450~{\rm MeV}$ is 
$SU(4|2)$ 
due to the vanishing electric charges of sea quarks.
Expanding  (valence)  isoscalar magnetic moments about the  
$SU(4|2)$ chiral limit, 
gives rise to linear pion-mass dependence~\cite{Beane:2002vq}.  
The choice of a linear extrapolation is not well motivated by such considerations, 
because the present calculations are far from the chiral limit.
Adopting a linear Ansatz, 
however, 
leads to consistent results for extrapolated values, 
albeit with somewhat increased uncertainties. 
}

\begin{figure}
\resizebox{0.9\linewidth}{!}{
\includegraphics{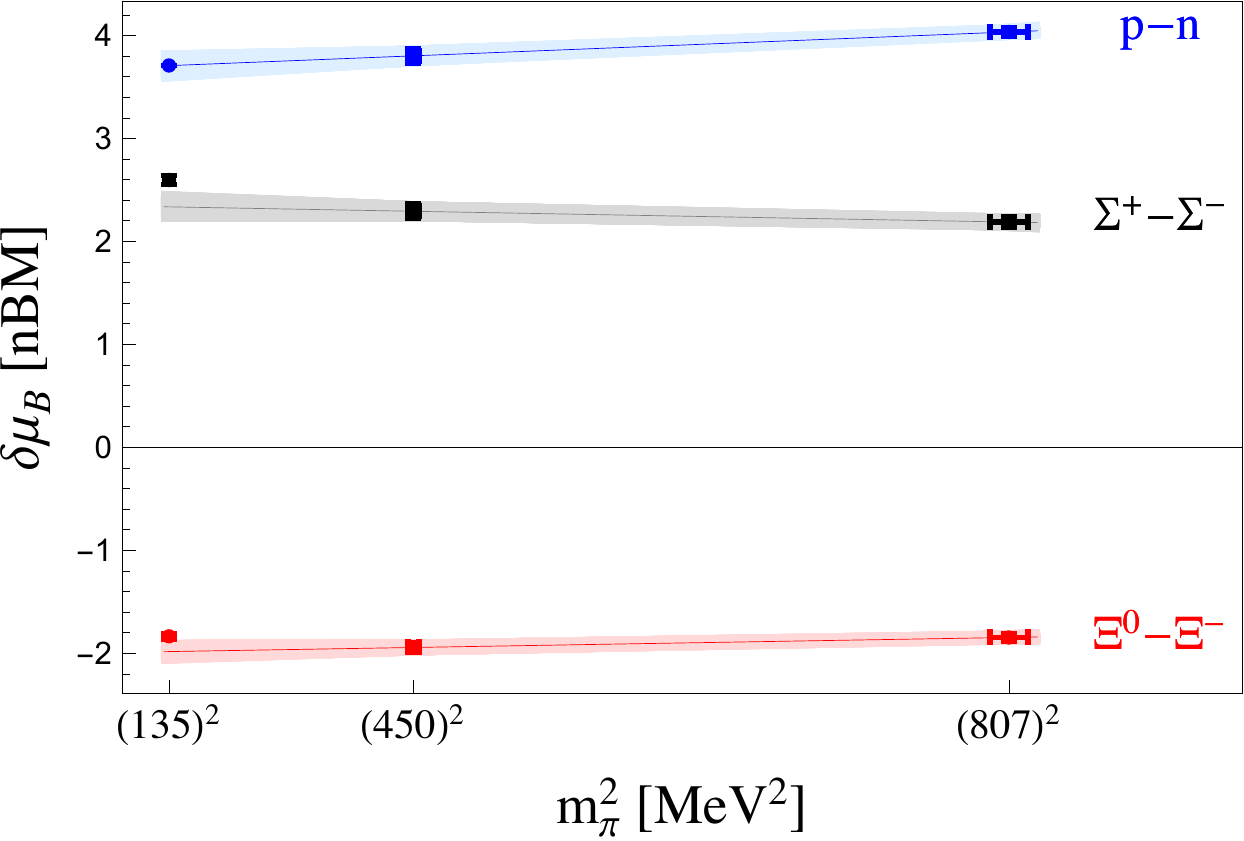}
}
\vskip 0.2in
\resizebox{0.9\linewidth}{!}{
\includegraphics{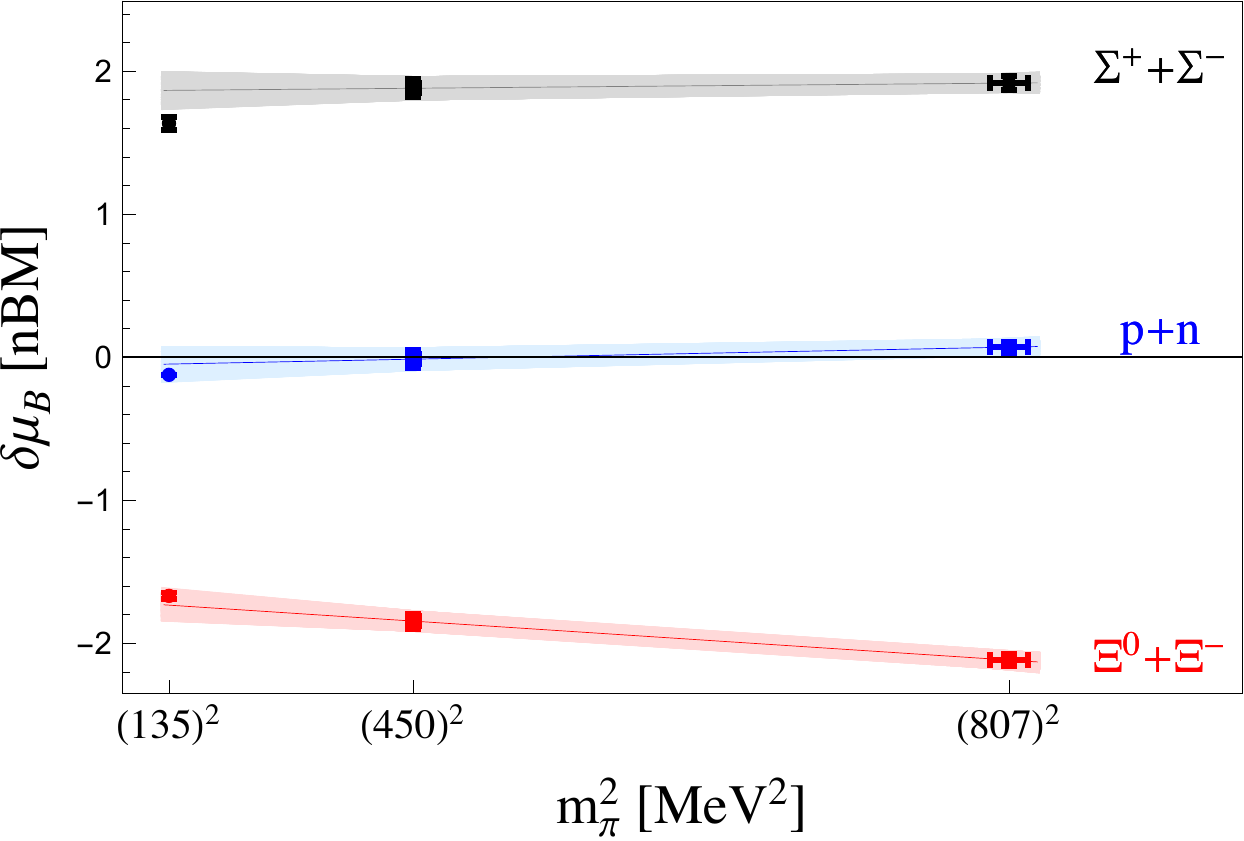}
}
\caption{
Quadratic pion-mass extrapolation of the anomalous part of the isovector and isoscalar magnetic moments from 
Ensembles I and III 
in $\texttt{[nBM]}$.
Isovector magnetic moments are free of quark-disconnected contributions;
subtracting the Dirac part does not change this because it arises solely from valence quarks. 
With 
$SU(3)_F$
breaking, 
the isoscalar moments on Ensemble III require disconnected contributions that  have not been determined.
Removing the Dirac part, 
moreover, 
makes the resulting moments more sensitive to these missing contributions.   
The shorthand
$\mu_{A \pm B} \equiv \mu_A \pm \mu_B$
for sums and differences of the baryon magnetic moments is used. 
Experimental values 
are given in 
$\texttt{[BM]}$,
and have not been included in any of these fits.}
\label{f:IMplot}
\end{figure}

Using the magnetic moments determined on Ensembles I--III, 
quadratic extrapolations to the physical pion mass, 
as in 
Eq.~\eqref{eq:Xtrap},
are performed for each anomalous isovector and isoscalar moment,
with the results given in 
Table~\ref{t:nBM_mpi}.
The extrapolations are performed using pairs of ensembles. 
Ensembles I and III represent the highest precision calculations, 
while
Ensembles II and III have the most similar lattice spacings. 
Despite the simplicity of the fits, 
the extrapolated values agree with experiment within uncertainties,
as exemplified in 
Fig.~\ref{f:IMplot}. 
The uncertainties on extrapolated values are obtained by resampling the 
LQCD results uniformly within 
$\pm 1\sigma$ 
and re-fitting.
While there appears to be a systematic trend in the extrapolated values using LQCD results from 
Ensembles II and III, 
namely they are larger in magnitude than those obtained from
Ensembles I and III, 
the level of precision currently prevents a definitive conclusion from being drawn.

The success of these simple pion-mass extrapolations is perhaps suggestive of an underlying expansion scheme. 
Rescaling the magnetic moments to
$\texttt{[nBM]}$ 
seems to account for most of the quark-mass dependence within each 
$U$-spin multiplet. 
The residual 
$U$-spin breaking might then be perturbative, 
depending on powers of the quark-mass difference
$m_d - m_s$.
The pion mass-squared extrapolations account for one insertion of this 
$U$-spin breaking quark-mass operator, 
but, 
with the assumption that the pion mass-squared remains linear in the quark mass. 
With the fixed strange-quark mass in our calculations, 
however, 
such a $U$-spin expansion cannot presently be tested. 
It would be quite interesting to study 
$U$-spin breaking in magnetic moments using the quark-mass tuning scheme of 
Refs.~\cite{Bietenholz:2010jr,Bietenholz:2011qq}, 
in which the singlet quark mass is held fixed. 
For this approach, 
polynomial 
$U$-spin breaking could be tested due to the variable light- and strange-quark masses, 
and without the rather long pion-mass extrapolation required here.

%
%
%
%
%
%

\subsubsection{Continuum Extrapolation}

The continuum limit of the magnetic moments can be investigated from the values computed on Ensembles I and II. 
The small difference $\sim 5\%$ in pion mass is expected to be entirely negligible for magnetic moments expressed in 
$\texttt{[nBM]}$. 
Bearing in mind the reduced statistics on Ensemble II, 
the magnetic moments in units of  
$\texttt{[nBM]}$   
are compared in  Table~\ref{t:nBM_a}  for Ensembles I and II.
Notice that the extracted magnetic moments from Ensemble II have statistical uncertainties which are 
$2$--$4$ times larger than those from Ensemble I. 
This scaling is  consistent with the differing sizes of the ensembles.

\begin{figure}
\resizebox{0.9\linewidth}{!}{
\includegraphics{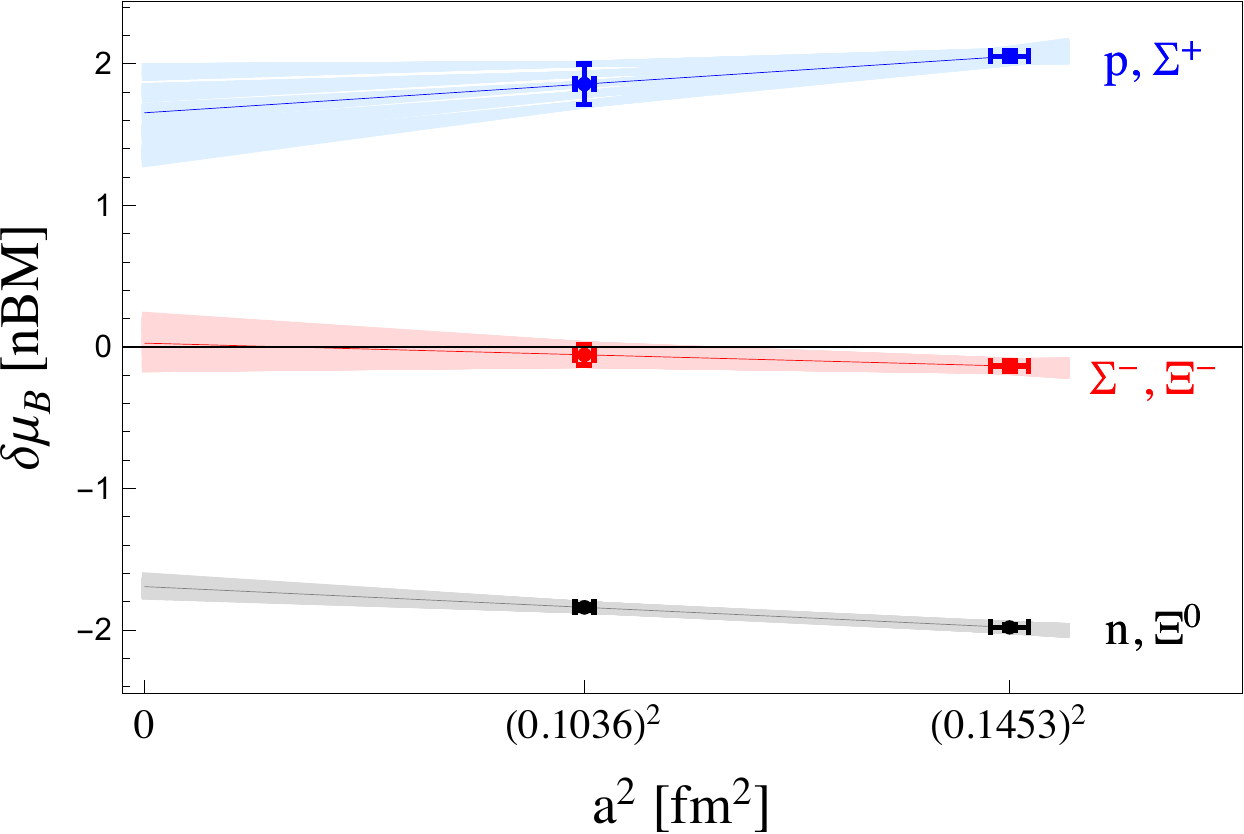}%
}
\caption{
Illustration of the lattice-spacing dependence of baryon anomalous magnetic moments from Table~\ref{t:nBM_a}. 
}\label{f:aplot}
\end{figure}

While the fermion action has only been perturbatively improved, 
with corrections na\"ively scaling as $\cO( \alpha_s^2 \, a)$,  
the value of the clover coefficient with tadpole improvement is 
consistent with that obtained from non-perturbative  $\cO(a)$ improvement, 
effectively leaving 
$\cO(a^2)$
uncertainties. 
Because the vector current is implemented through the link fields, Eq.~(\ref{eq:U1}),  it inherits the same level of discretization effects. 
Thus the magnetic moments are assumed to have quadratic dependence on the lattice spacing near the continuum limit, 
of the form
\begin{equation}
\delta \mu_B(a) =  \delta \mu_B(0) +  C_B \, a^2
\label{eq:asq}
.
\end{equation}
Results of continuum extrapolations using  Eq.~\eqref{eq:asq} are given in Table~\ref{t:nBM_a}, 
and  shown in  Fig.~\ref{f:aplot}. 
With the Dirac contributions removed, 
the anomalous magnetic moments should be more sensitive to the lattice spacing,
but it is found that the continuum extrapolated values for the charged baryons are consistent 
with those computed on the coarse ensemble.

%
\begin{table}
\caption{%
Lattice-spacing dependence of baryon anomalous magnetic moments, 
$\delta \mu_B$, 
determined in $\texttt{[nBM]}$, 
Eq.~\eqref{eq:BM}, 
at a pion mass of 
$m_\pi \sim 800 \, \texttt{MeV}$.
The first uncertainty quoted is statistical;
the second is systematic, while the uncertainty on the extrapolated values
combines the statistical and systematic uncertainties in quadrature.
}
\begin{center}
\resizebox{\linewidth}{!}{
\begin{tabular}{|c|cc|c|}
\hline
\hline
& 
\multicolumn{3}{|c|}{$\delta \mu _B\, \texttt{[nBM]}$}
\\
&  I  &  II  &  Extrapolation
\tabularnewline
\hline
\hline
$p, \Sigma^+$
&
$\phantom{-} 2.052(14)(34)$
&
$\quad
\phantom{-} 1.86(07)(13) \phantom{0}
\quad$
&
$\phantom{-} 1.67(34) \phantom{0}$
\tabularnewline
$n, \Xi^0$
&
$-1.982 (03)(19)$
&
$-1.840(10)(19)$
&
$- 1.705(76)$
\tabularnewline
$\Sigma^-, \Xi^-$
&
$-0.136(14)(32)$
&
$-0.056(28)(67)$
&
$ - 0.02(19) \phantom{1}$
\tabularnewline
\hline
\hline
\end{tabular}
}
\end{center}
\label{t:nBM_a}
\end{table}

The magnetic moment of the  $U$-spin triplet containing the neutron and $\Xi^0$
exhibits the strongest lattice-spacing dependence in absolute terms. 
The difference between magnetic moments on the coarse ensemble and the continuum-extrapolated value is 
relatively large
and the coarse result is more than $3\sigma$ from the extrapolated result. 
The anomalous magnetic moment of the 
$U$-spin doublet consisting of 
$\Sigma^-$
and
$\Xi^-$
baryons, 
however,
exhibits the greatest 
\emph{relative} 
change because the values are quite small and the extrapolated result is consistent with zero. 
This is surprising and suggests that the deviation from point-like magnetic moments computed on 
Ensembles I and II 
could just be a lattice-spacing artifact. 
Better statistics and computations at an additional lattice spacing are needed to support this conclusion.
It would additionally be interesting to compute the magnetic form factors of these baryons, 
in order to understand the distributions of charged currents that ultimately give rise to magnetic moments that are close to those of point-like particles.

%
%
%
%
%
%

\section{Magnetic Moment Relations}
\label{s:relations}

The curious pattern of baryon anomalous magnetic moments exhibited in 
Fig.~\ref{f:BMplot}
motivates further investigation of the relations between them. 
An examination of the  deviations from the Coleman-Glashow relations  
leads us to consider relations between the magnetic moments that hold in the NRQM and/or in the
large-$N_c$ limit of QCD, and deviations therefrom.

\subsection{$SU(3)_F$ Symmetry and the 
Coleman-Glashow Relations}
\label{s:CG}

In the limit of  $SU(3)_F$  symmetry, 
the lightest spin-half baryons form an octet, 
where the states are conventionally embedded as 
\begin{equation}
B_i \, {}^j
= 
\begin{pmatrix}
\frac{1}{\sqrt{2}} \Sigma^0 + \frac{1}{\sqrt{6}} \Lambda
& 
\Sigma^+
& 
p
\\
\Sigma^-
& 
- \frac{1}{\sqrt{2}} \Sigma^0 + \frac{1}{\sqrt{6}} \Lambda
& 
n
\\
\Xi^-
&
\Xi^0
&
- \frac{2}{\sqrt{6}} \Lambda
\end{pmatrix}_{i}^j
,
\end{equation} 
which  transforms as  $B \to V B V^\dagger$
under a transformation parametrized by 
$V \in SU(3)_F$. 
Electromagnetic interactions break the  
$SU(3)_F$ 
symmetry due to the different quark electric charges, 
which appear in the matrix 
\begin{equation}
Q_i \, {}^j 
= 
\text{diag} \left( \frac{2}{3}, - \frac{1}{3}, - \frac{1}{3} \right)_{i}^j
\label{eq:Q}
.
\end{equation} 
As a result,  
the baryon magnetic moment operators, 
which contain one insertion of  $Q$,  are not  $SU(3)_F$
invariant. 
Such symmetry breaking is most easily accounted for by promoting the charge matrix to a spurion field 
transforming as $Q \to V Q V^\dagger$, 
forming invariant operators using this field,  and then allowing  $Q$ to pick up the value in 
Eq.~\eqref{eq:Q}.

With
$SU(3)_F$ symmetry
there are only two independent magnetic moment operators in the Hamiltonian density
\begin{equation}
\mathcal{H}
=
-
\frac{e \, \bm{\sigma} \cdot \bm{B}}
{2 M_B}
\Big[
\mu_D
\left\langle \overline{B} \{ Q, B \} \right\rangle
+
\mu_F
\left\langle \overline{B} [ Q, B ] \right\rangle
\Big]
,
\label{eq:muDmuF} 
\end{equation}
where the angled brackets denote the trace over $SU(3)_F$
indices,  namely
$\langle A \rangle \equiv A_i {}^i$. 
For the six octet baryons with  $I_3 \neq 0$, 
there are  four relations between their magnetic moments resulting from this 
Hamiltonian density.
The remaining two baryons with  $I_3 = 0$  will be discussed  in  Sec.~\ref{s:LamSig}. 
Magnetic moment relations which emerge from  Eq.~\eqref{eq:muDmuF}
were first obtained by  Coleman and Glashow~\cite{Coleman:1961jn},  
and should describe exactly the LQCD results obtained on the 
$SU(3)_F$-symmetric ensembles.  
From Eq.~\eqref{eq:muDmuF},  
there are three  $U$-spin symmetry relations, 
see Eq.~\eqref{eq:Uspin},  
which dictate the equalities
\begin{eqnarray}
\mu_p 
&=&  
\mu_{\Sigma^+} 
,
\quad
\mu_n 
=
\mu_{\Xi^0}
,
\quad 
\text{and}
\quad
\mu_{\Sigma^-} 
= 
\mu_{\Xi^-}
.
\end{eqnarray}
The correlation functions from which these moments are extracted satisfy analogous relations configuration-by-configuration
 on Ensembles I and II. 
 Additionally, there is the non-trivial constraint
\begin{equation}
\mu_p + \mu_n + \mu_{\Sigma^-}
=
0
,
\label{eq:SU3check}
\end{equation}
that emerges on Ensembles I and II  after averaging over gauge configurations 
and is a useful check of the lattice results. 
As there is additional $SU(3)_F$ breaking due to quark mass differences on Ensemble III, as well as in nature, 
we investigate the size of deviations from the Coleman-Glashow relations by computing sums and differences of magnetic moments that vanish in the 
$SU(3)_F$-symmetric limit. 
Results are tabulated in  Table~\ref{t:CGnNM}.

%
\begin{table}
\caption{
The sums and differences of 
magnetic moments  in units of $\texttt{[nNM]}$ and $\texttt{[nBM]}$
that vanish in the 
$SU(3)_F$-symmetric limit.
The first uncertainty  is statistical, while the second is systematic. 
The abbreviation ``C-P'' indicates the sum of the six baryon magnetic moments in 
Eq.~\eqref{eq:CP}.
}
\begin{center}
\begin{tabular}{|c|cc|c|}
\hline
\hline
& 
\multicolumn{2}{c|}{$\mu _B\, \texttt{[nNM]}$}
&
$\mu_B \, \texttt{[NM]}$
\\
& I &  III &  Experiment
\tabularnewline
\hline
\hline
$p-\Sigma^+$
&
$0$
&
$\phantom{-} 0.081(15)(34)$
&
$\phantom{-} 0.33(1) \phantom{34}$
\tabularnewline
$\Xi^0 -n$
&
$0$
&
$\phantom{-} 0.264(10)(41)$
&
$\phantom{-} 0.663(14)$
\tabularnewline
$\Xi^- - \Sigma^-$
&
$0$
&
$\phantom{-} 0.274(20)(42)$
&
$\phantom{-} 0.509(26)$
\tabularnewline
$p + n + \Sigma^-$
&
$- 0.065(20)(49)$
&
$- 0.112(29)(72)$
&
$- 0.280(25)$
\tabularnewline
C-P
&
$- 0.065(20)(49)$
&
$\phantom{-} 0.116(22)(54)$
&
$\phantom{-} 0.139(26)$
\tabularnewline
\hline
\hline
& 
\multicolumn{2}{c|}{$\mu _B\, \texttt{[nBM]}$}
&
$\mu_B \, \texttt{[BM]}$
\\
& I &  III  &  Experiment
\tabularnewline
\hline
\hline
$p-\Sigma^+$
&
$0$
&
$- 0.192(15)(34)$
&
$- 0.323(13)$
\tabularnewline
$\Xi^0 -n$
&
$0$
&
$\phantom{-} 0.014(11)(43)$
&
$\phantom{-} 0.164(20)$
\tabularnewline
$\Xi^- - \Sigma^-$
&
$0$
&
$\phantom{-} 0.255(23)(47)$
&
$\phantom{-} 0.564(35)$
\tabularnewline
$p + n + \Sigma^-$
&
$- 0.065(20)(49)$
&
$- 0.219(31)(74)$
&
$- 0.603(32)$
\tabularnewline
C-P
&
$- 0.065(20)(49)$
&
$\phantom{-} 0.011(24)(58)$
&
$- 0.078(34)$
\tabularnewline
\hline
\hline
\end{tabular}
\end{center}
\label{t:CGnNM}
\end{table}

\begin{figure}
\resizebox{0.9\linewidth}{!}{
\includegraphics{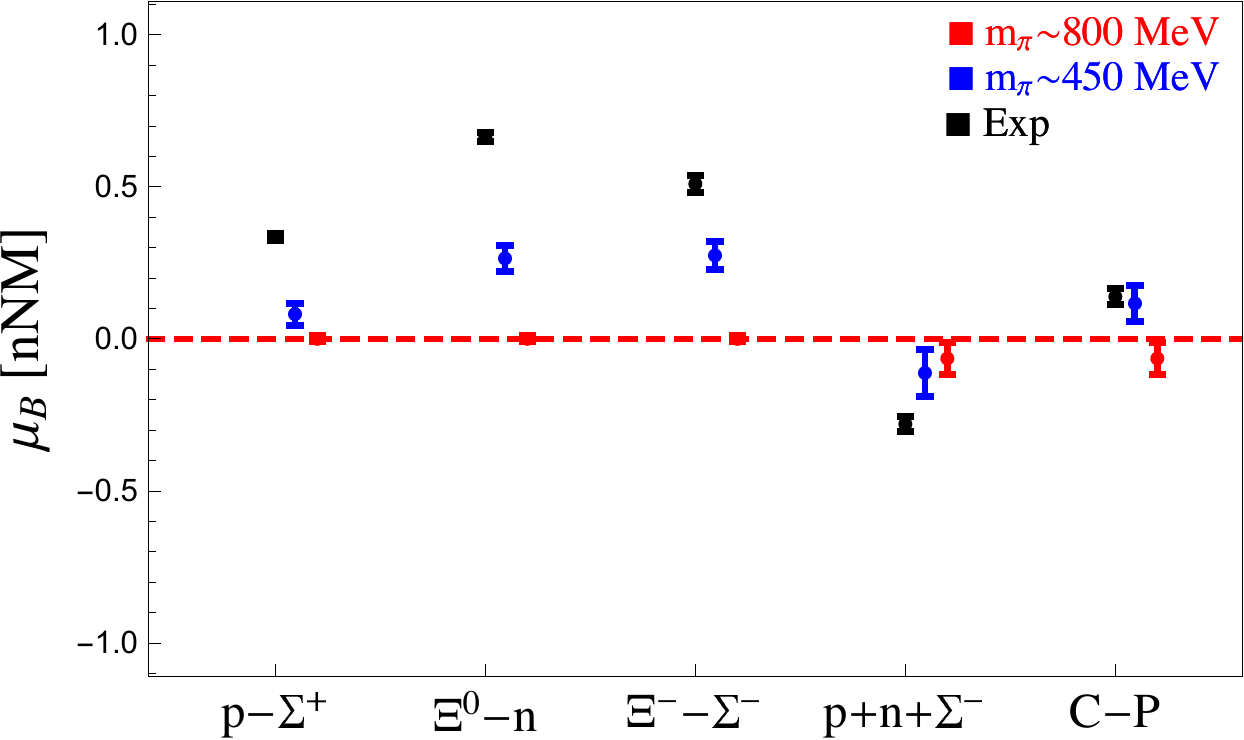}
}
\vskip 0.2in
\resizebox{0.9\linewidth}{!}{
\includegraphics{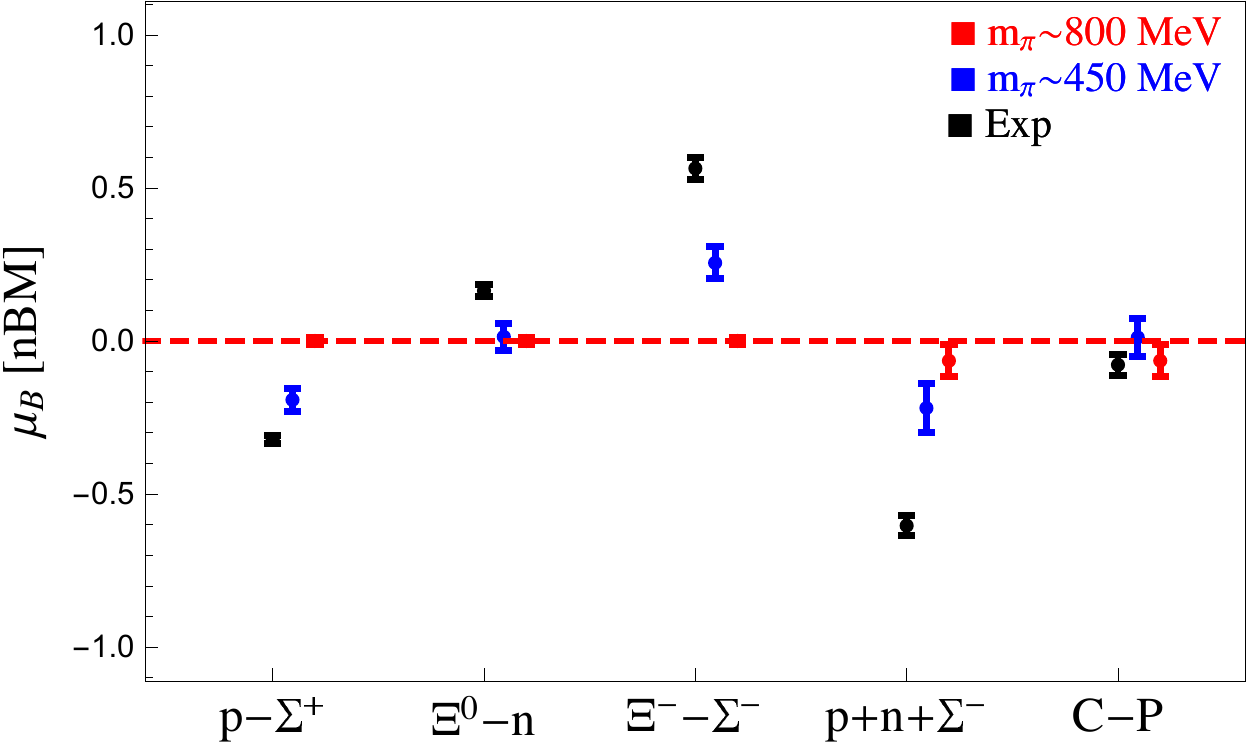}%
}
\caption{
The four Coleman-Glashow relations between magnetic moments,  
and the Caldi-Pagels relation are compared with experiment. 
For the  
$m_\pi \sim 450 \, \texttt{MeV}$
data, 
the quark-disconnected contributions have been neglected.
The results of the LQCD calculations  are presented in both 
$\texttt{[nNM]}$ and $\texttt{[nBM]}$, 
while the corresponding experimental results are given in 
$\texttt{[NM]}$ and $\texttt{[BM]}$,  respectively. 
}
\label{f:CGCPplot}
\end{figure}

The Coleman-Glashow  relations obviously also emerge in the 
$SU(3)_L \times SU(3)_R$ 
chiral limit in which $m_u = m_d = m_s  = 0$, 
with corrections occurring at next-to-leading order (NLO) in the chiral expansion. 
Such NLO corrections can be eliminated in forming  the smaller set of so-called 
Caldi-Pagels relations \cite{Caldi:1974ta}. 
Of interest here is the sum of all six  $I_3 \neq 0$ baryon magnetic moments 
\begin{eqnarray}
\mu_{\text{C-P}}
\equiv
\frac{1}{2}
\Big[
\mu_p + \mu_n + \mu_{\Sigma^+} + \mu_{\Sigma^-} + \mu_{\Xi^0} + \mu_{\Xi^-}
\Big]
\label{eq:CP}
.
\end{eqnarray}
This sum vanishes up to next-to-next-to-leading order (NNLO) corrections in the chiral expansion,
which scale parametrically as  $m_{K}^2 / \Lambda_\chi^2 \sim 15 \%$. 
The factor of  $\frac{1}{2}$  has been chosen so that  $\mu_{\text{C-P}}$  reduces to the relation in 
Eq.~\eqref{eq:SU3check}  with unit normalization in the limit of  $U$-spin symmetry. 
Results for 
$\mu_{\text{C-P}}$
are also given in 
Table~\ref{t:CGnNM}.
For each magnetic moment relation, 
the results are given in $\texttt{[nNM]}$, as well as $\texttt{[nBM]}$,
and are shown in Fig.~\ref{f:CGCPplot}. 
The LQCD results at  $m_\pi \sim 450 \, \texttt{MeV}$
are found to be closer to the $SU(3)_F$ limit than to experiment,  as expected.
Surprisingly, using $\texttt{[nBM]}$ units  substantially reduces  $SU(3)_F$ breaking only in the case of 
$\mu_{\Xi^0} - \mu_n$.

%
%
%
%
%
%

\begin{figure}
\begin{center}
\resizebox{0.9\linewidth}{!}{
\includegraphics{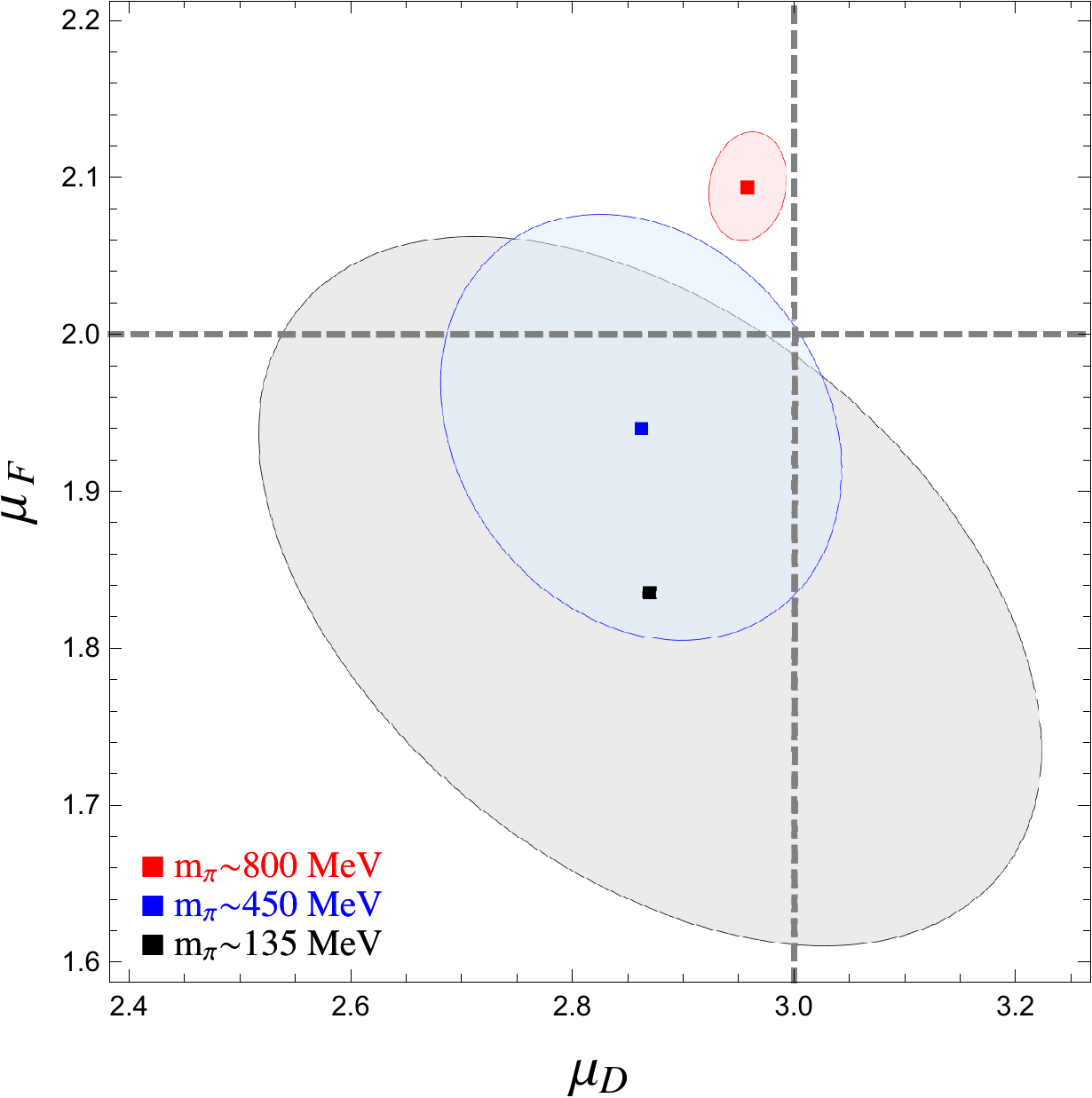}%
}
\end{center}
\caption{
Extracted values for the Coleman-Glashow moments, 
$\mu_D$
and
$\mu_F$,
from Ensembles I and III,  and experiment. 
The error ellipses represent the  uncertainty in these extractions due to both statistical and systematic uncertainties.
For Ensemble III and experiment, the latter includes an estimate 
$SU(3)_F$ 
breaking due to the quark sea. 
Integer values that are suggestive of the NRQM are shown with the dashed lines. 
}
\label{f:muDmuF}
\end{figure}

From the LQCD results on Ensemble I, 
the two independent magnetic moments, 
$\mu_D$ and $\mu_F$,
which we refer to as Coleman-Glashow moments, 
can be determined at 
$m_\pi\sim800 \, \texttt{MeV}$. 
Using the Hamiltonian density in 
Eq.~\eqref{eq:muDmuF}, 
three baryon magnetic moments not related by 
$U$-spin symmetry are
\begin{eqnarray}
\mu_p
&=&  
\left(
\frac{1}{3}
\mu_D
+
\mu_F
\right)
\texttt{[nBM]}
,\notag \\
\mu_n
&=&
- \frac{2}{3} \mu_D
\, \texttt{[nBM]}
,\notag \\
\mu_{\Sigma^-}
&=&
\left(
\frac{1}{3} \mu_D
- \mu_F 
\right)
\texttt{[nBM]}
.
\label{eq:CGBs}
\end{eqnarray}
A correlated fit to results for these magnetic moments leads to 
\begin{eqnarray}
\mu_D(m_\pi=800 \, \texttt{MeV})
&=& 
2.958(35)
\texttt{[nNM]}
, \notag \\
\mu_F(m_\pi=800 \, \texttt{MeV})
&=&
2.095(34)
\texttt{[nNM]}
,
\end{eqnarray}
where the quoted uncertainties are quadrature-combined statistical and systematic uncertainties.
These values and their correlated uncertainties are shown in 
Fig.~\ref{f:muDmuF}.

The values of  $\mu_D$  and  $\mu_F$  can be estimated at 
$m_\pi \sim 450 \, \texttt{MeV}$, 
after making additional assumptions.  
While  
$SU(3)_F$   
symmetry is not exact, 
there are only small deviations from the Coleman-Glashow relations on Ensemble III.  
To estimate the couplings,  
the proton and neutron magnetic moments are best;
because, 
for these baryons,  
$SU(3)_F$
breaking arises only from the quark sea. 
This effect from the strange sea quark is proportional to 
$ (m_s - m_l)/N_c$, 
where  
$N_c = 3$ 
is the number of colors;
hence,  
the expected size on Ensemble III is 
$\sim 6\%$, 
which reflects 
$\sim 50 \%$
reduction in flavor-symmetry breaking on Ensemble III compared to experiment. 
The extraction of Coleman-Glashow moments on Ensemble III using the proton and neutron magnetic moments 
results in
\begin{eqnarray}
\mu_D(m_\pi=450 \, \texttt{MeV})
&=& 
2.86(08)(16)
\texttt{[nNM]}
, \notag \\
\mu_F(m_\pi=450 \, \texttt{MeV})
&=&
1.94(08)(11)
\texttt{[nNM]}
,
\end{eqnarray}
where the first uncertainty reflects the quadrature-combined statistical and systematic uncertainty, 
and the second estimates the effect due to the non-degenerate strange sea quark. 
These values are included in  Fig.~\ref{f:muDmuF}.

%
%
%
%
%
%

A similar analysis of the experimentally measured proton and neutron magnetic moments, 
leads to
\begin{eqnarray}
\mu_D(m_\pi=135 \, \texttt{MeV})
&=& 
2.87(32)
\texttt{[nNM]}
, \notag \\
\mu_F(m_\pi=135 \, \texttt{MeV})
&=&
1.84(20)
\texttt{[nNM]}
,
\end{eqnarray}
where the leading 
$SU(3)_F$-breaking 
effects from the quark sea are estimated to be  
$\sim 11 \%$.
Interestingly,  the values of the Coleman-Glashow moments 
are found to exhibit only mild quark-mass dependence. 
Moreover, 
the nearness of these coefficients to integer values  is intriguing and highly suggestive of the NRQM.

$\chi$PT
 can be used to estimate the 
$SU(3)_L \times SU(3)_R$ 
chiral-limit values of the Coleman-Glashow moments. 
Beyond LO,  however, these calculations subsume quark-mass dependence into the couplings, 
which consequently become scale and scheme dependent. 
This can be remedied with future LQCD computations in which the pion-mass dependence is accounted for, 
however,  phenomenological analyses cannot resolve this dependence. 
As a result, 
we use the determinations of  $\mu_D$ and $\mu_F$ from Ref.~\cite{Puglia:1999th}, 
which employs a scheme in which the extracted parameters are relatively stable between 
the NLO and NNLO calculations,
and estimates of the nucleon mass in the three-flavor chiral limit from the BMW collaboration~\cite{Durr:2011mp}. 
We use values obtained from the NNLO calculation~\cite{Puglia:1999th}, 
which includes the decuplet degrees of freedom, 
and arrive at
\begin{eqnarray}
\mu_D(m_\pi=0 )
&=& 
3.8(1.1)
\texttt{[nNM]}
, \notag \\
\mu_F(m_\pi=0 )
&=&
2.5(0.6)
\texttt{[nNM]}
,\end{eqnarray}
where the quoted uncertainties arise from the difference between the NNLO and NLO values, 
and also the uncertainty in the 
three-flavor chiral limit value of the nucleon mass. 
These chiral-limit estimates are consistent with those obtained from a similar analysis of magnetic moments treating only the 
kaon-mass dependence~\cite{Durand:1997ya}, 
as well as computations without explicit decuplet baryons~\cite{Meissner:1997hn}.
Unfortunately, 
the relatively large uncertainties preclude definite conclusions from being drawn about the values, 
other than that they are consistent with the Coleman-Glashow moments extracted from the  
$SU(3)_F$-symmetric ensemble.   
LQCD calculations of both the nucleon mass and their magnetic moments at very light pion and kaon masses
would greatly reduce these uncertainties, 
and 
LQCD appears to be the only reliable tool with which to make such refinements.

\subsection{The NRQM}
\label{s:NRQM}

The baryon magnetic moments can  be compared to those in the NRQM.
Assuming strong isospin symmetry, 
there are two constituent quark masses, 
$M_Q$ 
for the light quarks
and 
$M_S$ 
for the strange quark, and 
hence there are only two independent magnetic moments. 
These moments can be utilized to write the NRQM predictions in terms of a constituent quark magneton unit,
which we define by
\begin{eqnarray}
\texttt{[cQM]}
=
\frac{e}{2 M_Q}, 
\end{eqnarray}
and the ratio of constituent quark masses
\begin{equation} 
\lambda
= 
M_Q  / M_S
.
\end{equation}
By virtue of their fractional electric charges, 
the magnetic moments of the up and down constituent quarks are written in terms of the constituent quark magneton simply as
$\mu_u = \frac{2}{3} \, \texttt{[cQM]}$
and 
$\mu_d = - \frac{1}{3} \, \texttt{[cQM]}$,
while for the strange constituent quark, 
$\mu_s = - \frac{1}{3} \lambda \, \texttt{[cQM]}$.  
With these definitions, 
the NRQM predictions for the nucleon magnetic moments take the form
\begin{eqnarray}
\mu_p 
&=&
\phantom{-}
\frac{4}{3} \mu_u
-
\frac{1}{3} \mu_d
=
\phantom{-}
1 \, \texttt{[cQM]}
,
\notag \\
\mu_n
&=& 
- \frac{1}{3} \mu_u 
+ \frac{4}{3} \mu_d
= 
-\frac{2}{3} \, \texttt{[cQM]}
.\end{eqnarray}

Comparing these model predictions with the Coleman-Glashow expectation for the neutron magnetic moment, 
Eq.~\eqref{eq:CGBs},
one can identify
\begin{equation}
\texttt{[cQM]} = \mu_D \texttt{[BM]},
\end{equation}
hence the NRQM gives rise to 
$\mu_D = \texttt{[cQM]} / \texttt{[BM]} = M_B / M_Q$, 
which is simply the ratio of the baryon mass to the constituent quark mass.
Furthermore, 
combining the NRQM prediction for the proton magnetic moment with 
expectations from $SU(3)_F$ symmetry yields the relation 
$\mu_F = \frac{2}{3} M_B / M_Q$.
While the baryon mass is generally less than the sum of its constituent quark masses due to binding, 
assuming that the constituent quarks in the NRQM are noninteracting leads to the integer values
$\mu_D = +3$ and $\mu_F = +2$,
which are indicated in  
Fig.~\ref{f:muDmuF}.

%
\begin{table}
\caption{%
Constituent quark masses and ratios extracted from the NRQM predictions of 
baryon magnetic moments, based upon the relations in Eqs.~\eqref{eq:Deltarels}, \eqref{eq:6lambdas}, and \eqref{eq:Rs}. 
To obtain these model parameters,  
 the magnetic moments determined on Ensembles I and III, 
as well as the experimental values are used. 
}
\begin{center}
\resizebox{\linewidth}{!}{
\begin{tabular}{|c|ccc|}
\hline
\hline
& 
\multicolumn{3}{c|}{$M_Q / M_N$}
\\
$\quad \texttt{[nNM]}^{-1}$
& 
I 
& 
III
& 
Experiment
\tabularnewline
\hline
\hline
$\frac{5}{3} \Delta \mu^{-1}_N
$
&
$\phantom{-} 0.3311(10)(26)$
&
$\phantom{-} 0.3471(18)(48)$
&
$\phantom{-} 0.3542(0)\phantom{1}$
\tabularnewline
$\frac{4}{3} \Delta \mu^{-1}_\Sigma
$
&
$\phantom{-} 0.3184(16)(37)$
&
$\phantom{-} 0.3409(23)(50)$
&
$\phantom{-} 0.3685(36)$
\tabularnewline
$\frac{1}{3} \Delta \mu^{-1}_\Xi
$
&
$\phantom{-} 0.395(07)(17) \phantom{1}$
&
$\phantom{-} 0.408(08)(20) \phantom{1}$
&
$\phantom{-} 0.556(15) \phantom{1}$
\tabularnewline
\hline
\hline
& 
\multicolumn{3}{c|}{$M_Q / M_B$}
\\
$\quad \texttt{[nBM]}^{-1}$
& 
I 
& 
III
& 
Experiment
\tabularnewline
\hline
\hline
$\frac{5}{3} \Delta \mu^{-1}_N
$
&
$\phantom{-} 0.3311(10)(26)$
&
$\phantom{-} 0.3471(18)(48)$
&
$\phantom{-} 0.3542(0) \phantom{1}$
\tabularnewline
$\frac{4}{3} \Delta \mu^{-1}_\Sigma
$
&
$\phantom{-} 0.3184(16)(37)$
&
$\phantom{-} 0.3107(21)(46)$
&
$\phantom{-} 0.2900(28)$
\tabularnewline
$\frac{1}{3} \Delta \mu^{-1}_\Xi
$
&
$\phantom{-} 0.395(07)(17) \phantom{1}$
&
$\phantom{-} 0.354(07)(17) \phantom{1}$
&
$\phantom{-} 0.396(11) \phantom{1}$
\tabularnewline
\hline
\hline
& 
\multicolumn{3}{c|}{$\lambda = M_Q / M_S$}
\\
& 
I 
& 
III
& 
Experiment
\tabularnewline
\hline
\hline
$\lambda_{N\Sigma}$
&
$\phantom{-} 0.776(30)(67)$
&
$\phantom{-} 0.640(42)(88)$
&
$\phantom{-} 0.147(64) \phantom{1}$
\tabularnewline
$\lambda_{\Sigma\Sigma}$
&
$\phantom{-} 0.747(31)(69)$
&
$\phantom{-} 0.630(44)(91)$
&
$\phantom{-} 0.153(68) \phantom{1}$
\tabularnewline
$\lambda_{\Xi\Sigma}$
&
$\phantom{-} 0.922(24)(59)$
&
$\phantom{-} 0.75(05)(11) \phantom{1}$
&
$\phantom{-} 0.23(11)\phantom{11}$
\tabularnewline
$\lambda_{N\Xi}$
&
$\phantom{-} 1.056(07)(17)$
&
$\phantom{-} 0.857(08)(19)$
&
$\phantom{-} 0.6777(50)$
\tabularnewline
$\lambda_{\Sigma\Xi}$
&
$\phantom{-} 1.015(05)(12)$
&
$\phantom{-} 0.842(08)(19)$
&
$\phantom{-} 0.705(12) \phantom{1}$
\tabularnewline
$\lambda_{\Xi \Xi}$
&
$\phantom{-} 1.260(30)(68)$
&
$\phantom{-} 1.009(28)(60)$
&
$\phantom{-} 1.064(37) \phantom{1}$
\tabularnewline
\hline
\hline
\end{tabular}
}
\\
\bigskip
\resizebox{\linewidth}{!}{
\begin{tabular}{|c|ccc|c|}
\hline
\hline
& 
\multicolumn{4}{c|}{Ratios}
\\
$R_X$
& 
I 
& 
III
& 
Experiment
& 
NRQM
\tabularnewline
\hline
\hline
$R_N$
&
$\phantom{-} 1.027(05)(15)$
&
$\phantom{-} 1.012(08)(25)$
&
$\phantom{-} 0.9732(0)\phantom{1}$
&
$1.00$
\tabularnewline
$R_{N\Sigma}$
&
$\phantom{-} 1.057(05)(15)$
&
$\phantom{-} 1.026(07)(22)$
&
$\phantom{-} 0.9456(91)$
&
$1.00$
\tabularnewline
$R_{N\Xi}$
&
$\phantom{-} 0.854(14)(33)$
&
$\phantom{-} 0.858(17)(44)$
&
$\phantom{-} 0.627(17) \phantom{1}$
&
$1.00$
\tabularnewline
$R_{\Sigma\Xi}$
&
$\phantom{-} 0.808(17)(40)$
&
$\phantom{-} 0.837(18)(43)$
&
$\phantom{-} 0.663(25) \phantom{1}$
&
$1.00$
\tabularnewline
$R_S$
&
$\phantom{-} 0.736(33)(75)$
&
$\phantom{-} 0.75(05)(11) \phantom{1}$
&
$\phantom{-} 0.216(96) \phantom{1}$
&
$1.00$
\tabularnewline
\hline
\hline
\end{tabular}
}
\end{center}
\label{t:cQM}
\end{table}

\begin{figure}
\resizebox{0.9\linewidth}{!}{
\includegraphics{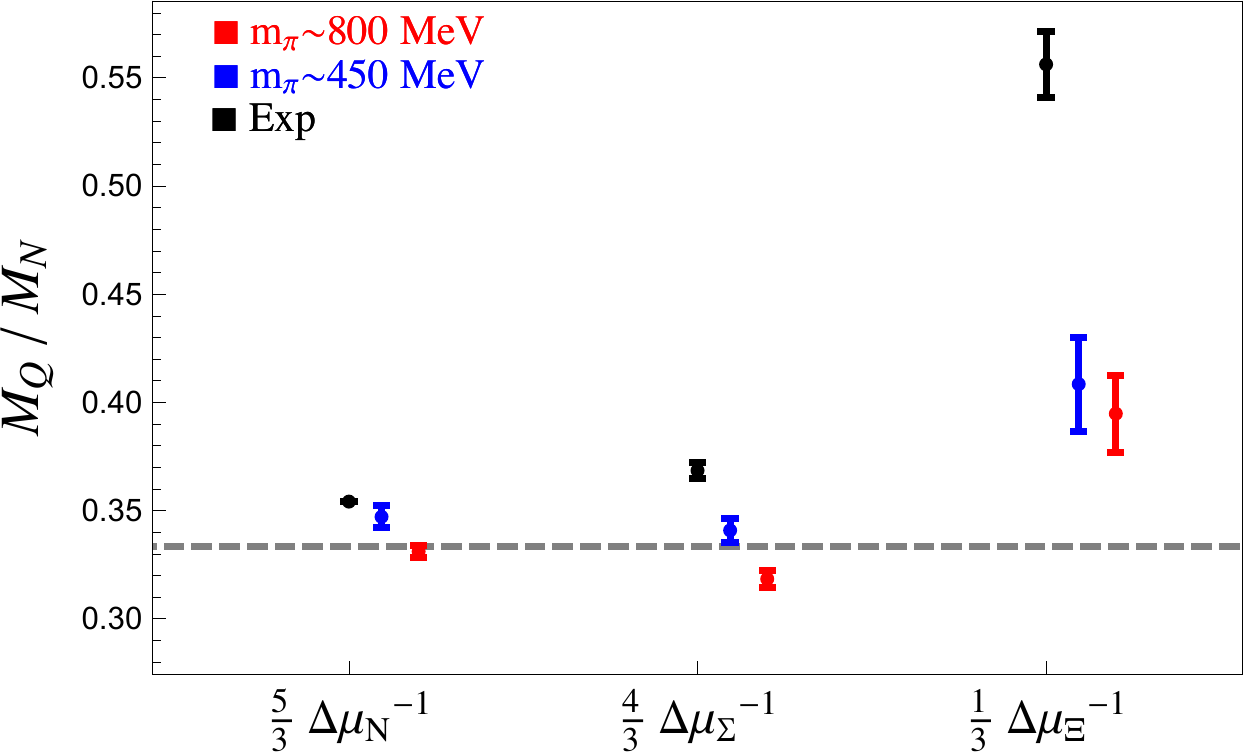}
}
\vskip 0.2in
\resizebox{0.9\linewidth}{!}{
\includegraphics{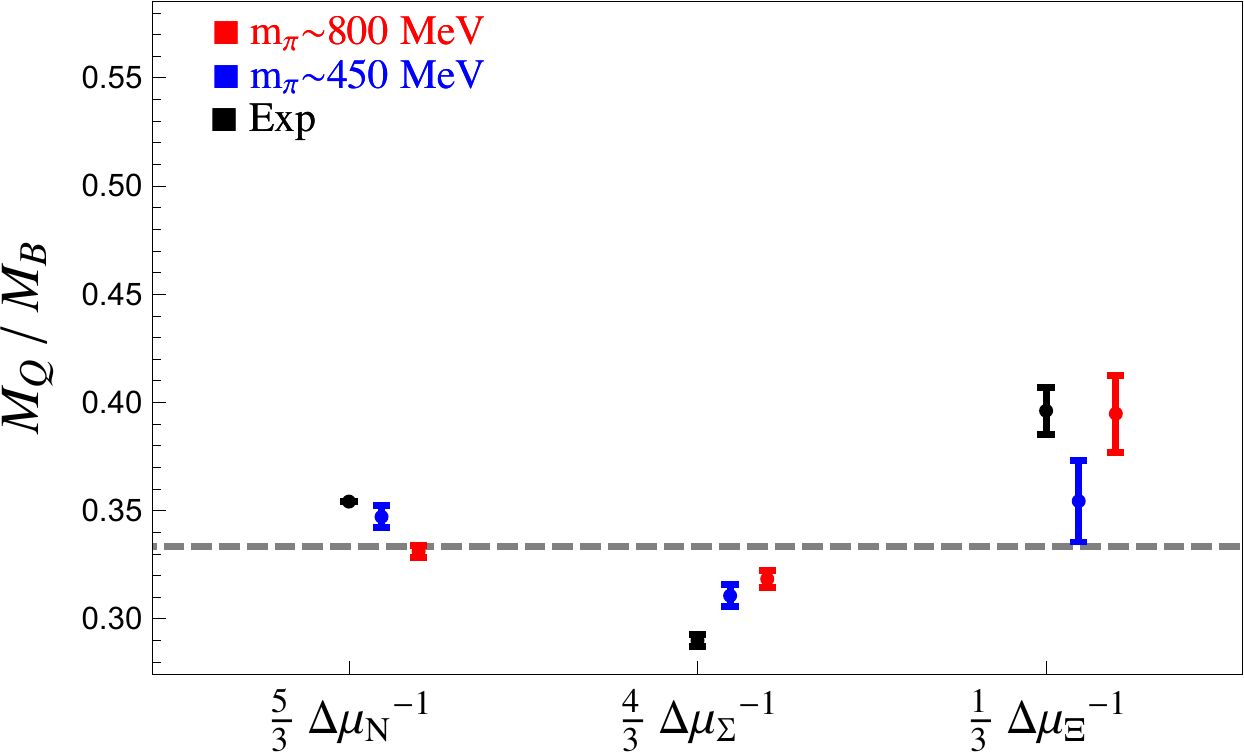}
}
\caption{
Ratios of the light constituent quark mass to baryon mass determined from the isovector  magnetic moment relations in Eq.~(\ref{eq:Deltarels}). 
Magnetic moments expressed in  $\texttt{[nNM]}$  permit an extraction of  $M_Q / M_N$ (upper panel), 
while moments expressed in  $\texttt{[nBM]}$ permit an extraction of  $M_Q / M_B$ (lower panel). 
The dashed horizontal line indicates a mass ratio of  $\frac{1}{3}$ that is expected in the NRQM. 
}
\label{f:MQ}
\end{figure}

The hyperon magnetic moments in the NRQM, 
moreover, 
are given by the expressions
\begin{eqnarray}
\mu_{\Sigma^+}
&=& 
\phantom{-}
\frac{4}{3} \mu_u
- 
\frac{1}{3} \mu_s
=
\phantom{-}
\left( 
\frac{8}{9} + \frac{1}{9} \lambda
\right) 
\, \texttt{[cQM]}
,\notag \\
\mu_{\Sigma^-}
&=&
\phantom{-}
\frac{4}{3} \mu_d
- 
\frac{1}{3} \mu_s
= 
-
\left(
\frac{4}{9} 
- 
\frac{1}{9} \lambda
\right)
\, \texttt{[cQM]}
,\notag \\
\mu_{\Xi^0}
&=& 
- \frac{1}{3} \mu_u
+ 
\frac{4}{3} \mu_s
= 
-
\left(
\frac{2}{9} 
+ \frac{4}{9} \lambda
\right)
\, \texttt{[cQM]}
,\notag \\
\mu_{\Xi^-}
&=&
- \frac{1}{3} \mu_d
+ 
\frac{4}{3} \mu_s
=
\phantom{-}
\left(
\frac{1}{9} 
- 
\frac{4}{9} \lambda
\right)
\, \texttt{[cQM]}
\label{eq:HyperQM}
,\end{eqnarray}
which depend on the strange constituent quark mass through  
$\lambda$. 
Confronting LQCD results with the NRQM predictions enables determinations of the constituent quark masses
at unphysical values of the 
(current)
quark masses. 
For example,  
the three isovector combinations of magnetic moments
\begin{eqnarray}
\Delta \mu_N 
&\equiv& 
\mu_p - \mu_n
,
\notag \\ 
\quad
\Delta \mu_\Sigma 
&\equiv& 
\mu_{\Sigma^+} - \mu_{\Sigma^-}
,
\notag \\ 
\quad
\Delta \mu_\Xi 
&\equiv& 
\mu_{\Xi^-} -  \mu_{\Xi^0}
,
\label{eq:Deltarels}
\end{eqnarray} 
are independent of the strange constituent quark mass,
from which the constituent quark magneton unit can be extracted,
\begin{equation}
\frac{e}{2 M_Q}
= 
\frac{3}{5}
\Delta \mu_N
=
\frac{3}{4} 
\Delta \mu_\Sigma
=
3 \,
\Delta \mu_\Xi
\label{eq:MQ}
.
\end{equation}
Values of these quantities in units of  
$\texttt{[nNM]}$ 
allow for various determinations of the mass ratio 
$M_N / M_Q$, 
while examining them in units of $\texttt{[nBM]}$ provide extractions of $M_B / M_Q$. 
The LQCD results for these quantities, along with their experimental values, are collected in  Table~\ref{t:cQM}, 
and displayed in Fig.~\ref{f:MQ}. 
The 
$M_Q$ 
values obtained from the isovector relations involving the nucleon and $\Sigma$ are similar in both units, 
but show a modest systematic trend in pion mass.
However, 
the corresponding ratios obtained from the isovector magnetic moment of the 
$\Xi$ 
exhibit much greater pion-mass dependence. 
The value 
$M_Q / M_N \sim 0.55$
inferred from the experimental determination of 
$\Delta \mu_\Xi$
suggests additional  
$SU(3)_F$  
breaking beyond the NRQM. 
Consistent with observations made above, 
the level of $SU(3)_F$ breaking is reduced by employing 
$\texttt{[nBM]}$ 
units.

%
%
%
%
%

\begin{figure}
\resizebox{0.9\linewidth}{!}{
\includegraphics{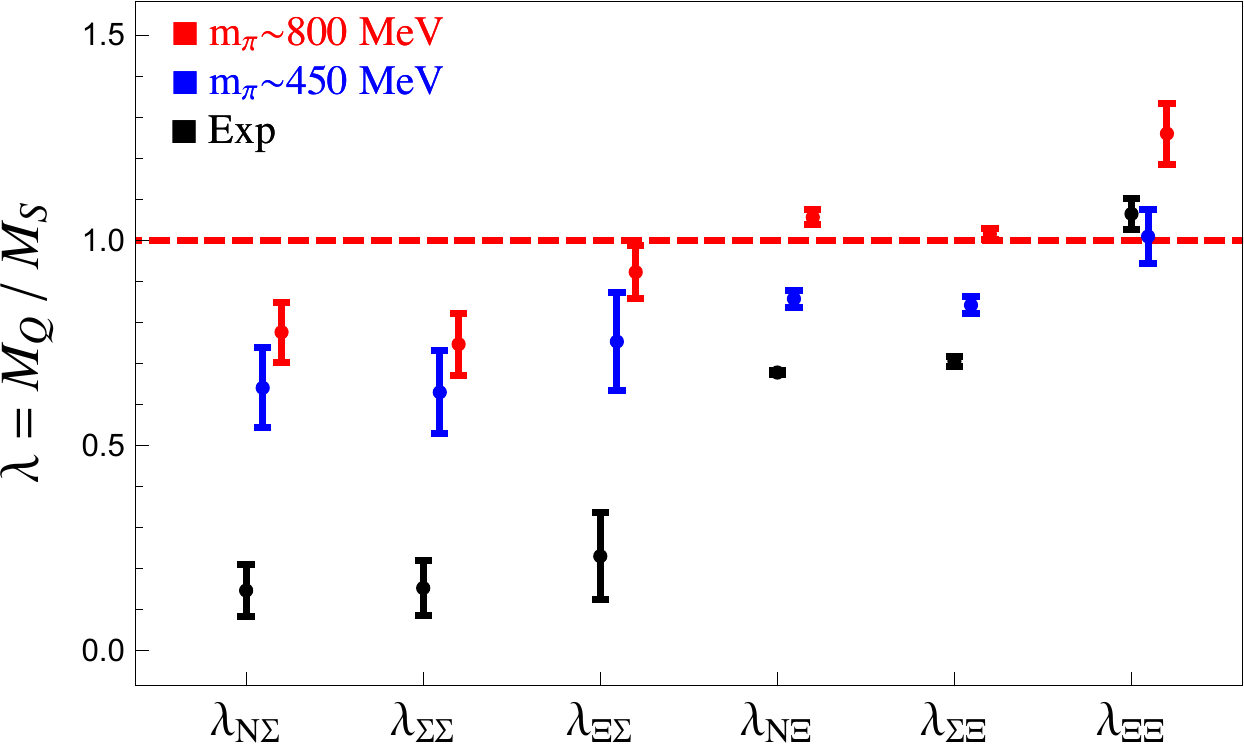}
}
\caption{
Values of 
$\lambda = M_Q / M_S$ 
determined 
from the relations in Eq.~\eqref{eq:6lambdas}. 
The dashed horizontal line indicates the expectation in the 
$SU(3)_F$-symmetric limit. 
}
\label{f:lambda}
\end{figure}

Values for 
$M_S$ 
can be isolated from linear combinations of hyperon magnetic moments. 
From 
Eq.~\eqref{eq:HyperQM},  
there are two such possibilities, 
\begin{equation}
\frac{e}{2 M_S}
=
3 \left( 
\mu_{\Sigma^+}
+ 2 \mu_{\Sigma^-}
\right)
=
- 
\frac{3}{4} 
\left(
\mu_{\Xi^0}
+ 
2 \mu_{\Xi^-}
\right)
\label{eq:MS}
.
\end{equation}
Ratios of these quantities to those appearing in  Eq.~\eqref{eq:MQ}
permit determinations of the constituent quark mass ratio  $\lambda$ in  the six  ways
\begin{eqnarray}
\lambda_{N\Sigma}
&=&
5 \,
\frac{\mu_{\Sigma^+} + 2 \mu_{\Sigma^-}}{\mu_p - \mu_n}
,
\quad
\lambda_{\Sigma\Sigma}
=
4 \,
\frac{\mu_{\Sigma^+} + 2 \mu_{\Sigma^-}}{\mu_{\Sigma^+} - \mu_{\Sigma^-}}
,\notag\\
\lambda_{\Xi\Sigma}
&=&
\frac{\mu_{\Sigma^+} + 2 \mu_{\Sigma^-}}{\mu_{\Xi^-} - \mu_{\Xi^0}}
,
\quad
\lambda_{N\Xi}
=
- \frac{5}{4} 
\frac{\mu_{\Xi^0} + 2 \mu_{\Xi^-}}{\mu_p - \mu_n}
,\notag \\
\lambda_{\Sigma\Xi}
&=&
-  \frac{\mu_{\Xi^0} + 2 \mu_{\Xi^-}}{\mu_{\Sigma^+} - \mu_{\Sigma^-}}
,
\quad
\lambda_{\Xi \Xi}
=
- \frac{1}{4} \frac{\mu_{\Xi^0} + 2 \mu_{\Xi^-}}{\mu_{\Xi^-} - \mu_{\Xi^0}}
\label{eq:6lambdas}
,
\end{eqnarray}
where the first subscript denotes the relation in Eq.~\eqref{eq:MQ} that gives $M_Q$,
and the second denotes the relation that gives $M_S$. 
The values of 
$\lambda$ 
determined from these relations are collected in 
Table~\ref{t:cQM}, and shown in  Fig.~\ref{f:lambda}.
A generic feature of the extracted values of 
$\lambda$ is that those 
obtained from Ensemble III are closer to unity than those obtained from experiment,
as expected from $SU(3)_F$ symmetry.  
Aside from this gross feature, 
differing results for the six ratios are suggestive of additional 
$SU(3)_F$ 
breaking beyond the NRQM predictions;
and, 
when considering the entire baryon octet,  
the constituent quark magnetic moments are not exactly identical to their Dirac moments.
The ratio  $M_Q / M_S \gtrsim 1$ determined exclusively from the  $\Xi$ magnetic moments is quite interesting.  
%
%
%
%
%
%

Notice that 
$SU(3)_F$-symmetric 
LQCD results in 
Fig.~\ref{f:lambda} 
can deviate from 
$\lambda = 1$. 
This indicates values of the magnetic moments that, 
while consistent with 
$SU(3)_F$
symmetry, 
are inconsistent with the NRQM at some level. 
While the quantities 
$\lambda_{\Xi\Sigma}$ 
and 
$\lambda_{\Xi\Xi}$ 
are equal to unity for any values of the Coleman-Glashow moments, 
the others can take a range of values.
Only for the integer values 
$\mu_D=3$ 
and 
$\mu_F=2$ 
will all 
$\lambda$ 
determinations equal unity.
The small deviations of 
$\mu_D$ 
and 
$\mu_F$ 
from these integer values, 
as shown in Fig.~\ref{f:muDmuF}, 
give rise to the deviations from unity that are noticeable in 
Fig.~\ref{f:lambda}.

\begin{figure}
\resizebox{0.9\linewidth}{!}{
\includegraphics{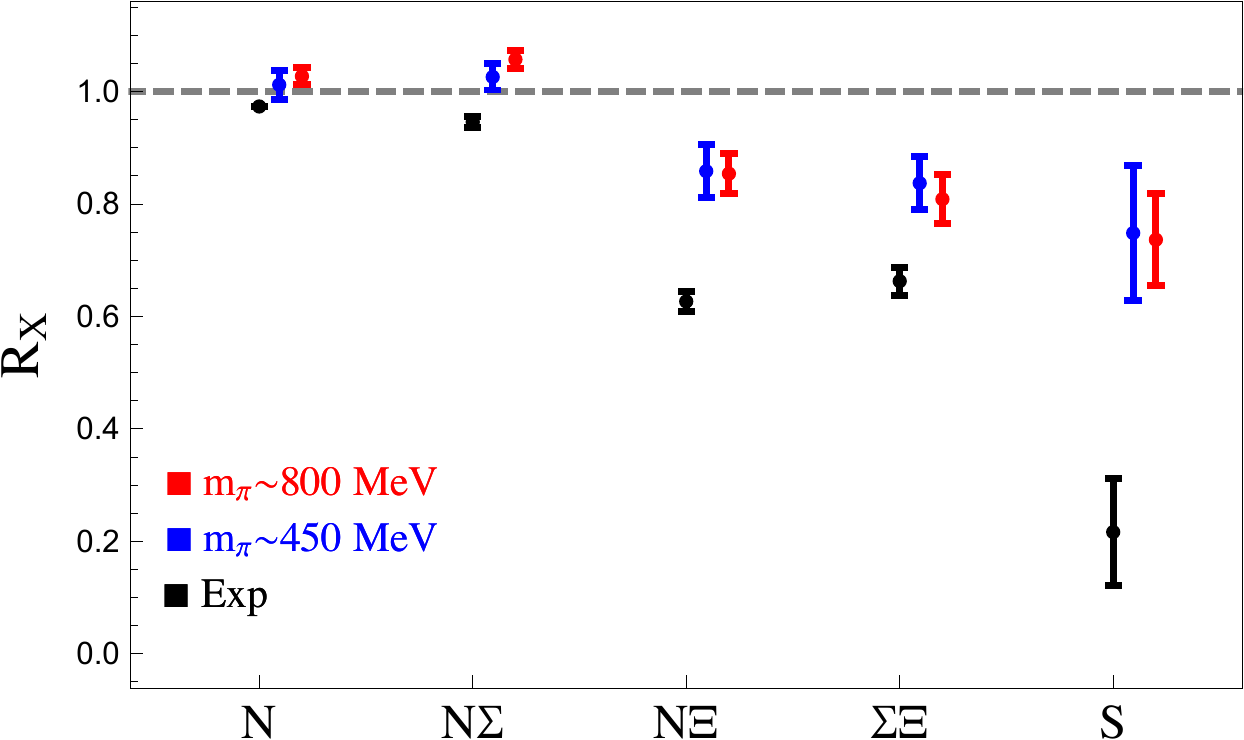}
}
\caption{
The magnetic moment ratios
$R_X$ for 
$X = N$,  $N\Sigma$,  $N\Xi$,  $\Sigma\Xi$,
and $S$, 
which are  predicted to be unity in the NRQM, 
see Eqs.~\eqref{eq:Rs} and \eqref{eq:RS}. 
}
\label{f:Rplot}
\end{figure}

Ratios of magnetic moments can be compared to NRQM predictions;
and, 
as they are insensitive to the overall choice of units, 
they provide complementary information to the previously considered relations.
Normalizing them so that predicted values in the NRQM are unity,
leads to the relevant combinations
\begin{eqnarray}
R_{N} 
&=& 
- \frac{2}{3} \, 
\frac{\mu_p}{\mu_n}, 
\quad
R_{N \Sigma} 
=
- \frac{1}{2} \,
\frac{\mu_{\Sigma^+} - \mu_{\Sigma^-} }{\mu_n}, 
\notag \\
R_{N \Xi} 
&=&
-
2 \, 
\frac{\mu_{\Xi^-} -  \mu_{\Xi^0}}{\mu_{n} }, 
\quad
R_{\Sigma \Xi} 
=
4 \,
\frac{\mu_{\Xi^-} - \mu_{\Xi^0}}{\mu_{\Sigma^+} - \mu_{\Sigma^-}}
\label{eq:Rs}
.
\end{eqnarray}
These ratios each compare two determinations of $M_Q$, 
and note that one of the three ratios 
$R_{N\Sigma}$,  
$R_{N\Xi}$,  
and 
$R_{\Sigma\Xi}$ is redundant. 
A further ratio,
\begin{eqnarray}
R_S
&=&
- 4 \,
\frac{\mu_{\Sigma^+} + 2 \, \mu_{\Sigma^-}}{\mu_{\Xi^0} + 2 \, \mu_{\Xi^-}}
\label{eq:RS}
,
\end{eqnarray}
compares the two possible determinations of $M_S$.   
Results for the ratios in Eqs.~(\ref{eq:Rs}) and (\ref{eq:RS}) are given in 
Table~\ref{t:cQM},  and shown  in    Fig.~\ref{f:Rplot}.  
The ratio  $R_N$
is  close to unity,  and is often touted as a success of the NRQM. 
The same applies equally well to the lesser-known ratio
$R_{N\Sigma}$.  When the  $\Xi$ magnetic moments enter into the determination of $M_Q$, 
however,  the situation is less clear.  
Finally, the ratio
$R_S$
highlights the inadequacy of  the NRQM in explaining
 the experimentally measured magnetic moments of the 
$\Xi$
baryons.  
Notice that the LQCD results on Ensembles I and III generally seem to agree better with the NRQM than experiment. 
The suggestive pattern among baryon isospin multiplets, 
however,  
is a puzzling feature that remains to be understood.

%
%
%
%
%
%

\subsection{The Large--$N_c$ Limit of QCD}
\label{s:LargeN}

%
\begin{table}
\caption{
Ratios of combinations of the baryon magnetic moments 
that are predicted to be unity in the large-$N_c$ limit%
~\cite{Jenkins:1994md},
as given in 
Eqs.~\eqref{eq:r10}--\eqref{eq:rv1}. 
The uncertainties in the LQCD results correspond to the statistical and systematic, respectively.
The order,  
$\mathcal{O}$, 
in the large-$N_c$ expansion is shown at which deviations from unity are expected, 
with  
$\Delta m_q$  
denoting corrections from $SU(3)_F$ breaking. 
Relations marked with a `` * ''  correspond to ratios that are also predicted to be unity in the NRQM. 
}
\begin{center}
\resizebox{\linewidth}{!}{
      \begin{tabular}{|c|ccc|c|}
\hline  
\hline
& 
\multicolumn{3}{c|}{$\mathcal{R}_X$}
&
\\
$X$
& 
I 
& 
III
& 
Exp
& 
$\mathcal{O}$
\tabularnewline
\hline
\hline
$S7^*$
&
$1.105(10)(24) \phantom{1}$
&
$1.080(17)(41)$
&
$1.213(36) \phantom{1}$
&
$1/N_c$
\tabularnewline
$V10_1$
&
$1.202(04)(10) \phantom{1}$
&
$1.228(07)(16)$
&
$1.301(13) \phantom{1}$
&
$1/N_c$
\tabularnewline
$V10_2$
&
$0.8014(28)(66)$
&
$0.818(05)(11)$
&
$0.8671(84)$
&
$\Delta m_q /N_c$
\tabularnewline
$V10_3$
&
$0.9016(31)(75)$
&
$0.921(05)(12)$
&
$0.9755(94)$
&
$\Delta m_q /N_c$
\tabularnewline
$S/V1$
&
$0.6369(16)(42)$
&
$0.710(06)(13)$
&
$0.893(17) \phantom{1}$
&
$\Delta m_q$,  $\Delta m_q/N_c$
\tabularnewline
$V1^*$
&
$0.904(08)(20) \phantom{1}$
&
$0.928(08)(20)$
&
$0.899(16) \phantom{1}$
&
$1/N_c^2$
\tabularnewline
\hline
\hline
\end{tabular}
}
\end{center}
\label{t:Nc}
\end{table}

Various relations between magnetic moments of the baryon octet  emerge in the limit of a large number of colors, 
$N_c$. 
A comprehensive set of  
large-$N_c$  
relations between moments was derived in 
Ref.~\cite{Jenkins:1994md}, 
and includes relations valid to different orders in the 
$1/N_c$
expansion, 
along with additional relations in a combined expansion about the 
$SU(3)_F$-symmetric limit. 
Using the experimentally measured magnetic moments, 
these  large-$N_c$
relations generally exhibit the expected pattern.
Moreover,  
this expansion
seems to indicate why certain NRQM predictions work better than others.

\begin{figure}
\resizebox{\linewidth}{!}{
\includegraphics{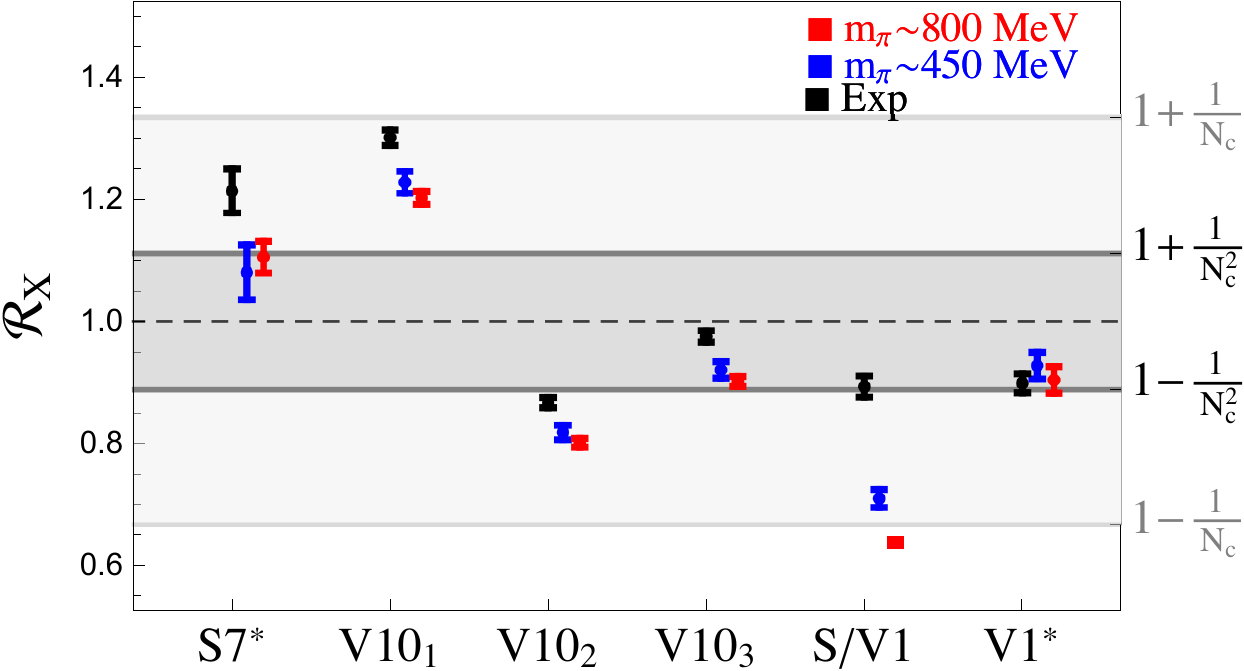}
}
\caption{
Relations between magnetic moments that are predicted to be unity in the large-$N_c$ limit~\protect\cite{Jenkins:1994md}.
They have been determined from LQCD results on Ensembles I and III, 
and from the experimental values. 
Uncertainties reflect quadrature-combined statistical and systematics. 
The light gray band shows a spread of $1/ N_c$, 
while the dark gray band shows a spread of  $1/N_c^2$,  around unity.  
The `` * '' 's denote ratios that are also predicted to be unity in the NRQM.  
}
\label{f:Nc}
\end{figure}

LQCD computations of magnetic moments allow for further tests of these large-$N_c$ relations,
as has been done with the experimental values.
We focus on the 
large-$N_c$  
relations for the  
$I_3 \neq 0$ 
octet baryons,  
for which there are six relations appearing in 
Ref.~\cite{Jenkins:1994md}. 
These are re-expressed in terms of a ratio that is predicted to be unity in the large-$N_c$ limit, 
using a naming convention that indicates the relation from which it is derived. 
The simplest ratios involve the isovector nucleon and  
$\Sigma$ 
magnetic moments, 
\begin{eqnarray}
\mathcal{R}_{V10_1}
&=&
\frac{\mu_p - \mu_n}{\mu_{\Sigma^+} - \mu_{\Sigma^-}}
=
1
+ 
\mathcal{O}(1/N_c)
,
\\
\mathcal{R}_{V10_2}
&=&
\frac{\left( 1 - \frac{1}{N_c} \right)  ( \mu_p - \mu_n )}{\mu_{\Sigma^+} - \mu_{\Sigma^-}}
=
1
+ 
\mathcal{O}(\Delta m_q /N_c)
,
\notag \\
\mathcal{R}_{V10_3}
&=&
\frac{\mu_p - \mu_n}{\left( 1 + \frac{1}{N_c} \right) (\mu_{\Sigma^+} - \mu_{\Sigma^-})}
=
1
+ 
\mathcal{O}(\Delta m_q /N_c)
,
\notag 
\label{eq:r10}
\end{eqnarray}
where 
$\Delta m_q = m_s-m$
denotes 
$SU(3)_F$ symmetry breaking due to different quark masses. 
In the 
$SU(3)_F$ limit,  
the second and third ratios have corrections that scale as  
$1/N_c^2$.
Notice that the difference between these two ratios is also of 
$\mathcal{O}(1/N_c^2)$. 
Experimentally, 
the second two relations, 
$V10_2$ and $V10_3$,
are satisfied better than the first,
$V10_1$,  
and in a way which is suggestive of 
$1/N_c^2$
corrections versus
$1/N_c$
corrections, 
respectively.
The remarkable proximity of the ratio
$\mathcal{R}_{V10_3}$
to unity appears to be accidental, 
due to 
higher-order terms in 
$1/N_c$. 
From the LQCD results collected in  
Table~\ref{t:Nc},  
and shown in  
Fig.~\ref{f:Nc},
their pattern does not appear consistent with large-$N_c$ predictions.  
Results for 
$\mathcal{R}_{V10_1}$
and
$\mathcal{R}_{V10_2}$
on 
Ensemble I
both deviate from unity by 
$\sim 20 \%$.
There does not appear to be an improvement in
$N_c$-scaling from these
$SU(3)_F$-symmetric results. 
The trend in
$SU(3)_F$ 
breaking, 
moreover, 
is opposite that predicted, 
as the more flavor-symmetric results are pulled away from unity rather than toward it.

%
%
%
%
%
%
%

This contrary pattern is more clearly observed in another large-$N_c$  relation. 
Combining all of the isoscalar moments and the nucleon isovector moment
leads to 
\begin{eqnarray}
\mathcal{R}_{S/V1}
&=&
\frac{\frac{1}{2} \left(\mu_p + \mu_n\right) 
+ 
3 \left( \frac{1}{N_c} - \frac{2}{N_c^2} \right) (\mu_p - \mu_n)}
{\mu_{\Sigma^+} + \mu_{\Sigma^-} - \frac{1}{2} ( \mu_{\Xi^0} + \mu_{\Xi^-})}
\label{eq:rsv1}
\\
&=&
1
+ 
\mathcal{O}(\Delta m_q)
+ 
\mathcal{O}(\Delta m_q / N_c)
,\notag
\end{eqnarray}
which is  predicted to be unity in the  $SU(3)_F$  limit up to  $1/N_c^2$ corrections. 
The experimentally measured moments give a ratio consistent with this scaling. 
Moving toward the 
$SU(3)_F$ 
point, 
however, 
introduces significant deviations,
ultimately producing a value on Ensemble I that is 
instead 
completely consistent with the NRQM, 
which predicts a ratio of
$\sim 0.62$
in the limit of 
$SU(3)_F$.
This may suggest that the good agreement at the physical point is an accident due to cancellation between higher-order contributions in
$SU(3)_F$ 
breaking and those that are purely 
$1/N_c$.  
This is the first relation we are aware of that strongly favors the 
NRQM at the expense of a large-$N_c$ relation.
Notice that while our computations on Ensemble III have omitted the quark-disconnected contributions, 
these missing contributions only affect the  
$\mathcal{O}(\Delta m_q / N_c)$
corrections.

The remaining two ratios are predicted to be unity in the large-$N_c$ limit independent of $SU(3)_F$ breaking,
and are also unity in the NRQM. 
 The first such ratio is formed from the isoscalar magnetic moments
\begin{eqnarray}
\mathcal{R}_{S7}
&=&
\frac{5(\mu_p + \mu_n) - (\mu_{\Xi^0} + \mu_{\Xi^-})}{4 (\mu_{\Sigma^+} + \mu_{\Sigma^-})}
\label{eq:rs7}
=
1 + \mathcal{O}(1/N_c)
. \quad
\end{eqnarray}
The values for this relation, 
given in  Table~\ref{t:Nc} and shown in Fig.~\ref{f:Nc}, 
appear consistent with  $1/N_c$ scaling, 
and the LQCD results are slightly closer to unity.
While the result from Ensemble III is missing the quark-disconnected contribution, 
the predicted scaling is unaffected. 
The final ratio is formed from the isovector magnetic moments,
\begin{eqnarray}
\mathcal{R}_{V1}
&=&
\frac{\mu_p - \mu_n - 3 (\mu_{\Xi^0} - \mu_{\Xi^-})}{2 (\mu_{\Sigma^+} - \mu_{\Sigma^-})}
=
1+ \mathcal{O}(1/N_c^2)
, \quad 
\label{eq:rv1}
\end{eqnarray}
for which there are no disconnected contributions and the LQCD calculations are complete. 
This ratio is consistent with corrections scaling as  
$1/N_c^2$, 
and appears insensitive to the pion mass.

\section{Coupled $\Lambda$--$\Sigma^0$ System}
\label{s:LamSig}

The two $I_3 = 0$ octet baryons, $\Lambda$ and $\Sigma^0$,
 mix in the presence of a magnetic field  as the quark charge assignments break isospin symmetry. 
As a result, 
the energy eigenstates in a background magnetic field are linear combinations of these isospin eigenstates,
and a coupled-channels analysis of the corresponding LQCD results is required.

\subsection{$SU(3)_F$-Symmetric Limit}

In the basis defined by 
$\begin{pmatrix}
\Sigma^0 \\ \Lambda 
\end{pmatrix}$,
the $2\times 2$ Hamiltonian resulting from the Coleman-Glashow effective interactions 
in Eq.~\eqref{eq:muDmuF} becomes
\begin{equation}
H_{I_3 =0}
=
+
\frac{e \, \mu_n}{2 M_B} \bm{\sigma} \cdot \bm{B}
\ {1\over 2}
\begin{pmatrix}
1 &  -\sqrt{3} \\
\sqrt{3} & -1
\end{pmatrix}
,
\label{eq:2x2Ham}
\end{equation} 
where
$\mu_n = - \frac{2}{3} \mu_D$
is the magnetic moment of the neutron in 
$\texttt{[BM]}$. 
In terms of these isospin eigenstates, the magnetic moments and dipole transition moment are given by
\begin{eqnarray}
\mu_\Lambda = \frac{1}{2} \mu_n, 
\quad  
\mu_{\Sigma^0} = - \frac{1}{2} \mu_n, 
\quad 
\text{and}
\quad
\mu_{\Lambda\Sigma} = - \frac{\sqrt{3}}{2} \mu_n
.\notag \\
\end{eqnarray}
In non-vanishing magnetic fields, 
the U-spin eigenstates, $\lambda_\pm$, are linear combinations of the isospin eigenstates,
and can be written in the form
\begin{equation}
\begin{pmatrix}
\lambda_+ 
\\
\lambda_-
\end{pmatrix}
=
\begin{pmatrix}
\phantom{-} \cos \theta 
& 
\sin \theta
\\
- \sin \theta
& 
\cos \theta
\end{pmatrix}
\begin{pmatrix}
\Sigma^0 \\
\Lambda 
\end{pmatrix}
.
\end{equation}
Diagonalizing the Hamiltonian in Eq.~(\ref{eq:2x2Ham}),
leads to a  mixing angle and eigenstate magnetic moments of 
\begin{equation}
\theta = 30^\circ,
\quad
\mu_{\lambda_\pm}  
=
\mp \mu_n
.
\label{eq:su3diag}
\end{equation}
Notice that the $\lambda_+$  eigenstate has a positive magnetic moment as
$\mu_n < 0$, and that the moments of the eigenstates are twice the flavor-diagonal moments.
Consequently, 
weak magnetic fields only partially lift the degeneracy between the 
$\Lambda$ and $\Sigma^0$
baryons because the opposite spin states come in nearly degenerate pairs, 
split by their magnetic polarizabilities that enter at 
$\cO(\bm{B}^2)$,
\begin{equation}
E^{(s)}_{\lambda_+} (\bm{B})
=
E^{(-s)}_{\lambda_-} (\bm{B})
+ 
\mathcal{O}(\bm{B}^2)
.
\end{equation} 
The magnetic polarizabilities arise from operators involving two insertions of the electric charge matrix $Q$.  
In the limit of  $SU(3)_F$ symmetry,  
there are four such independent operators that appear in the effective Hamiltonian density in the form,
\begin{equation}
\Delta \mathcal{H}
=
- \frac{1}{2} 4 \pi \bm{B}^2 
\,
\sum_{j=1}^4 
\beta_j  \mathcal{O}_j
,
\end{equation}
where the $\beta_j$ are numerical coefficients, and the operators  $\cO_j$ can be conveniently taken 
to be~\footnote{Notice that the additional operator
$\langle \overline{B} Q \rangle \langle B Q \rangle$
is redundant because of the Cayley-Hamilton identity. 
} 
\begin{eqnarray}
\cO_1
&=&
\left\langle \overline{B} B \rangle \langle Q^2 \right\rangle, 
\notag \\
\cO_2
&=&
\left\langle \overline{B} \left\{ Q, \{ Q, B \} \right\} \right\rangle,
\notag \\
\cO_3 
&=&
\left\langle \overline{B} \left\{ Q,  [ Q, B ] \right\} \right\rangle
\equiv
\left\langle \overline{B} \left[Q, \{ Q, B \} \right] \right\rangle,
\notag \\
\cO_4  
&=&
\left\langle \overline{B} \left[ Q, [ Q, B ] \right] \right\rangle
\label{eq:BBHam}
.\end{eqnarray}
This basis has been chosen because the operators 
$\cO_3$ and $\cO_4$ do not contribute to the magnetic polarizaibilities of electrically neutral octet baryons. 
The contribution from  $\cO_1$ is the same for all octet baryons, 
and therefore does not contribute to the electromagnetic mixing of $\Lambda$  and $\Sigma^0$ baryons. 
The resulting contributions to the effective Hamiltonian  in the $I_3 = 0$ sector are, 
\begin{equation}
\Delta H_{I_3=0}
=
- \frac{1}{2} 4 \pi  \bm{B}^2
\left[ 
\frac{2}{3} \beta_1 \, \mathbb{1}
+ 
\frac{2}{9}
\beta_2
\begin{pmatrix}
5 & \phantom{-} \sqrt{3} \\
\sqrt{3} & \phantom{-} 3 
\end{pmatrix}
\right]
.\end{equation}
The magnetic polarizabilities of flavor basis states are readily identified as the linear combinations,
\begin{eqnarray}
\beta_\Lambda
=
\frac{2}{3} \beta_1 
+
\frac{2}{3} \beta_2
,  
\quad
\beta_{\Sigma^0}
=
\frac{2}{3} \beta_1 
+
\frac{10}{9} \beta_2
,
\end{eqnarray}
along with their magnetic transition polarizability 
\begin{equation}
\beta_{\Lambda\Sigma} 
=
\frac{2 \sqrt{3}}{9}
\beta_2 
.
\end{equation}
Due to the structure of 
$\Delta H_{I_3=0}$, 
the eigenstates 
$\lambda_\pm$  necessarily remain eigenstates in its presence, 
and have magnetic polarizabilities given by
\begin{equation}
\beta_{\lambda_+}
=
\beta_n 
+
\frac{4}{\sqrt{3}}
\beta_{\Lambda\Sigma}, 
\quad 
\text{and}
\quad
\beta_{\lambda_-} 
= 
\beta_n
,\end{equation}
where these results are expressed in terms of the magnetic polarizability of the neutron, 
which, from Eq.~\eqref{eq:BBHam},  is $\beta_n = \frac{2}{3} \beta_1 +  \frac{4}{9} \beta_2$. 
Therefore,  the four 
eigenstates have energies
\begin{eqnarray}
E_{\lambda_-}^{(- \frac{1}{2})}
(B_z)
&=&
M_B
+
\mu_n
\frac{e B_z}{2 M_B}
- 
2\pi 
\beta_n
B_z^2
,\notag \\
E_{\lambda_+}^{(+ \frac{1}{2})}
(B_z)
&=&
M_B
+
\mu_n
\frac{e B_z}{2 M_B}
- 
2 \pi
\left(
\beta_n
+ 
\frac{4}{\sqrt{3}}
\beta_{\Lambda\Sigma}
\right)
B_z^2
,\notag \\
E_{\lambda_-}^{(+ \frac{1}{2})}
(B_z)
&=&
M_B
-
\mu_n
\frac{e B_z}{2 M_B}
- 
2 \pi
\beta_n
B_z^2
,\notag \\
E_{\lambda_+}^{(- \frac{1}{2})}
(B_z)
&=&
M_B
- 
\mu_n
\frac{e B_z}{2 M_B}
- 
2 \pi
\left(
\beta_n
+ 
\frac{4}{\sqrt{3}}
\beta_{\Lambda\Sigma}
\right)
B_z^2, 
\notag \\
\label{eq:LamSigSpec}
\end{eqnarray}
up to quadratic order in the magnetic field. 
These have been listed in order of increasing energy, from the ground state upwards, 
with the assumption that the magnetic field points along the positive 
$z$-axis, i.e., $B_z > 0$, 
and that the transition polarizability  $\beta_{\Lambda \Sigma}$ is negative leading to the inequality
$\beta_{\lambda_-} > \beta_{\lambda_+}$.

The above discussion must be adapted to our LQCD calculations, in which the electric charges of sea quarks vanish. 
As shown in Appendix~\ref{s:B},  the only modification required to  Eq.~\eqref{eq:LamSigSpec} 
is the replacement of the neutron's magnetic polarizability by its quark-connected part, 
$\beta_n \to \beta_n^{(c)}$. 
The magnetic transition polarizability is unchanged because, 
in the mass-symmetric limit, 
it arises  only from quark-connected contributions, 
$\beta_{\Lambda \Sigma} = \beta_{\Lambda \Sigma}^{(c)}$. 
Thus,  the ordering of energy eigenstates depends on the sign of 
$\beta_{\Lambda \Sigma}$,
which shall be seen to be negative.

\subsection{Coupled-Channels Analysis}
\label{s:CCA}

At the level of baryon two-point correlation functions, 
the coupled-channels 
$\Lambda$--$\Sigma^0$
system requires the matrix of correlation functions
\begin{equation}
\mathbb{G}^{(s)} (t, n_\Phi)
=
\begin{pmatrix}
G^{(s)}_{\Sigma\Sigma}(t,n_\Phi)
&
G^{(s)}_{\Sigma\Lambda}(t,n_\Phi)
\\
G^{(s)}_{\Lambda\Sigma}(t,n_\Phi)
&
G^{(s)}_{\Lambda \Lambda}(t,n_\Phi)
\end{pmatrix}
\label{eq:Gmat}
,
\end{equation}
for each value of spin,  $s = \pm \frac{1}{2}$.  
The matrix of correlation functions is diagonal when the magnetic field vanishes, i.e. $n_\Phi =0$;  and, 
when the magnetic field is non-vanishing, 
the principal correlators obtained from diagonalization contain information about the energy eigenstates of the coupled system.  
In our production,   the transition correlators
$G^{(s)}_{\Sigma\Lambda}(t,n_\Phi)$ and $G^{(s)}_{\Lambda\Sigma}(t,n_\Phi)$
were not computed, however, they can 
be obtained on Ensembles I and II by appealing to  $SU(3)_F$  symmetry. 
In this limit, the transition correlation functions can be shown to be
\begin{eqnarray}
G^{(s)}_{\Sigma\Lambda}(t,n_\Phi)
&=& 
G^{(s)}_{\Lambda\Sigma}(t,n_\Phi)
\notag \\
&=& 
\frac{\sqrt{3}}{2} 
\left[
G^{(s)}_{\Sigma \Sigma} (t,n_\Phi) 
-  
G^{(s)}_{\Lambda \Lambda} (t,n_\Phi)
\right]
,
\end{eqnarray}
as derived in Appendix~\ref{s:B} by capitalizing on the mass degeneracy of the 
$\Lambda$ and $\Sigma^0$ baryons. 
This coupled-channels system can then be solved~\footnote{As the exact solution is known, 
i.e.~the electromagnetic mixing angle is  $\theta = 30^\circ$,  the principal correlators are readily found to be 
\begin{eqnarray}
G^{(s)}_{\lambda+} (t, n_\Phi)
&=&
\frac{1}{2} 
\left[
3 G^{(s)}_{\Sigma \Sigma} (t, n_\Phi) 
- 
G^{(s)}_{\Lambda \Lambda} (t, n_\Phi)
\right]
,
\notag \\
G^{(s)}_{\lambda-} (t, n_\Phi)
&=&
\frac{1}{2}
\left[
3 G^{(s)}_{\Lambda \Lambda} (t, n_\Phi) 
- 
G^{(s)}_{\Sigma \Sigma} (t, n_\Phi)
\right]
\notag
.
\end{eqnarray}
These solutions compare favorably with the numerically determined solution of the generalized eigenvalue problem, 
Eq.~\eqref{eq:GEVP}. 
}
by obtaining the principal corrrelators, 
$G^{(s)}_{\lambda} (t, t_0, n_\Phi)$, 
which are solutions to the generalized eigenvalue problem~\cite{Luscher:1990ck},
\begin{eqnarray}
\mathbb{G}^{(s)} (t, n_\Phi)
\,
| \lambda \rangle
=
G^{(s)}_{\lambda} (t, t_0,n_\Phi)
\,
\mathbb{G}^{(s)} (t_0, n_\Phi)
\,
| \lambda \rangle
\label{eq:GEVP}
.
\end{eqnarray}
A time-offset parameter $t_0$ has been introduced, 
and can be varied to stabilize extraction of the ground-state contribution to the principal correlators, 
which appears in the long-time limit as
\begin{equation}
G^{(s)}_{\lambda_\pm} (t, t_0, n_\Phi)
\sim
e^{- E_{\lambda_\pm}^{(s)} (B_z)  \, (t - t_0) }
+
\cdots
\, .
\end{equation}

%
%
%
\begin{figure}
\resizebox{\linewidth}{!}{
\includegraphics{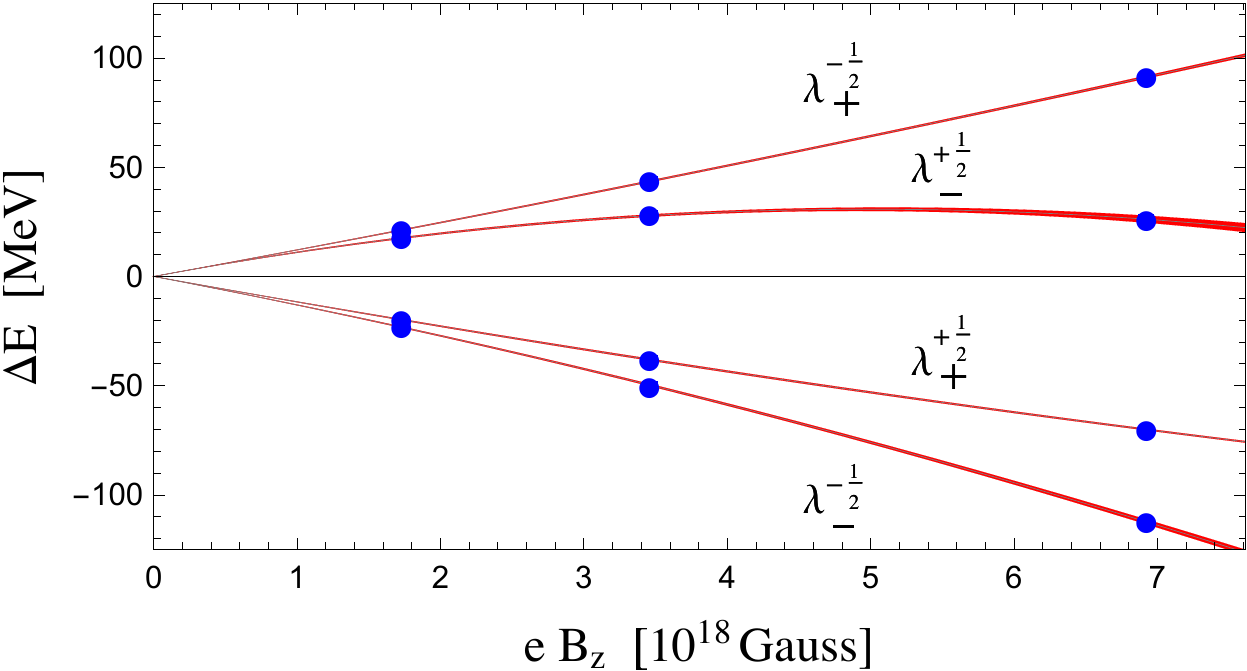}
}
\caption{
Energy eigenvalues,  
$\Delta E = E(B_z) - E(0)$, 
of the  
$\Lambda$--$\Sigma^0$
system as a function of the magnetic field, 
$e \, B_z$, 
calculated with LQCD on Ensemble I. 
The spectrum is consistent with the analytic expectation given in 
Eq.~\eqref{eq:LamSigSpec} with   
$\beta_{\Lambda \Sigma} < 0$. 
Fits to the magnetic-field dependence of each eigenstate are shown as bands, 
and include linear and quadratic magnetic field terms.}
\label{f:LamSigB}
\end{figure}
%
%
%

Using a bootstrap ensemble of SS correlation functions for the
$\Lambda$
and 
$\Sigma^0$
baryons, 
the matrix of correlations functions,  
Eq.~\eqref{eq:Gmat}, 
needed to solve the generalized eigenvalue problem posed in 
Eq.~\eqref{eq:GEVP} is formed. 
For each spin, 
$s$, 
this is then used to extract the ground-state energy of the two principal correlation functions  by inspecting 
the plateau region of their effective masses. 
Results for the spectrum are shown in 
Fig.~\ref{f:LamSigB}.
The effect of varying
$t_0$
is numerically insignificant on the extraction of ground-state energies. 
Further details about these fits appear in  Appendix~\ref{s:B}, along with the tabulated energies. 
Superposed on the numerically determined spectrum are fits to the magnetic field-strength dependence of the energies, 
from which the magnetic moments are determined.
These are found to be consistent with 
$\mp \mu_n$
for 
$\lambda_{\pm}$. 
More precise values are obtained, 
however, 
by determining the Zeeman splittings for each principal correlator from taking ratios of the two spin projections.  
These results are also given in Appendix~\ref{s:B}.

The ordering of energy levels in this system follows that of 
Eq.~\eqref{eq:LamSigSpec}, 
which anticipates a negative value for the transition polarizability. 
From fits to the spin-averaged principal correlators, 
the value of the transition magnetic polarizability is found to be
\begin{equation}
\beta_{\Sigma\Lambda} = - 1.82(06)(12)(02) \,
\texttt{[10}^{\texttt{-4}} \,  \texttt{fm}^{\texttt{3}} \texttt{]}
,
\label{eq:transmagpol}
\end{equation}
where the uncertainties reflect in order:
statistics, 
systematics, 
and the determination of the lattice spacing. 
Results for the quark-connected part of the neutron polarizability 
extracted from the 
$\Lambda$--$\Sigma^0$
system are consistent with the direct calculation of 
$\beta_n^{(c)}$ in Ref.~\cite{Chang:2015qxa}.

\subsection{$SU(3)_F$ Breaking and the Physical Point}

Without off-diagonal $\Lambda$--$\Sigma^0$ correlators, it is not possible  to investigate the mixing of 
the $I_3  = 0$ baryons on Ensemble III.
Nonetheless it is instructive to anticipate the behavior of this system with  $SU(3)_F$ breaking,
which can be accomplished using the experimentally measured magnetic moments.
Elements of the magnetic moment matrix,
\begin{equation}
\mathbb{M}
= 
\begin{pmatrix}
\mu_{\Sigma^0} 
& 
\mu_{\Lambda \Sigma}
\\
\mu_{\Lambda \Sigma}
&
\mu_{\Lambda}
\end{pmatrix}
,
\end{equation}
have not been completely determined experimentally. 
In particular,  the sign of the transition moment is not known
and the magnetic moment of the $\Sigma^0$
baryon has not been measured. 
The former only affects the magnetic mixing angle. 
Given the magnitude of the transition moment, 
$|\mu_{\Lambda \Sigma}| =  1.61(8) \, \texttt{[NM]}$, 
and the proximity of nature to the $SU(3)_F$-symmetric limit where
$\mu_{\Lambda \Sigma} > 0$, it is reasonable to assume that $\mu_{\Lambda \Sigma} > 0$  holds elsewhere.
The value of 
$\mu_{\Sigma^0}$
can be fixed from the assumption of isospin symmetry. 
In the limit of 
$SU(2)_F$ symmetry,  
the  
$\Sigma$ 
baryons form an isotriplet,
and by considering the two independent magnetic moment operators that act on this triplet in the isospin limit, 
for example Ref.%
~\cite{Jiang:2009jn}, 
the magnetic moment of 
$\Sigma^0$ 
is found to be
\begin{equation}
\mu_{\Sigma^0} 
= 
\frac{1}{2} \left( \mu_{\Sigma^+} + \mu_{\Sigma^-} \right)
+ 
\cO \left(
\alpha = \frac{e^2}{4\pi}, \, m_d - m_u
\right)
,
\end{equation}
because a single insertion of the charge matrix, 
with isoscalar and isovector components, 
is unable to induce an isotensor magnetic moment.
Using the experimentally measured magnetic moments of the   
$\Sigma^\pm$  
baryons leads to the value of 
$\mu_{\Sigma^0}$, 
up to isospin-breaking corrections 
(standardly estimated to be 
$\sim 1 \%$),
and therefore a magnetic moment matrix of
\begin{equation}
\mathbb{M}
= 
\begin{pmatrix}
\phantom{-}
0.649(14)(06)
& 
\phantom{-} 1.61 (8)
\\
1.61 (8)
&
-0.613(4)
\end{pmatrix}
\, \texttt{[NM]}
.
\end{equation}
The uncertainties quoted are experimental, 
with the exception of the second uncertainty given for 
$\mu_{\Sigma^0}$, 
which is an estimate of isospin breaking effects. 
Diagonalizing this matrix gives,
\begin{eqnarray}
\theta 
&=&
\phantom{-}
34.30 (62)(15)
{}^\circ
,\notag \\
\mu_{+}
&=&
\phantom{-}
1.747(88)(13)
\, \texttt{[NM]}
,\notag \\
\mu_{-}
&=&
-
1.711 (82)(06)
\, \texttt{[NM]}
\label{eq:MM}
,
\end{eqnarray}
which are within $\sim 15\%$ of their values in the limit of $SU(3)_F$,
see 
Eq.~(\ref{eq:su3diag}).

Breaking of  $SU(3)_F$ symmetry by the  baryon masses further complicates the 
$\Lambda$--$\Sigma^0$ system. 
The $\lambda_\pm$ are no longer simply the linear combinations of states that diagonalize the magnetic moment matrix,
and the Hamiltonian,
\begin{equation}
H_{I_3 = 0}
= 
\Delta_{\Lambda \Sigma} 
\begin{pmatrix}
1 & 0 \\
0 & 0 
\end{pmatrix}
- \frac{e \, \bm{\sigma} \cdot \bm{B}}{2 M_N}
\, \mathbb{M}
,
\end{equation}
must be diagonalized, 
where the mass splitting is defined to be 
$\Delta_{\Lambda \Sigma} = M_{\Sigma^0} - M_{\Lambda}$.
Using 
$\lambda_\pm^{(s)}$
to denote the eigenstates in the presence of a magnetic field, 
 the corresponding energy eigenvalues are given by
\begin{eqnarray}
E_\pm^{(s)}
&=&
\frac{1}{2}
\Bigg[
\Delta_{\Lambda \Sigma} 
 - \frac{e s B_z}{M_N}
\left(
\mu_{\Sigma_0} + \mu_{\Lambda}
\right) 
\notag \\
&&
\pm
\sqrt{
\left( 
\Delta_{\Lambda \Sigma} 
- \frac{e s B_z}{M_N}
\Delta \mu 
\right)^2 +
\left(
\mu_{\Lambda\Sigma}
\frac{e B_z }{M_N}
\right)^2}
\,
\Bigg]
, \quad \,
\label{eq:physmix}
\end{eqnarray}
%
with $s = \pm \frac{1}{2}$ and 
$\Delta \mu = \mu_{\Sigma^0} - \mu_\Lambda$. 
Contributions from the magnetic polarizabilities are omitted  because there is a lack of experimental information to constrain them.

%
%
%
\begin{figure}
\resizebox{0.9\linewidth}{!}{
\includegraphics{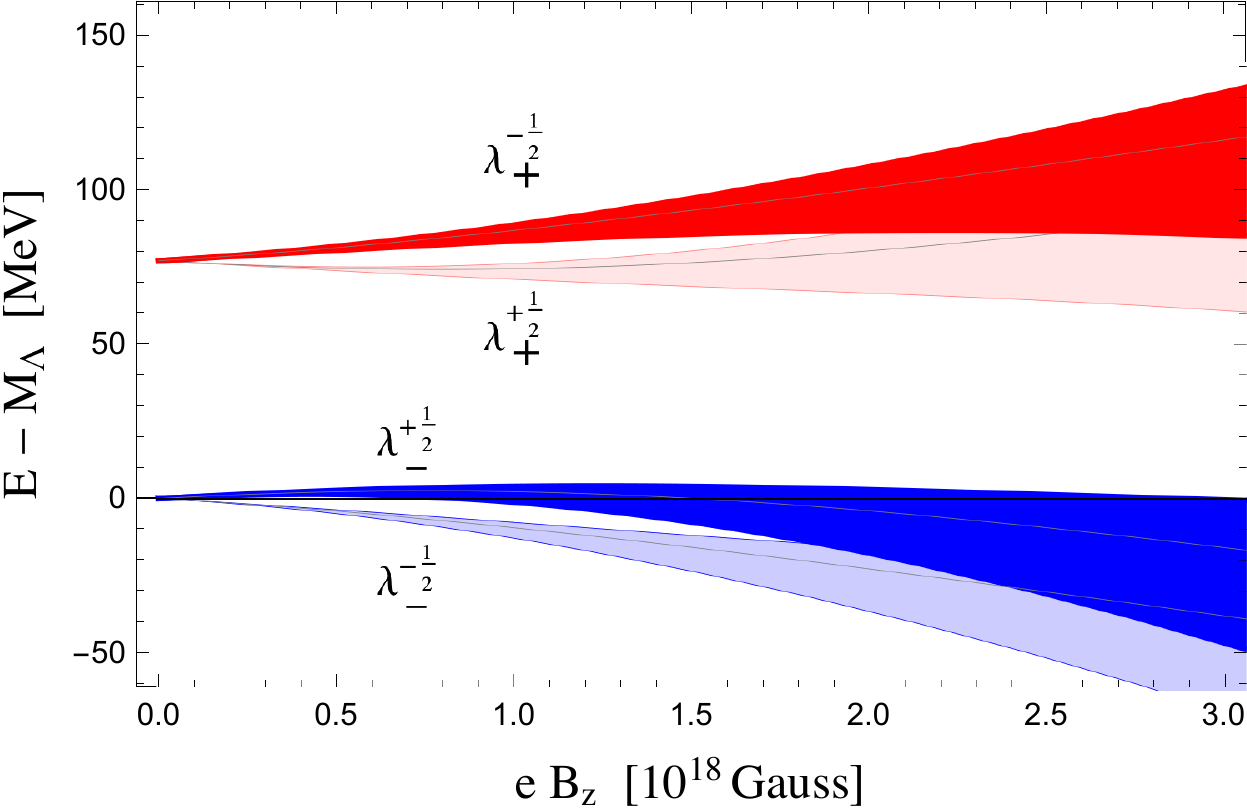}
}
\caption{
Anticipated energy levels of the  $\Lambda$--$\Sigma^0$ system as a function of magnetic field, 
$e \, B_z$
at the physical quark masses. 
The energy relative to the mass of the $\Lambda$  baryon is determined from Eq.~\eqref{eq:physmix}
using experimental data for the magnetic moments. 
To limit the effect from magnetic polarizabilities and higher-order terms, 
the field strength is restricted to smaller values compared to those in
Fig.~\ref{f:LamSigB}.
The bands  arise from including a polarizability term in each energy, 
and varying each $\beta$  between $-5$ and  $+10 \, \, \texttt{[10}^{\texttt{-4}} \,  \texttt{fm}^{\texttt{3}} \texttt{]}$. 
}
\label{f:physmix}
\end{figure}
%
%
%

Using the experimentally measured mass splitting  and the magnetic moments from  Eq.~\eqref{eq:MM},
the magnetic-field dependence of the energy eigenstates is shown in Fig.~\ref{f:physmix},
and should be contrasted with that in the  $SU(3)_F$-symmetric case shown in Fig.~\ref{f:LamSigB}. 
As Ensemble III is closer to the $SU(3)_F$ limit than nature, 
we expect the $\Lambda$--$\Sigma^0$ system to behave more like that 
found in the $SU(3)_F$ limit.
Contrasting expectations at the physical quark masses with the behavior found at the 
$SU(3)_F$-symmetric point, 
we see that extraction of moments in the 
$\Lambda$--$\Sigma^0$ 
system from 
Eq.~\eqref{eq:physmix}
will be challenging, 
as will the polarizabilities.

\section{Strange Matter In Large Magnetic Fields}
\label{sec:LargeB}

The above study focuses on the patterns and scaling of the baryon magnetic moments,
which, 
along with the electric charges,  
dominate the response of the baryons to small applied magnetic fields.
The calculations presented herein allow exploration of the behavior of baryons in 
very large magnetic fields, 
up to field strengths, 
$B \, \lsim 10^{19}$ Gauss, 
comparable to fields conjectured to exist in the context of astrophysical environments, 
in particular, the interiors of magnetars%
~\cite{Harding:2006qn}.
It is interesting to consider the  neutron and hyperon states
to address the possibility of a large magnetic field stabilizing strange baryons in dense matter, 
and consequently softening the nuclear equation of state.

\begin{figure}
\resizebox{0.9\linewidth}{!}{
\includegraphics{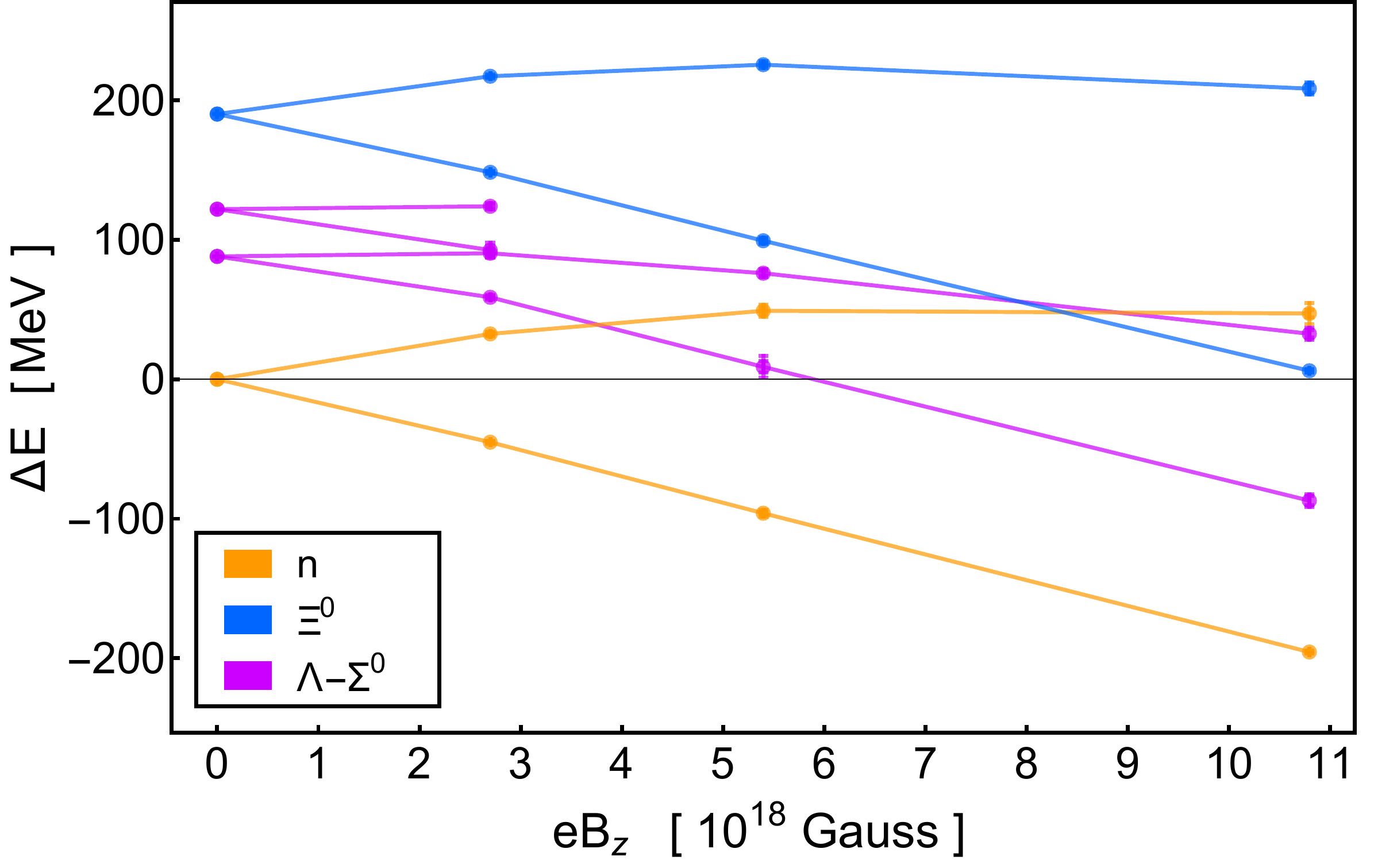} 
}
\caption{
Energies of the electrically neutral  baryons in magnetic fields relative to the nucleon mass in zero field for  
$m_\pi \sim 450 \, \texttt{MeV}$.
These results do not include contributions from quark-disconnected diagrams.
Not all relevant energy levels could be extracted from the LQCD calculations,  
and is the reason for termination 
of two of the levels at 
$e B_z=3\times 10^{18}~{\rm Gauss}$ in the plot.
 }
  \label{fig:largefields}
\end{figure}

The composition of dense hadronic matter is determined by an interplay between the hadron masses, 
Pauli-blocking, conserved 
charges, and by the interactions between hadrons. 
In the presence of a magnetic field, 
the relative energies of the hadrons change, 
as do the interactions between them.
Addressing the composition of dense hadronic matter in the presence of a magnetic field is a very complex task that is beyond the scope of this work.
Here, 
one part of this question is highlighted, 
namely the way in which the hadron masses change in large magnetic fields.
For simplicity, 
the focus is on the neutral baryons, 
as shown in 
Figs.~\ref{f:LamSigB} and \ref{fig:largefields}.

In the  $SU(3)_F$-symmetric limit
and in the absence of electromagnetism, 
there are sixteen degenerate states corresponding to the two spin states associated with each octet baryon.  
Including the leading QED self-energy correction increases the masses of the charged baryons by 
$\sim 1 \, \texttt{MeV}$.
Therefore, 
at zero density and in the absence of a background magnetic field, 
the lowest-lying state consists of the degenerate 
$n$, 
$\Lambda$, 
$\Sigma^0$, 
and
$\Xi^0$
baryons, 
each with two spin degrees of freedom, 
while the 
$p$, 
$\Sigma^\pm$, 
and 
$\Xi^-$, 
are degenerate, 
but higher in energy by 
$\sim 1 \, \texttt{MeV}$.
With the addition of a background magnetic field,  
the 
$n$, 
$\lambda_- \equiv \frac{\sqrt{3}}{2} \Lambda - \frac{1}{2} \Sigma^0$,  
and  
$\Xi^0$ 
remain degenerate for all values of the magnetic field due to 
$U$-spin symmetry, 
under which they form a triplet. 
As we shall argue, 
their spin-down components are the octet baryon states of lowest energy, 
and this is confirmed for  
$B \, \lsim 10^{19}$ Gauss 
at the quark masses used on Ensemble I, 
see Fig.~\ref{f:LamSigB}.
While this computation omits the quark-disconnected contributions, 
the valence $U$-spin symmetry
(see Appendix~\ref{s:B}) 
leads to degeneracies between the spin-projected
$\lambda_-$ 
and
$n$, 
$\Xi^0$ 
states, 
which we have also verified numerically.

While also negatively shifted, 
the spin-up component of the 
$U$-spin singlet state,
$\lambda_+ \equiv \frac{1}{2} \Lambda + \frac{\sqrt{3}}{2} \Sigma^0$, 
is found to be higher in energy than the triplet.
This can be attributed at 
$\cO(B^2)$ 
to the smaller magnetic polarizability of the singlet versus the triplet states, 
and occurs due to the sign of transition polarizability, 
$\beta_{\Lambda \Sigma} < 0$, 
see Eq.~\eqref{eq:LamSigSpec}. 
Because the transition polarizability does not receive contributions from quark-disconnected diagrams, 
the level ordering between singlet and triplet states is robust against partial quenching of the magnetic field. 
Therefore, 
in the absence of strong interactions between baryons,
the ground state of dense hadronic matter at these quark masses would have an equal number density of the 
$U$-spin triplet states,
($n$, 
$\lambda_-$, 
and 
$\Xi^0$).
Not enough is presently known about the interactions between baryons in dense magnetized matter to determine how this conclusion will be modified.   
The higher-order energy shifts, 
moreover, 
are compromised by the omission of quark-disconnected diagrams;  
and, 
a thorough investigation of their importance would be required in larger magnetic fields.

On Ensemble III, 
$U$-spin is explicitly broken by the quark masses. 
For this 
$N_f=2+1$ 
case with 
$m_\pi\sim 450 \, \texttt{MeV}$,  
the two neutron spin states remain the states of lowest energy for 
$B \, \lsim 4\times 10^{18}$ Gauss.
Therefore,  
in the absence of interactions,  
spontaneous generation of magnetic fields of this size is unlikely to 
stabilize strange baryons in dense matter, 
continuing to prefer neutron matter.  
Above this critical value, 
however,
one of the components of the coupled 
$\Lambda$--$\Sigma^0$ 
system becomes lighter than the spin-up neutron,
see  
Fig.~\ref{fig:largefields}.
Consequently, 
the lowest-energy configuration in such large fields is likely to have non-zero strangeness.
Na\"ively extrapolating these results to the physical point, 
where the 
$SU(3)_F$-breaking  differences between baryons masses are larger,
and considering the energetics of such systems
(the energy density in large magnetic fields and the energy recovered from 
lowering the energy of the baryon states), 
suggests that it is unlikely that a spontaneously generated magnetic field could 
stabilize strange matter in any known astrophysical setting.

\section{Summary}
\label{s:end}

The magnetic moments of the lowest-lying octet of baryons are calculated with LQCD including uniform and constant 
background magnetic fields.
This technique allows for determinations of the energies of each baryon spin state 
 as a function of  applied magnetic field, while 
the corresponding Zeeman splittings allow for extractions of the magnetic moments. 
These calculations are performed on three ensembles of gauge configurations, 
from which the pion-mass and lattice-spacing dependences of the magnetic moments are explored. 
Several interesting observations are made based on these results.
\begin{itemize}

\item
Baryon magnetic moments are consistent with  mild pion-mass dependence 
($\lesssim 10\%$ from 
$m_\pi \sim 800 \, \texttt{MeV}$
down to the physical point)
when expressed in units of natural baryon magnetons, 
$\texttt{[nBM]}$
defined in Eq.~\eqref{eq:BM}. 
This feature is shown in 
Fig.~\ref{f:Mplot}. 

\item
In such natural units, 
the baryon anomalous magnetic moments  take essentially only three values: 
$\delta \mu_B \sim 0,\pm 2$, 
see Fig.~\ref{f:BMplot}.
The vanishing anomalous moments imply nearly point-like magnetic structure for the 
$\Sigma^-$
and
$\Xi^-$
hyperons. 

\item
The values of baryon anomalous magnetic moments are  consistent with the
$SU(3)_F$-symmetric limit  
with  Coleman-Glashow moments taking the values $\mu_D \sim + 3$ and $\mu_F \sim + 2$
over a wide range of pion masses,  see Fig.~\ref{f:muDmuF}.

\item 
These particular values of the Coleman-Glashow moments are consistent with the NRQM, 
however, 
careful scrutiny of NRQM predictions reveals further features, 
see 
Figs.~\ref{f:MQ}--\ref{f:Rplot}.

\item
In most cases, 
the magnetic moments are consistent with relations derived from the large-$N_c$ limit of QCD.
At the $SU(3)_F$-symmetric point, 
however,
there is a notable exception that is more consistent with the NRQM than the  large-$N_c$ limit.

\item
A coupled-channels analysis is required to extract magnetic moments and transition moments 
from the $\Lambda$--$\Sigma^0$ system, 
because the magnetic field induces mixing between the 
$\Lambda$ and $\Sigma^0$.
At the $SU(3)_F$-symmetric point, such an analysis produces energy levels in the $\Lambda$--$\Sigma^0$ sector
consistent with expectations based upon $SU(3)_F$ symmetry, see Fig.~\ref{f:LamSigB}, 
and permits the first determination of the transition magnetic polarizability, given in Eq.~(\ref{eq:transmagpol}).

\item
Spin-dependent energy levels of the baryons in large magnetic fields are obtained, 
see Fig.~\ref{fig:largefields}, 
from which we conclude that it is unlikely that such fields stabilize strange matter in 
astrophysical objects at realistic densities.

\end{itemize}

After the decades that have passed since the discovery 
that nature is in close proximity to an exact flavor symmetry among the three lightest quarks,
the magnetic moments of the lowest-lying baryons continue to provide (increasingly subtle) glimpses into their structure.
Enabled by the largest supercomputers, 
LQCD calculations have the ability to explore the structure of matter in unphysical situations, 
and thereby provide new insights that cannot be gained through laboratory experiments. 
On the basis of what we find,  
another generation of more precise LQCD calculations 
over a broader range of light-quark masses is warranted.  
The scientific impact of such a series of LQCD calculations would be enhanced by improved
precision in the experimental determination of the strange baryon magnetic moments and, 
if possible, 
measurement of their polarizabilities.


\begin{acknowledgments}
We would like to thank Silas R.~Beane, Zohreh Davoudi,  Amol Deshmukh,  David Kaplan, Sanjay Reddy, and Phiala Shanahan 
for several interesting discussions and comments.
We are especially grateful to David Kaplan for first suggesting that 
\emph{natural baryon magnetons} 
might be an interesting unit to consider
during our analysis that appeared in 
Ref.~\cite{Beane:2014ora}.
WD, MJS, and BCT would like to thank the KITP at UC Santa Barbara for kind hospitality during the completion of this work, 
and partial support from the U.S.~National Science Foundation under Grant No. NSF PHY11-25915.
Calculations were performed using computational resources provided by the 
Extreme Science and Engineering Discovery Environment (XSEDE), 
which is supported by U.S. National Science Foundation grant number OCI-1053575, 
and NERSC, which is supported by U.S. Department of Energy Grant Number DE-AC02-05CH11231.
The PRACE Research Infrastructure resources Curie based in France at the Tr\`es Grand Centre de Calcul and MareNostrum-III
based in Spain at the Barcelona Supercomputing Center
were also used.
Development work required for this project was carried out on the Hyak High Performance Computing and Data Ecosystem at 
the University of Washington, 
supported, in part, by the U.S. National Science Foundation Major Research Instrumentation Award, 
Grant Number 0922770.
Parts of the calculations used the Chroma software
suite~\cite{Edwards:2004sx}.  
WD was supported in part by the
U.S. Department of Energy Early Career Research Award DE-SC00-10495 and by Grant Number DE-SC0011090.
EC was supported in part by the USQCD SciDAC project, 
the U.S. Department of Energy through Grant Number DE-SC00-10337, 
and by 
U.S. Department of Energy grant No. DE-FG02- 00ER41132.
KO was supported by the 
U.S. Department of Energy through Grant Number DE- FG02-04ER41302 and by the 
U.S. Department of Energy through Grant Number DE-AC05-06OR23177,
under which JSA operates the Thomas Jefferson National Accelerator Facility.  
The work of AP was supported by the contract
FIS2011-24154 from MEC (Spain) 
and 
FEDER. 
MJS was supported in part by U.S. Department of Energy grants No. DE-FG02-00ER41132 and DE-SC00-10337.
BCT was supported in part by a joint 
The City College of New York-RIKEN/Brookhaven Research Center fellowship, 
a grant from the Professional Staff Congress of The CUNY, 
and by the U.S. National Science Foundation, under Grant No. PHY15-15738.
BCT also acknowledges the INT for its hospitality and partial support during the intermediate stages of this work, 
and the organizers of the program 
``INT-16-1: Nuclear Physics from Lattice QCD''
for providing a stimulating environment. 
\end{acknowledgments}


\appendix


\section{Analysis---Fits to Zeeman Splittings and Extraction of Magnetic Moments}
\label{s:A}

Fits to the two-point correlation functions calculated with LQCD,
which lead to  extractions of the baryon Zeeman splittings as a function of the background magnetic field,
are described here. 
Subsequent fits to the field-strength dependence of these splittings,
which enable the determination of baryon magnetic moments, 
are also described.
While multiple independent analyses of the LQCD correlation functions have been performed in the present work, 
only one analysis is detailed, as the values extracted agree between analyses within the quoted uncertainties.

Calculating quark propagators on each QCD gauge configuration with background magnetic fields enables computation of  
 spin-projected baryon two-point functions.  
 The smeared and point baryon interpolating operators employed in this work are 
discussed in Ref.~\cite{Chang:2015qxa}. 
With the exception of the coupled  $\Lambda$--$\Sigma^0$ system,  which is detailed separately in  Appendix~\ref{s:B}, 
the main analysis utilizes both SS and SP correlation functions in linear combinations chosen to minimize uncertainties on the extraction of energies. 
From each ensemble of $N_\text{cfg}$ correlation functions (see Table~\ref{t:configs}) associated with a given baryon channel,
the average correlation functions from blocks of size $N_{\text{block}}$, 
where $N_{\text{block}} = 7$,  $1$,  and $6$, are calculated  for Ensembles I, II, and III, respectively.  
These block-averaged correlation functions are labeled by
$G^{(s)}_{i} (t, n_\Phi)$,  where  $s$  denotes the projection of baryon spin along the $z$-axis,  and  $i$ 
is the blocked ensemble index, $i = 1, \ldots, N_{\text{cfg}} / N_{\text{block}}$. 
Correlators are computed for the various baryons, for each value of the magnetic flux quantum. 
We use the quanta $n_\Phi = 0$,  $3$,  $-6$,  and $12$
($n_\Phi = -6$ is treated as  $n_\Phi = 6$ by reversing the spin axis).   
The block-averaged correlation functions are
 used to create bootstrap ensembles of size
$N_{\text{BS}} = N_{\text{boot}} N_{\text{cfg}} / N_{\text{block}}$, 
where each member of the bootstrap ensemble consists of an average of 
$M_{\text{BS}} = N_{\text{cfg}} / N_{\text{block}}$
random samples of the blocked data. 
The bootstrap factor, 
$N_{\text{boot}}$,
has the value 
$4$, 
$3$, 
and
$4$
on Ensembles I, II, and III, respectively.

%
%
%
\begin{figure}
\resizebox{0.725\linewidth}{!}{
\includegraphics{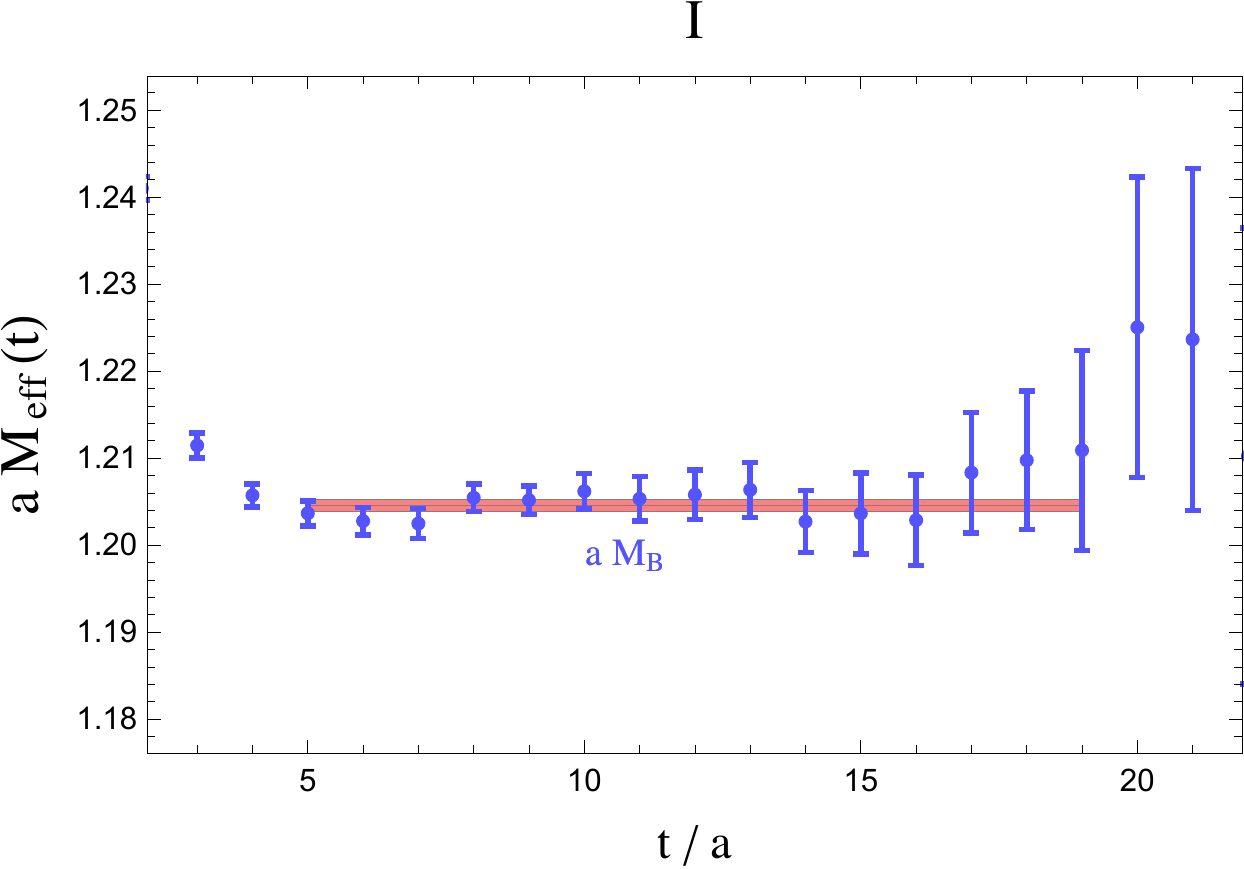}
}
\vskip 0.05in
\resizebox{0.725\linewidth}{!}{
\includegraphics{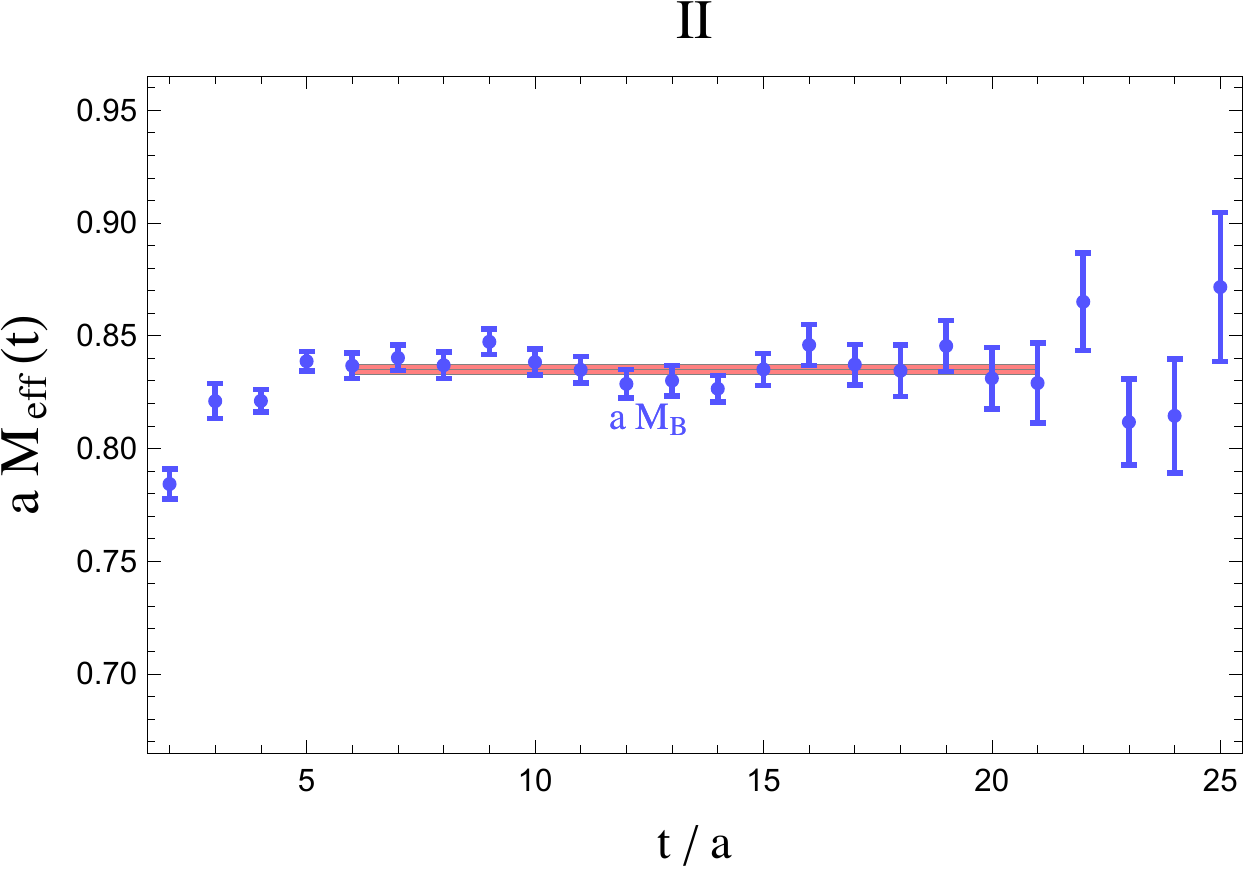}
}
\vskip 0.05in
\resizebox{0.725\linewidth}{!}{
\includegraphics{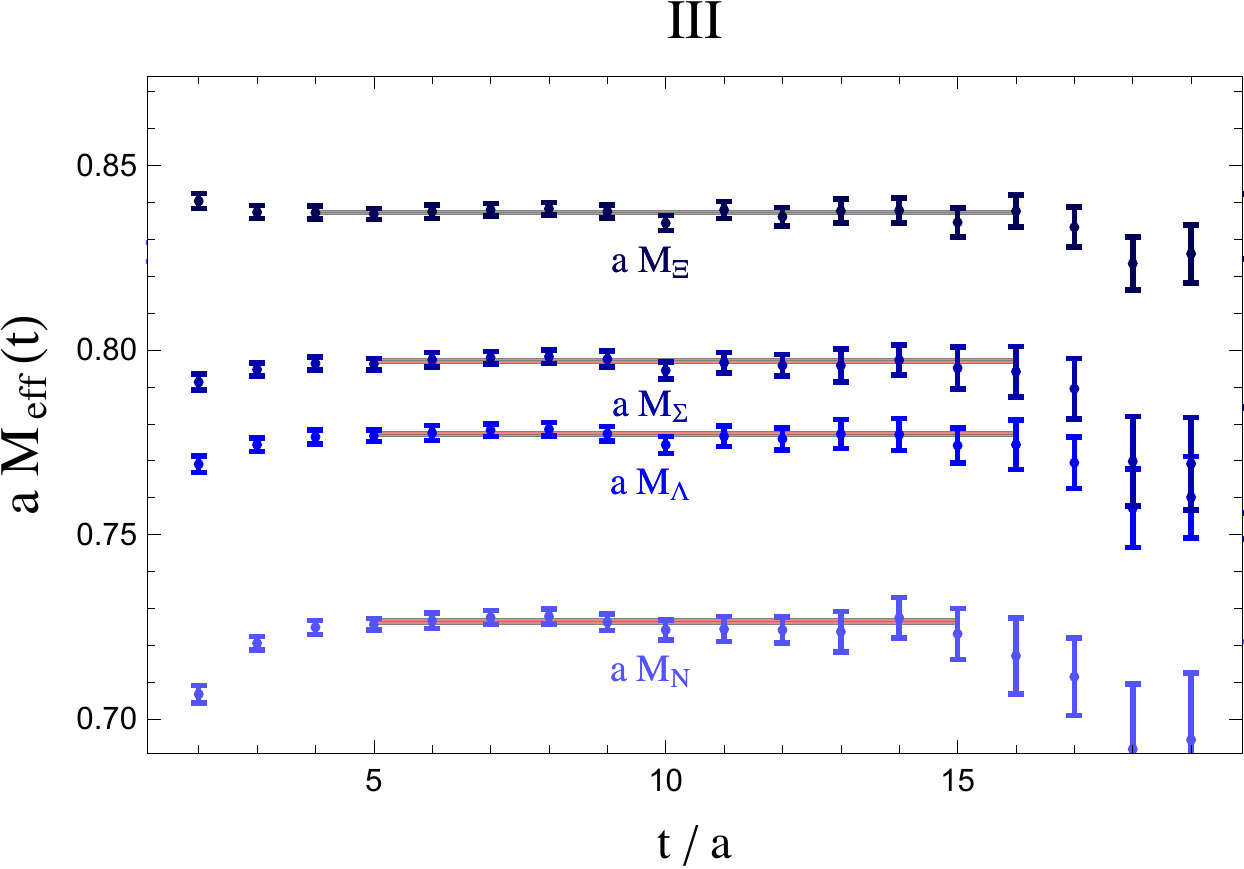}
}
\caption{
Baryon effective-mass plots for Ensembles I--III and  fits to the baryon masses, 
with bands depicting the quadrature-combined statistical and systematic uncertainties. 
The latter arise from both the fit and the choice of fit window. 
} 
\label{f:Meff}
\end{figure}
%
%
%

To determine baryon masses,  a bootstrap ensemble
from block and spin-averaged correlation functions
\begin{equation}
G_{i}(t)
\equiv
\frac{1}{2}
\left[
G^{(+\frac{1}{2})}_{i}(t,0) 
+ 
G^{(-\frac{1}{2})}_{i}(t,0)
\right]
,
\end{equation}
in vanishing magnetic field,  $n_\Phi = 0$, is used. 
To aid in the analysis,  the effective-mass function for each member of the bootstrap ensemble
\begin{equation}
\left\{
a M_{\text{eff}} (t)
\right\}_i
\equiv
- \log \frac{G_i(t+a)}{G_i(t)}
,
\end{equation}
is formed, and the  effective masses are shown in  Fig.~\ref{f:Meff}. 
Fits to  $G_i(t)$
enable extraction of the baryon ground-state energies
through the long-time behavior of the ensemble 
average 
$\sim Z \exp (- E t)$,  and fit windows are chosen to maximize the correlated $\chi^2$-probability, 
also known as the integrated $\chi^2$.
The window is then varied over the eight adjacent fit windows obtained by adjusting the starting and ending time by one 
lattice unit in either direction.  
Values of the baryon masses extracted from the fits are collected in 
Table~\ref{t:bigtable}.

To determine the baryon Zeeman splittings,  $a \Delta E$,  double ratios of block-averaged, spin-projected baryon correlation functions are constructed. 
These have the form
\begin{equation}
R_i
(t)
=
\frac{G^{(+\frac{1}{2})}_i(t,n_\Phi)} 
{G^{(-\frac{1}{2})}_i(t,n_\Phi)}
\Bigg/
\frac{G^{(+\frac{1}{2})}_i(t,0)}
{G^{(-\frac{1}{2})}_i(t,0)}
\label{eq:Ratio}
,
\end{equation}
where, for simplicity, the dependence on the flux quantum in  $R_i(t)$ is implicit. 
The bootstrap ensemble of double ratios suppresses statistical fluctuations. 
These ratios are shown in 
Fig.~\ref{f:EeffB}, 
and the long-time behavior of their ensemble average, 
$\sim Z \exp \left( - \Delta E \, t \right)$, 
enables extraction of the Zeeman splitting 
$\Delta E$
defined in 
Eq.~\eqref{eq:Zee}.
The fitting strategy employed for 
$R_i(t)$
is the same as that utilized for
$G_i(t)$. 
There is one important restriction on establishing fit windows for  $R_i(t)$:
windows must begin only after each of the individual correlators in the double ratio,
Eq.~\eqref{eq:Ratio}, exhibits ground-state saturation. 
The extracted Zeeman splittings are given in  Table~\ref{t:bigtable}.

The magnetic moments can be extracted from the Zeeman splittings in sufficiently weak magnetic fields, 
Eq.~\eqref{eq:Linear}, and 
consequently fits of various functions of the magnetic field to these splittings are considered. 
Three functional forms are fit to the bootstrap  ensembles,
\begin{eqnarray}
F_1(B) 
&=& 
- \mu B
, 
\notag \\
F_2(B) 
&=&
- \mu B 
+  f_2 \,  B | B | 
,
\notag \\
F_3(B) 
&=&
- \mu B 
+ 
f_3 \, B^3
\label{eq:BFits}
,
\end{eqnarray}
where the second fit function is motived by relativistic corrections to the Zeeman splittings due to Landau levels, 
see Ref.%
~\cite{Lee:2014iha}. 
Accordingly $F_2(B)$  is used only to fit the Zeeman splittings of charged particles. 
When   lattice units are used for the magnetic field,  $a^2 e B_z$,  the subsequent extraction of magnetic moments
are in lattice magnetons,
%
$\texttt{[LatM]}
= 
\frac{1}{2} e \, a
$.
%
The magnetic field dependence of the extracted Zeeman splittings is shown in 
Fig.~\ref{f:BfitB}, 
along with two representative fits:
a fit to all three splittings using 
$F_3(B)$, 
and a fit using 
$F_1(B)$ 
omitting the largest magnetic field. 
To quantify the uncertainty due to the choice of fit function, 
we also use fits to all three splittings using 
$F_1(B)$,
and 
$F_2(B)$ 
for charged particles. 
The values extracted for magnetic moments appear in 
Table~\ref{t:bigtable}. 
In many cases, 
the dominant uncertainty in determining magnetic moments arises from the systematics of the fit. 
This can be remedied in the future by performing computations at additional  magnetic field strengths.

\begin{widetext}

%
\begin{table}
\caption{%
The left panel shows  baryon masses,  $a M_B$,  determined on the three ensembles.
The first uncertainty shown on masses is statistical,  
while the second is the fit systematic, 
including the choice of fit window. 
Notice that the octet baryons are degenerate on Ensembles I and II.
Additionally, in the left panel, Zeeman splittings, 
$a \, \Delta E$, 
computed on the three ensembles, 
for three values of the magnetic flux quantum, 
$n_\Phi$, are shown. 
The first uncertainty  is statistical, 
while the second is the systematic due to the fit, 
including the choice of fit window. 
Zeeman splittings on Ensembles I and II are paired due to    
$U$-spin symmetry. 
On the right appear 
baryon magnetic moments, 
$\mu_B$, 
determined in lattice magnetons
$\texttt{[LatM]}$, 
$\texttt{[NM]}$ 
and 
$\texttt{[nNM]}$, 
see Eq.~\eqref{eq:nNM}.  
The first uncertainty quoted on magnetic moments is statistical, 
while the second is the systematic due to the fit, 
including the choice of fit function. 
The third is the uncertainty due to the determination of the lattice spacing, 
which is only present for quantities reported in 
$\texttt{[NM]}$. 
Ensembles I and II necessarily maintain exact 
$U$-spin symmetry leading to repeated entries for magnetic moments. 
}
\begin{center}
\resizebox{\linewidth}{!}{
\begin{tabular}{|l|ccc|}
\hline
\hline
& 
\multicolumn{3}{c|}{$a M_B$} 
\tabularnewline
\hline
\hline
I
&
\multicolumn{3}{c|}{$1.20456(13)(67)$} 
\tabularnewline
II
&
\multicolumn{3}{c|}{$0.8352(05)(20) \phantom{8}$} 
\tabularnewline 
\hline
\hline
& 
$a M_N$  
&
$a M_\Lambda$
&
\tabularnewline
\hline
\hline
III
&
$0.72649(17)(82)$ 
&
$0.77729(14)(69) \phantom{8}$
&
\tabularnewline
\hline
&
&
$a M_\Sigma$ 
& 
$a M_\Xi$
\tabularnewline
\hline
\hline
III
&
&
$0.79708(13)(71) \phantom{8}$ 
&
$0.83732(10)(53)$ 
\tabularnewline
\hline
\hline
\multicolumn{4}{c}{ }
\tabularnewline
\hline
\hline
I
& 
\multicolumn{3}{c|}{$a \, \Delta E$ }
\\
& 
$n_\Phi = 3$
& 
$n_\Phi = 6$
& 
$n_\Phi = 12$
\tabularnewline
\hline
\hline
$p, \, \Sigma^+$
&
$- 0.04664(18)(50) \phantom{8}$
&
$- 0.0933(04)(13) \phantom{8}$
&
$- 0.1804(12)(52) \phantom{8}$
\tabularnewline
$n, \, \Xi^0$
&
$\phantom{-} 0.029999(23)(77)$
&
$\phantom{-} 0.05792(07)(23)$
&
$\phantom{-} 0.10127(07)(37)$
\tabularnewline
$\Sigma^-,  \Xi^-$
&
$\phantom{-} 0.01793(26)(70) \phantom{8}$
&
$\phantom{-} 0.0334(07)(15) \phantom{8}$
&
$\phantom{-} 0.0695(13)(26) \phantom{8}$
\tabularnewline
\hline
\hline
II
& 
\multicolumn{3}{c|}{$a \, \Delta E$ }
\\
& 
$n_\Phi = 3$
& 
$n_\Phi = 6$
& 
$n_\Phi = 12$
\tabularnewline
\hline
\hline
$p, \, \Sigma^+$
&
$- 0.0280(08)(21) \phantom{8}$
&
$- 0.0559(08)(23) \phantom{8}$
&
$- 0.0986(34)(88) \phantom{8}$
\tabularnewline
$n, \, \Xi^0$
&
$\phantom{-} 0.01794(04)(15)$
&
$\phantom{-} 0.03515(10)(31)$
&
$\phantom{-} 0.06594(17)(60)$
\tabularnewline
$\Sigma^-, \Xi^-$
&
$\phantom{-} 0.01099(26)(99)$
&
$\phantom{-} 0.0195(05)(19) \phantom{8}$
&
$\phantom{-} 0.0417(13)(35) \phantom{8}$
\tabularnewline
\hline
\hline
III
& 
\multicolumn{3}{c|}{$a \, \Delta E$ }
\\
& 
$n_\Phi = 3$
& 
$n_\Phi = 6$
& 
$n_\Phi = 12$
\tabularnewline
\hline
\hline
$p$
&
$- 0.07250(23)(94)$
&
$- 0.1393(08)(26)$
&
$- 0.2353(22)(64)$
\tabularnewline
$\Sigma^+$
&
$- 0.07058(20)(69)$
&
$- 0.1352(06)(20)$
&
$- 0.2314(19)(50)$
\tabularnewline
$n$
&
$\phantom{-} 0.04746(09)(35)$
&
$\phantom{-} 0.0903(03)(14)$
&
$\phantom{-} 0.1399(08)(28)$
\tabularnewline
$\Xi^0$
&
$\phantom{-} 0.04099(10)(28)$
&
$\phantom{-} 0.0771(05)(14)$
&
$\phantom{-} 0.1224(05)(15)$
\tabularnewline
$\Sigma^-$
&
$\phantom{-} 0.02836(29)(98)$
&
$\phantom{-} 0.0538(04)(16)$
&
$\phantom{-} 0.1054(18)(39)$
\tabularnewline
$\Xi^-$
&
$\phantom{-} 0.02083(37)(97)$
&
$\phantom{-} 0.0418(05)(19)$
&
$\phantom{-} 0.0962(23)(73)$
\tabularnewline
\hline
\hline
\end{tabular}
$\quad$
\begin{tabular}{|l|ccc|}
\hline
\hline
& 
\multicolumn{3}{c|}{$\mu _B\, \texttt{[LatM]}$}
\\
$B$
& 
I 
& 
II
& 
III
\tabularnewline
\hline
\hline
$p$
&
$\quad
\phantom{-} 2.534(12)(28)
\quad$
&
$\phantom{-} 3.42(08)(15) \phantom{8}$
&
$\quad
\phantom{-} 3.984(30)(69)
\quad$
\tabularnewline
$\Sigma^+$
&
$\quad
\phantom{-} 2.534(12)(28)
\quad$
&
$\phantom{-} 3.42(08)(15) \phantom{8}$
&
$\phantom{-} 3.873(23)(55)$
\tabularnewline
$n$
&
$-1.645 (03)(16)$
&
$-2.203(12)(22)$
&
$-2.626(11)(51)$
\tabularnewline
$\Xi^0$
&
$-1.645 (03)(16)$
&
$-2.203(12)(22)$
&
$-2.262(12)(39)$
\tabularnewline
$\Sigma^-$
&
$-0.943(12)(26)$
&
$-1.264(34)(80)$
&
$-1.513(26)(54)$
\tabularnewline
$\Xi^-$
&
$-0.943(12)(26)$
&
$-1.264(34)(80)$
&
$-1.136(20)(41)$
\tabularnewline
\hline
\hline
& 
\multicolumn{3}{c|}{$\mu _B\, \texttt{[NM]}$}
\tabularnewline
$B$
& 
I 
&
II
& 
III
\tabularnewline
\hline
\hline
$p$
&
$\phantom{-} 1.752(08)(19)(19)$
&
$\phantom{-} 1.640(39)(72)(23)$
&
$\phantom{-} 2.213(17)(39)(30)$
\tabularnewline
$\Sigma^+$
&
$\phantom{-} 1.752(08)(19)(19)$
&
$\phantom{-} 1.640(39)(72)(23)$
&
$\phantom{-} 2.151(13)(31)(29)$
\tabularnewline
$n$
&
$- 1.138(02)(11)(12)$
&
$- 1.056(06)(10)(15)$
&
$- 1.458(06)(28)(20)$
\tabularnewline
$\Xi^0$
&
$- 1.138(02)(11)(12)$
&
$- 1.056(06)(10)(15)$
&
$- 1.256(07)(30)(12)$
\tabularnewline
$\Sigma^-$
&
$- 0.652(08)(18)(07)$
&
$- 0.606(16)(38)(09)$
&
$- 0.840(15)(30)(12)$
\tabularnewline
$\Xi^-$
&
$- 0.652(08)(18)(07)$
&
$- 0.606(16)(38)(09)$
&
$- 0.631(11)(23)(09)$
\tabularnewline
\hline
\hline
& 
\multicolumn{3}{c|}{$\mu _B\, \texttt{[nNM]}$}
\tabularnewline
$B$
& 
I 
&
II
& 
III
\tabularnewline
\hline
\hline
$p$
&
$\phantom{-} 3.052(14)(34)$
&
$\phantom{-} 2.86(07)(13) \phantom{8}$
&
$\phantom{-} 2.895(22)(51)$
\tabularnewline
$\Sigma^+$
&
$\phantom{-} 3.052(14)(34)$
&
$\phantom{-} 2.86(07)(13) \phantom{8}$
&
$\phantom{-} 2.813(17)(40)$
\tabularnewline
$n$
&
$- 1.982(03)(19)$
&
$- 1.840(10)(19)$
&
$- 1.908(08)(37)$
\tabularnewline
$\Xi^0$
&
$- 1.982(03)(19)$
&
$- 1.840(10)(19)$
&
$- 1.643(09)(29)$
\tabularnewline
$\Sigma^-$
&
$- 1.136(14)(32)$
&
$- 1.056(28)(67)$
&
$- 1.099(19)(39)$
\tabularnewline
$\Xi^-$
&
$- 1.136(14)(32)$
&
$- 1.056(28)(67)$
&
$- 0.825(14)(30)$
\tabularnewline
\hline
\hline
\end{tabular}
}
\end{center}
\label{t:bigtable}
\end{table}

%
%
%
%
%
%
\begin{figure}
\resizebox{\linewidth}{!}{
\includegraphics{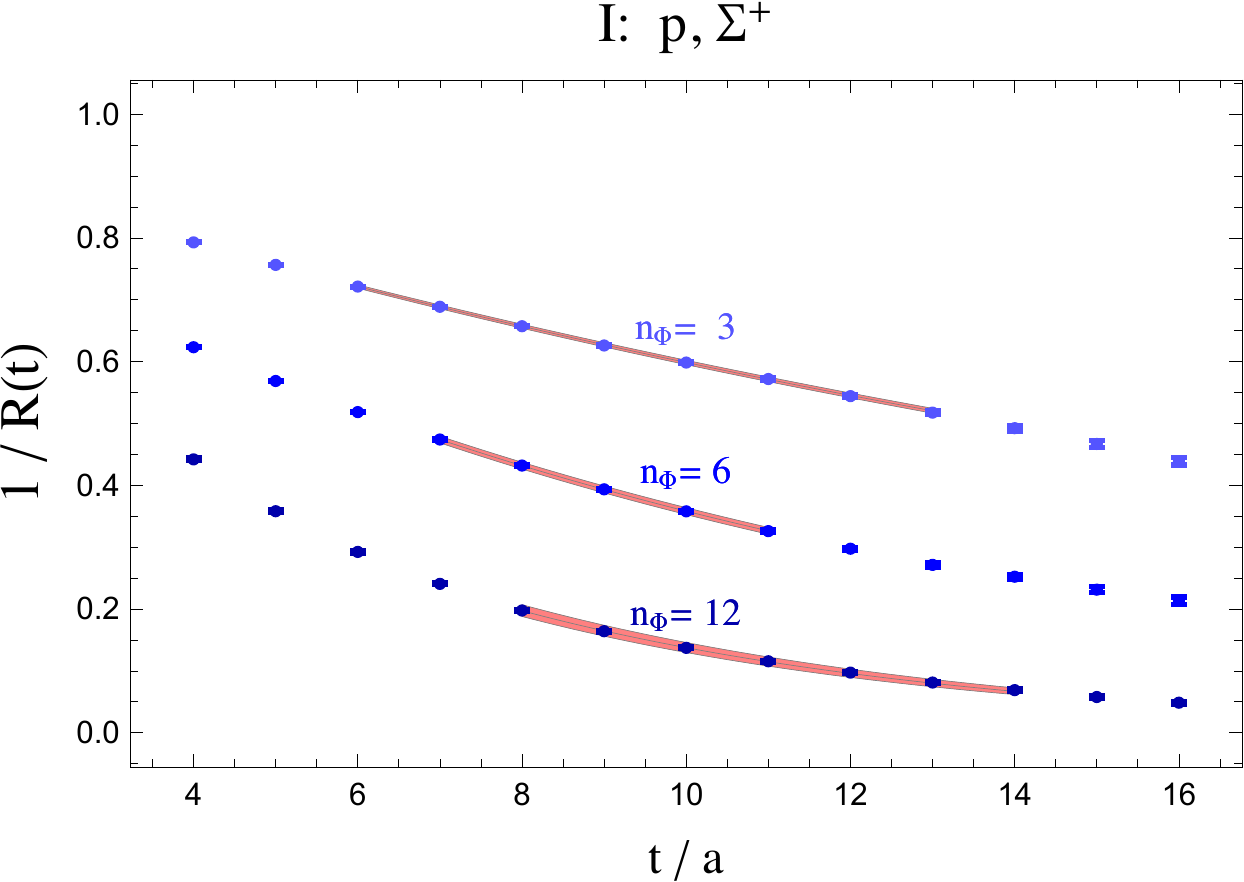}
$\quad$
\includegraphics{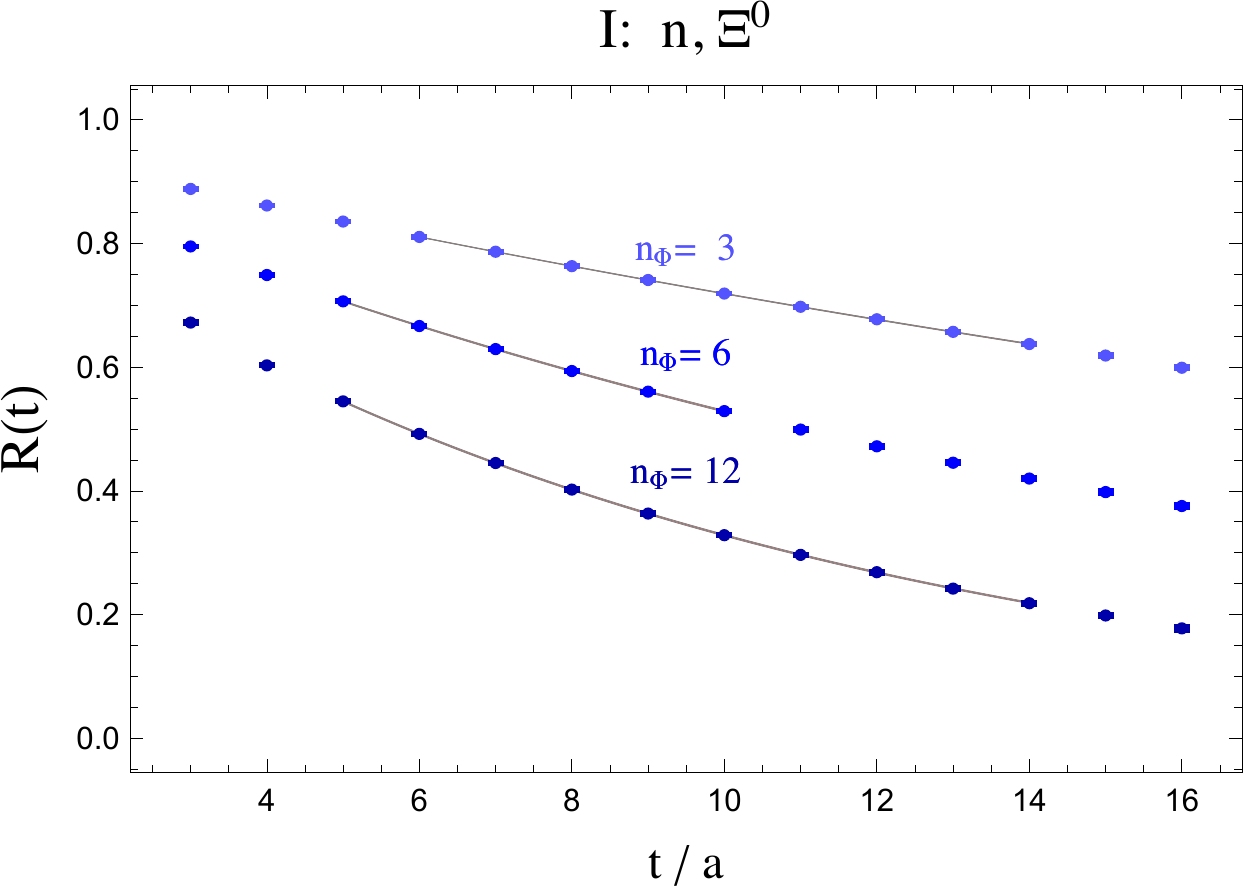}
$\quad$
\includegraphics{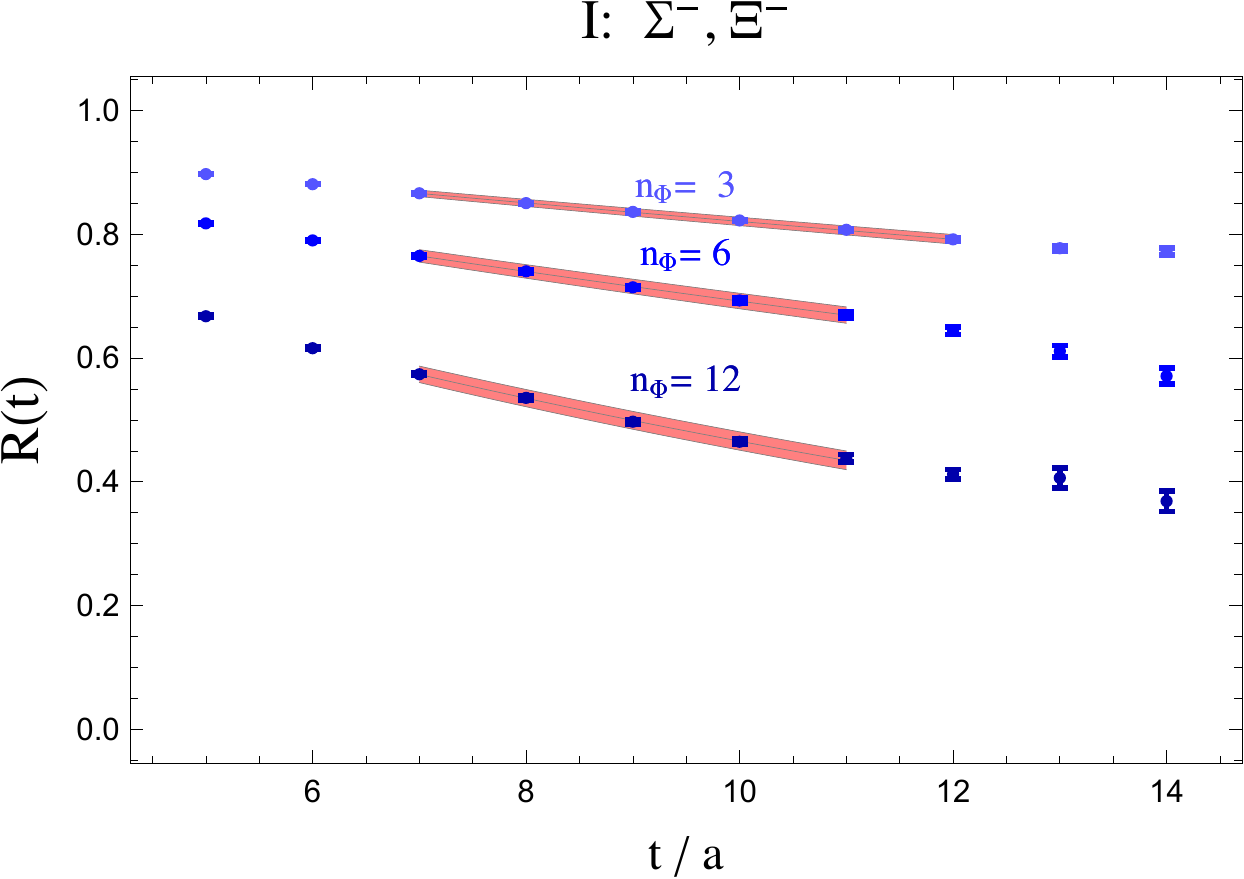}
}
\\
\medskip
\resizebox{\linewidth}{!}{
\includegraphics{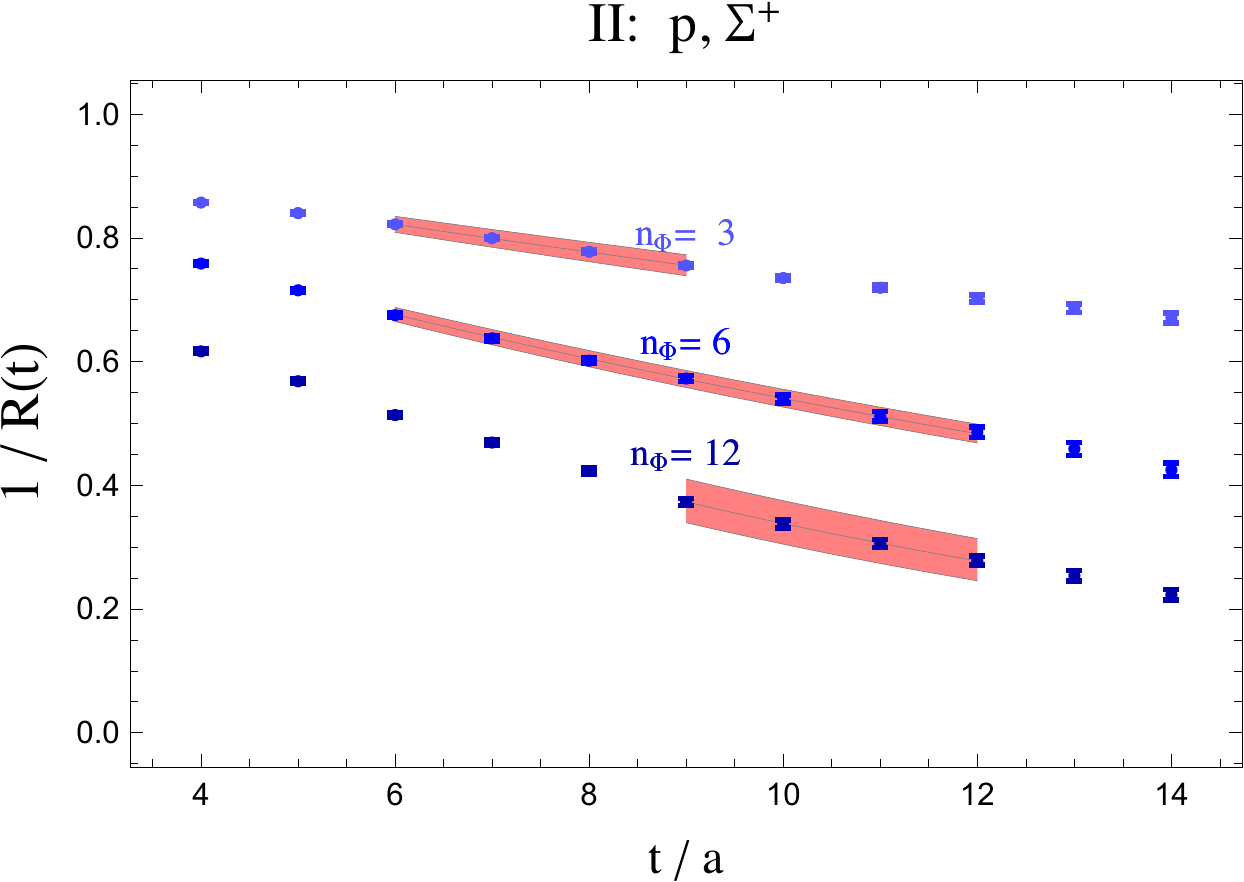}
$\quad$
\includegraphics{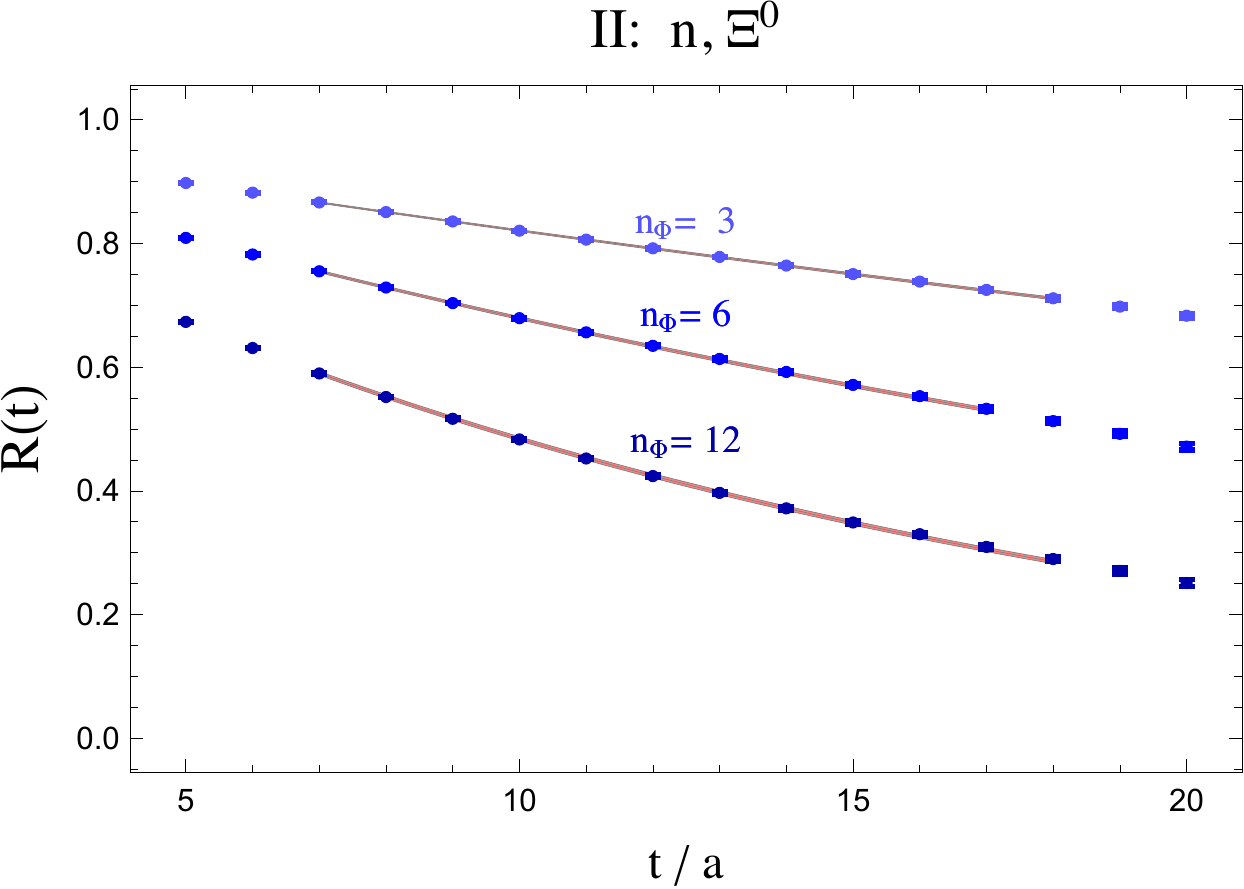}
$\quad$
\includegraphics{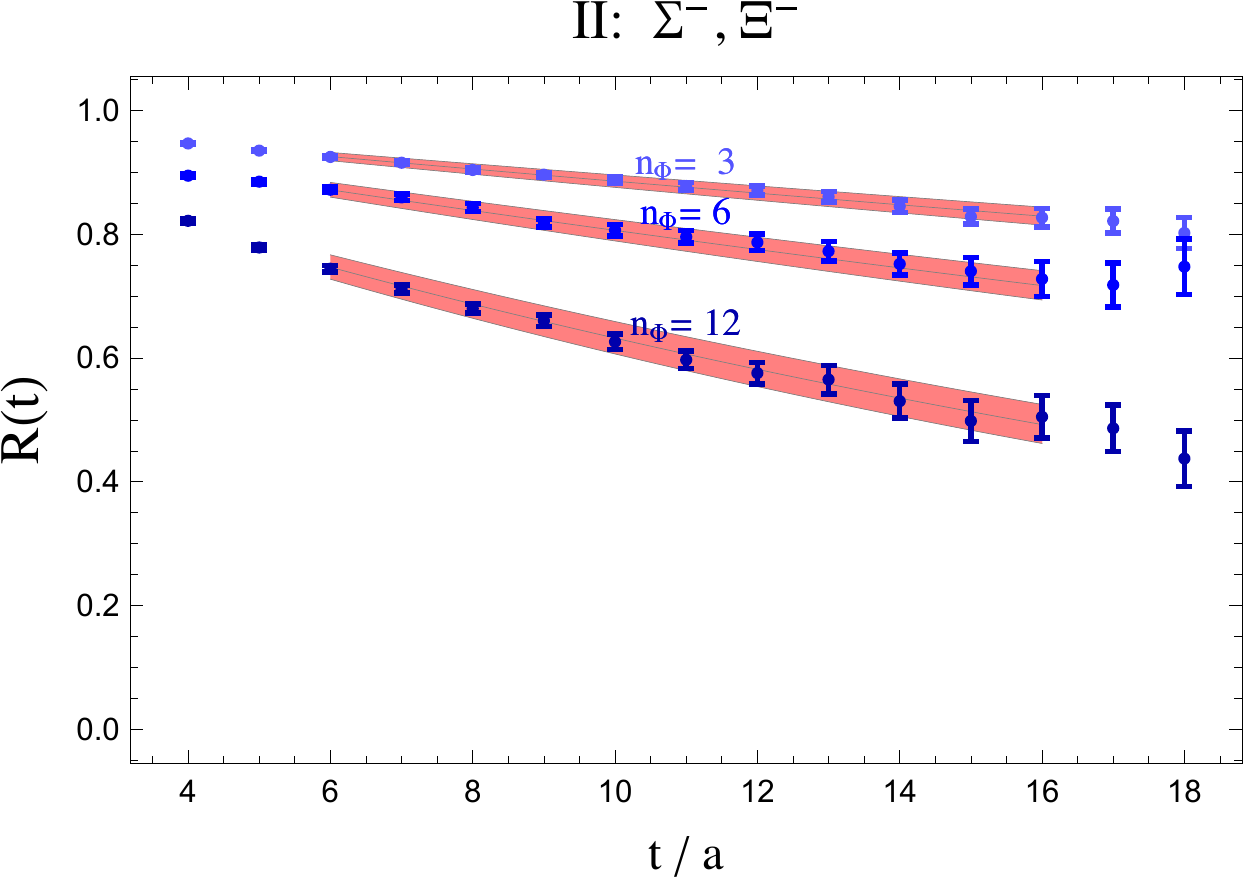}
}
\\
\medskip
\resizebox{\linewidth}{!}{
\includegraphics{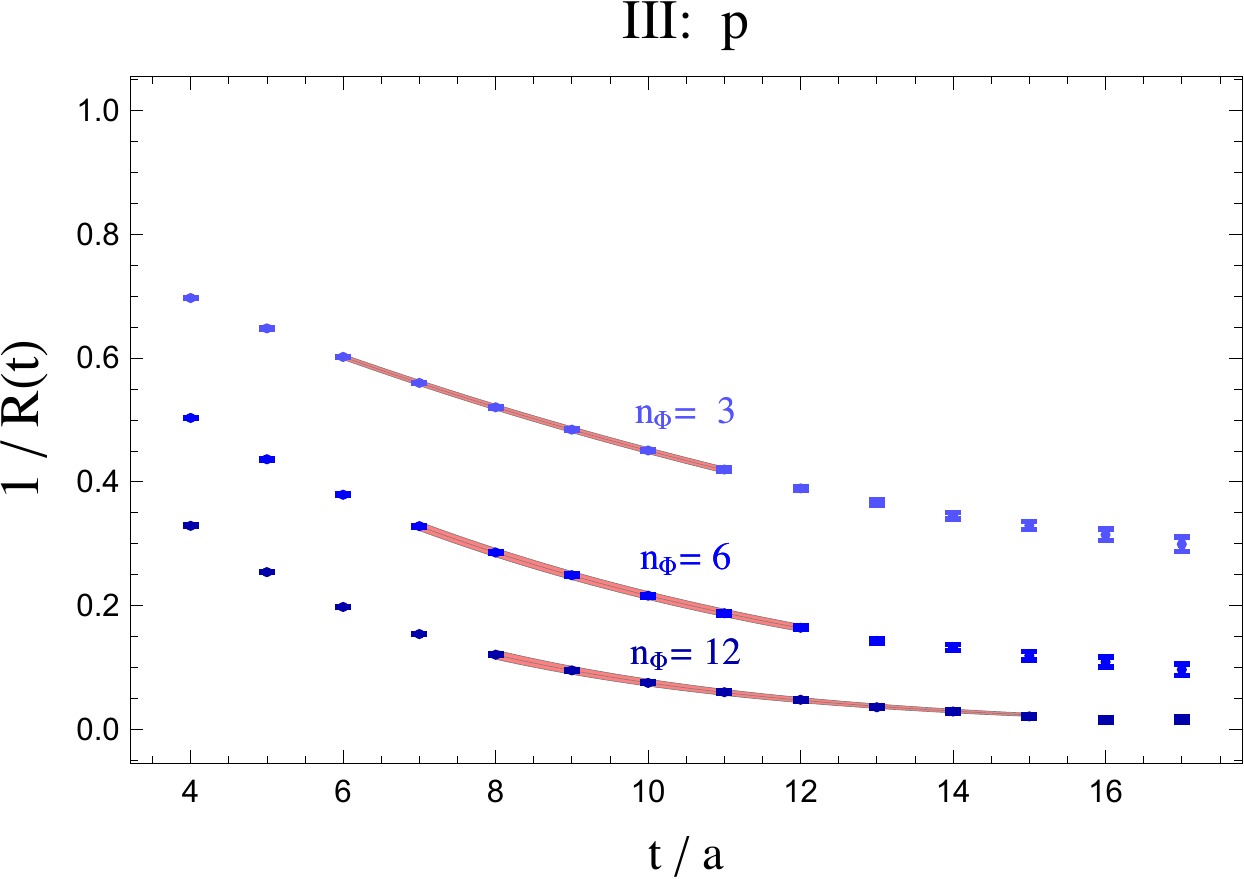}
$\quad$
\includegraphics{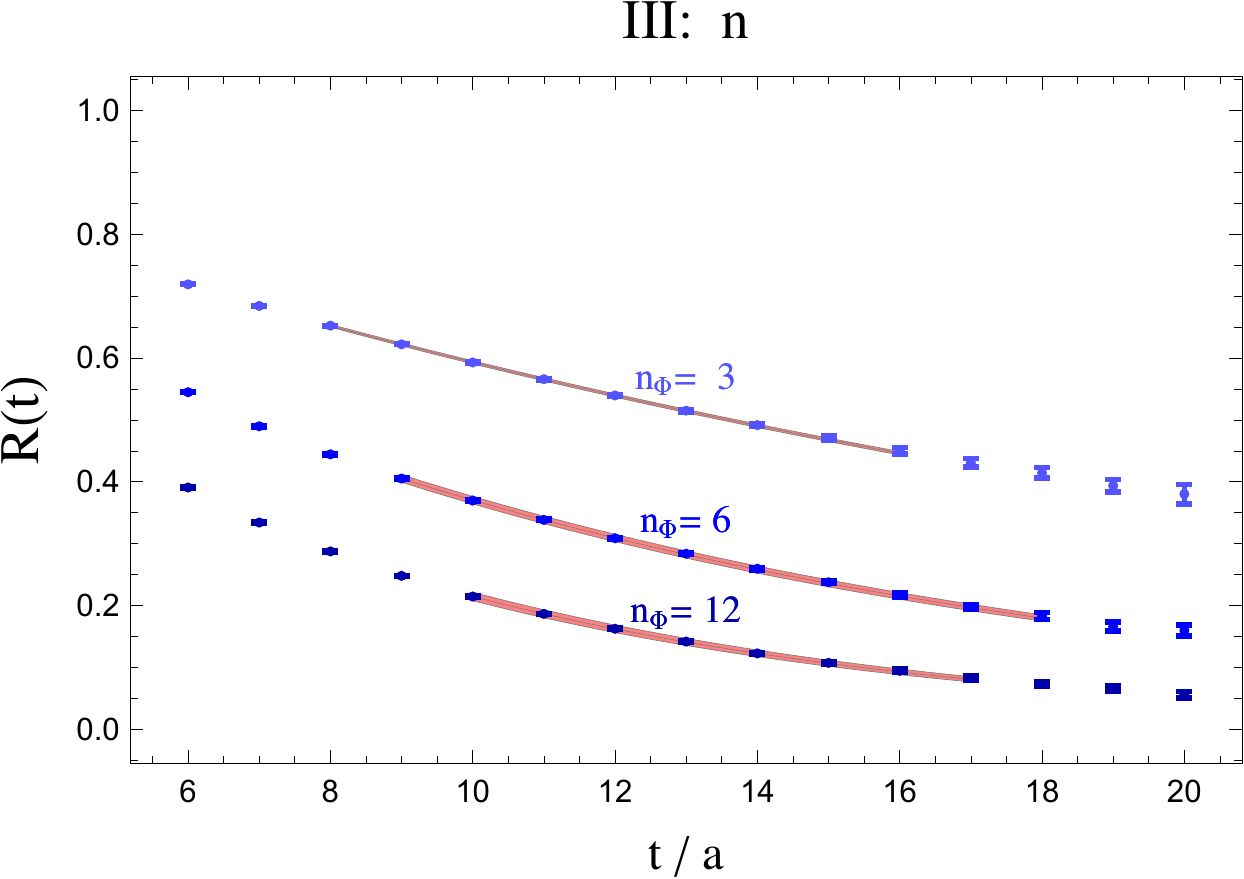}
$\quad$
\includegraphics{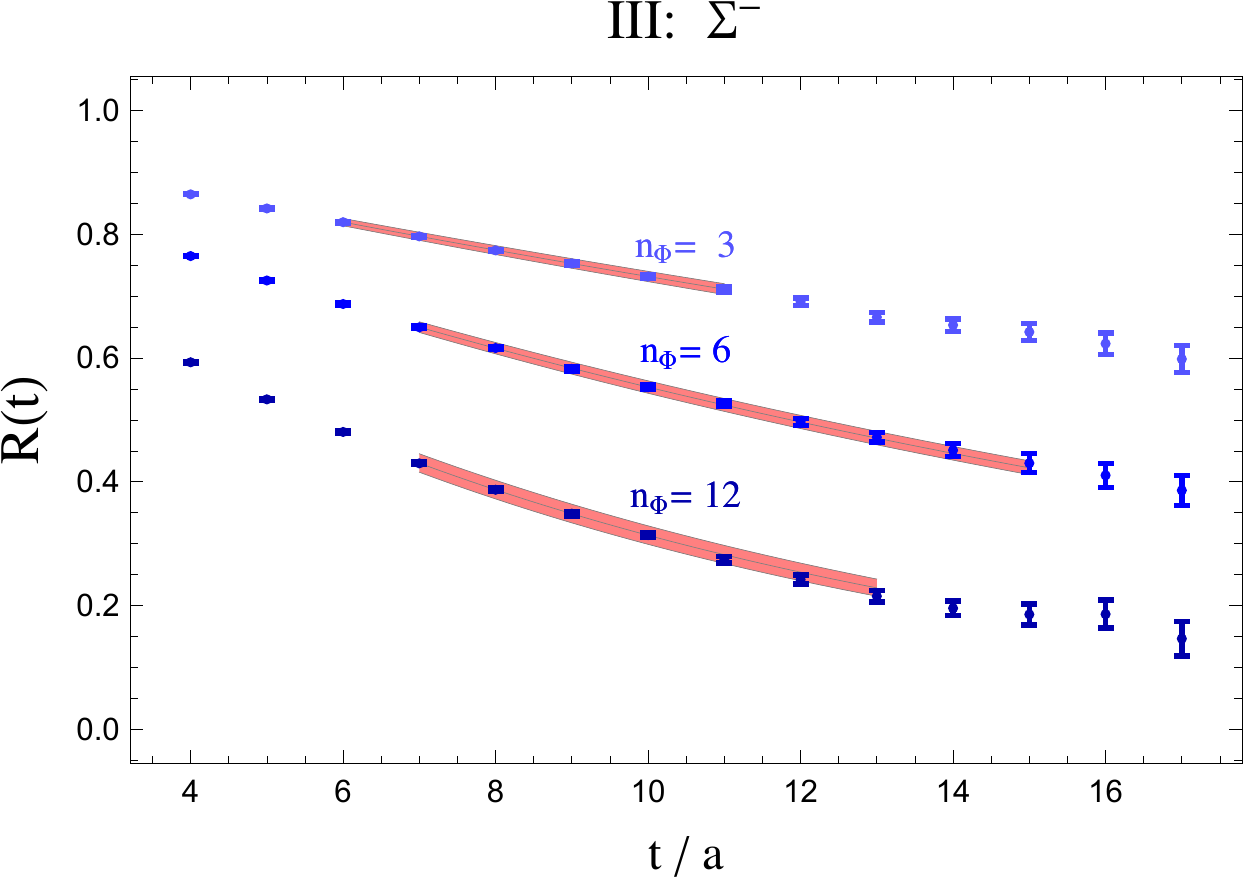}
}
\\
\medskip
\resizebox{\linewidth}{!}{
\includegraphics{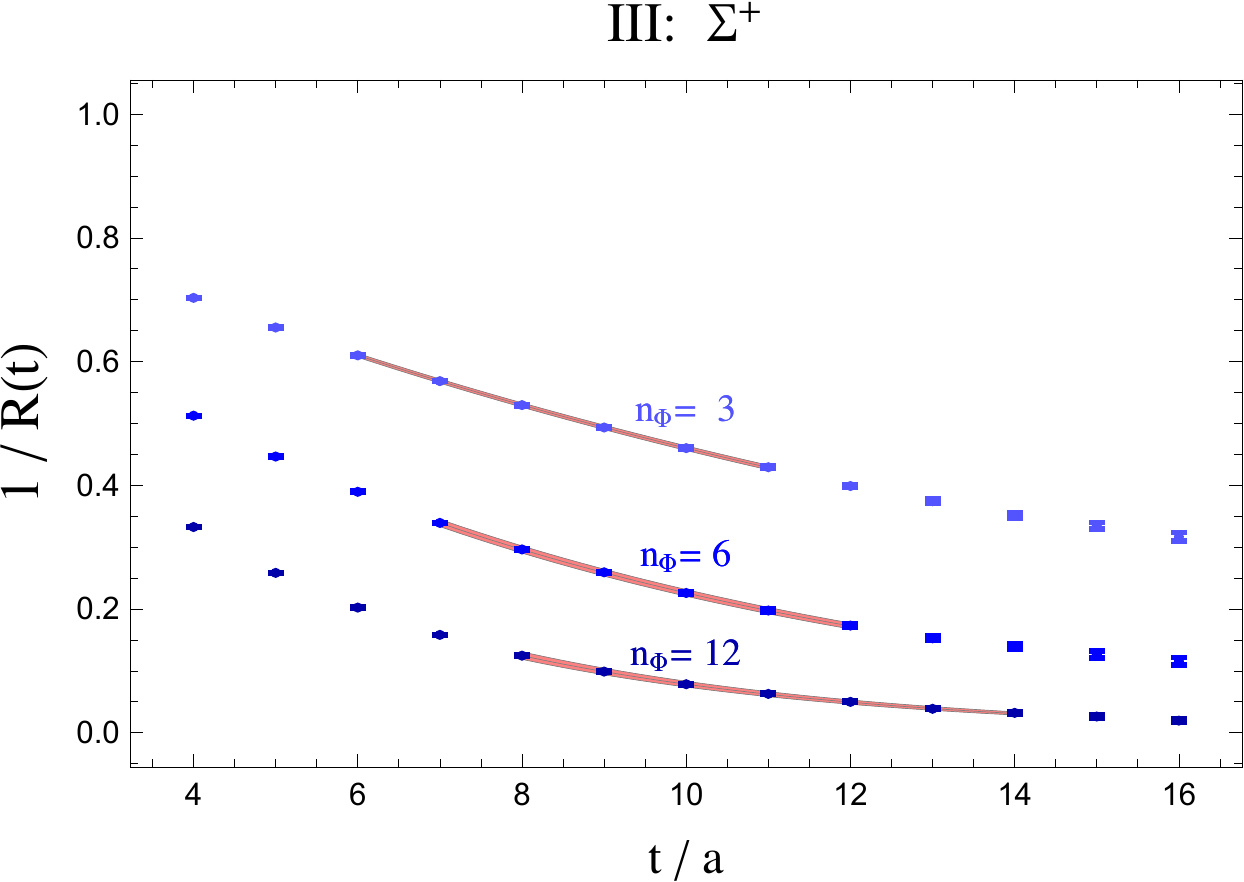}
$\quad$
\includegraphics{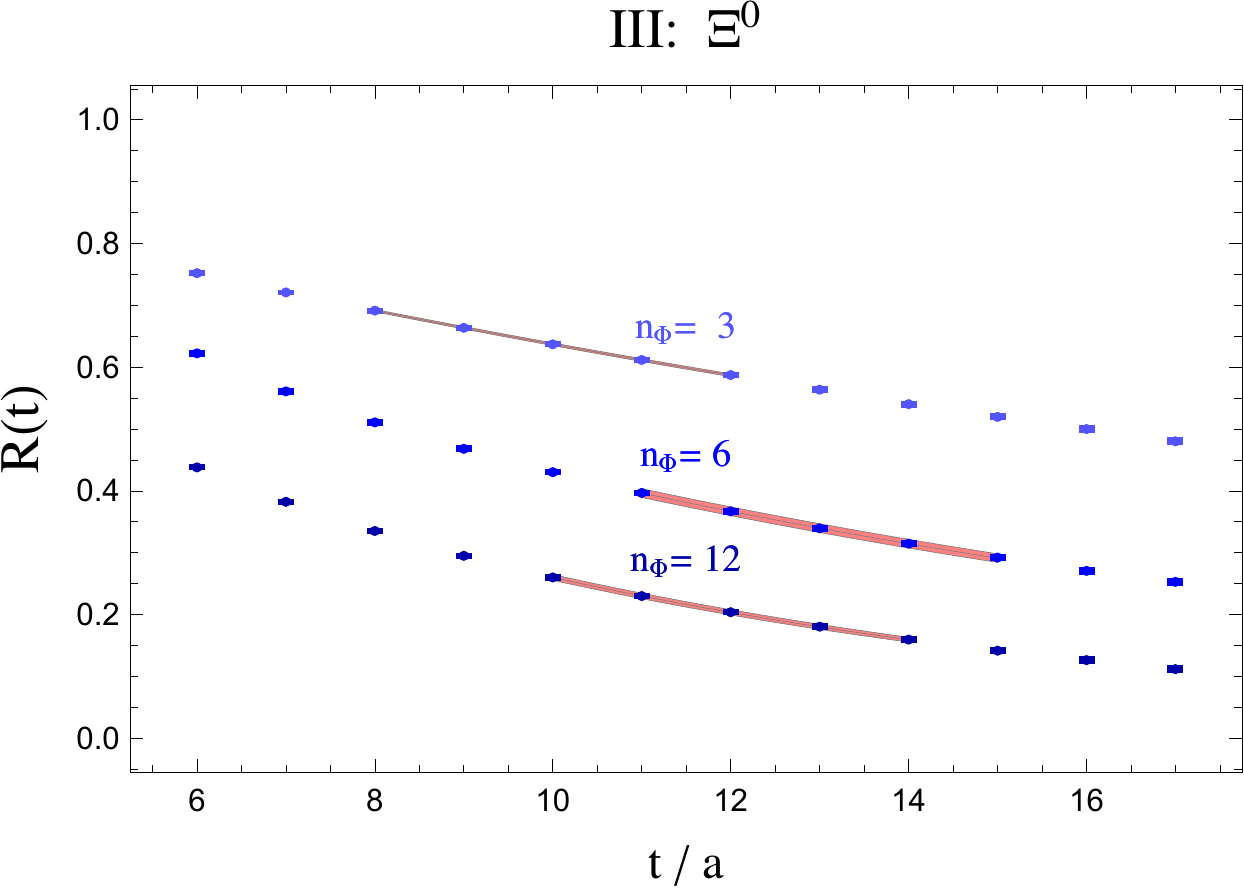}
$\quad$
\includegraphics{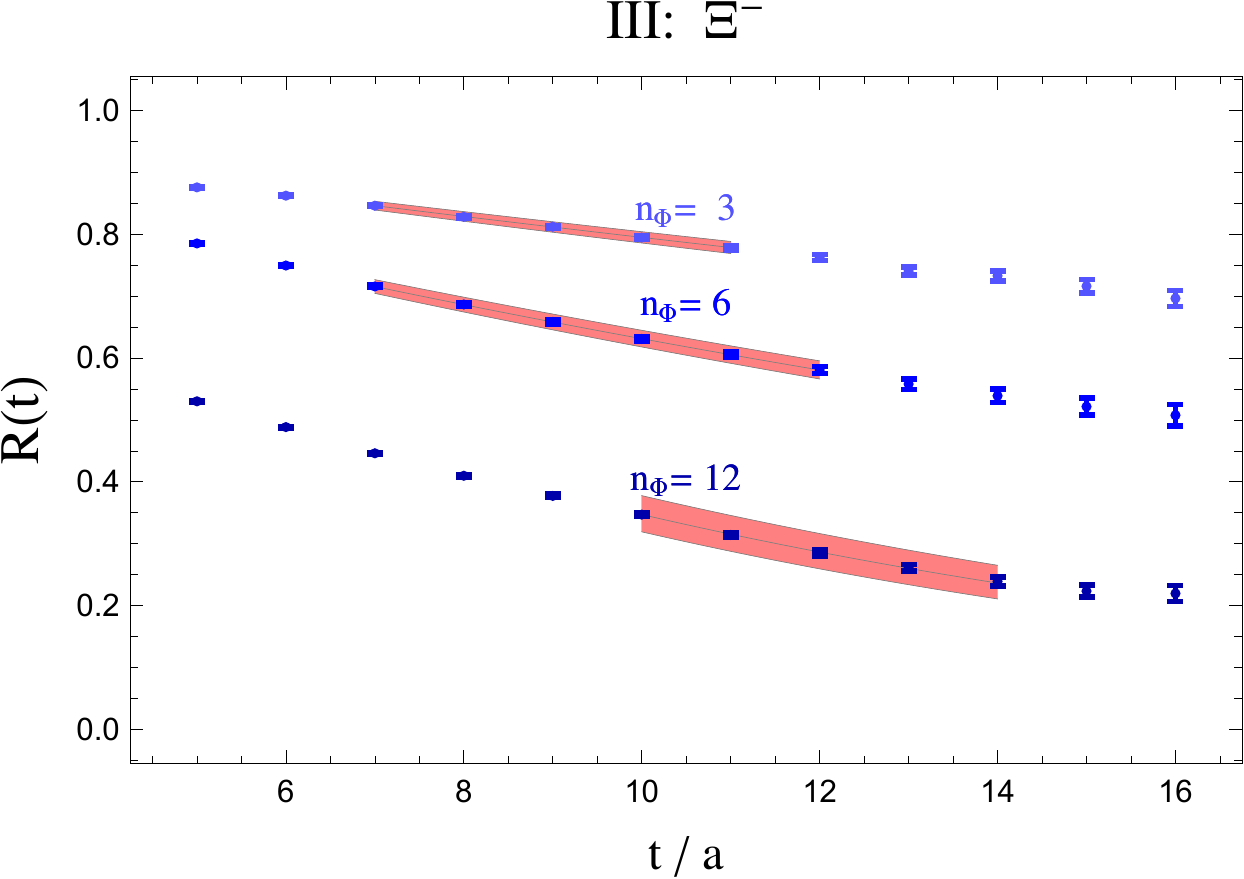}
}
\\
\medskip
\caption{
Ratios of spin-projected baryon correlation functions, 
$R(t)$  in Eq.~\eqref{eq:Ratio}, computed on Ensembles I--III,
and associated fits.
The shaded bands depict the uncertainty in the extracted energy, 
and include quadrature-combined statistical and systematic uncertainties, with the latter arising from the fit and  
choice of fit window.  
For positively charged baryons,  the inverse ratios are shown.  
}
\label{f:EeffB}
\end{figure}
%
%
%
%
%
%
%
%
%
%
%
%
%
%
%
%
\begin{figure}
\resizebox{\linewidth}{!}{
\includegraphics{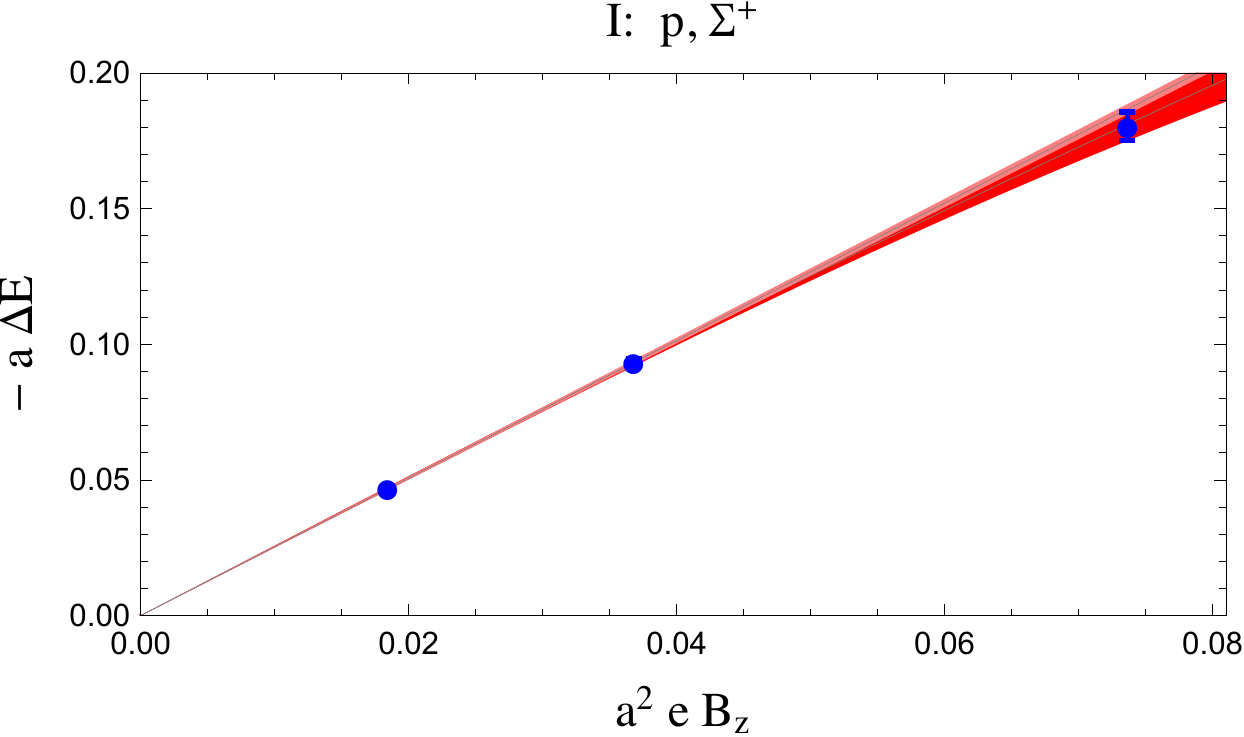}
$\quad$
\includegraphics{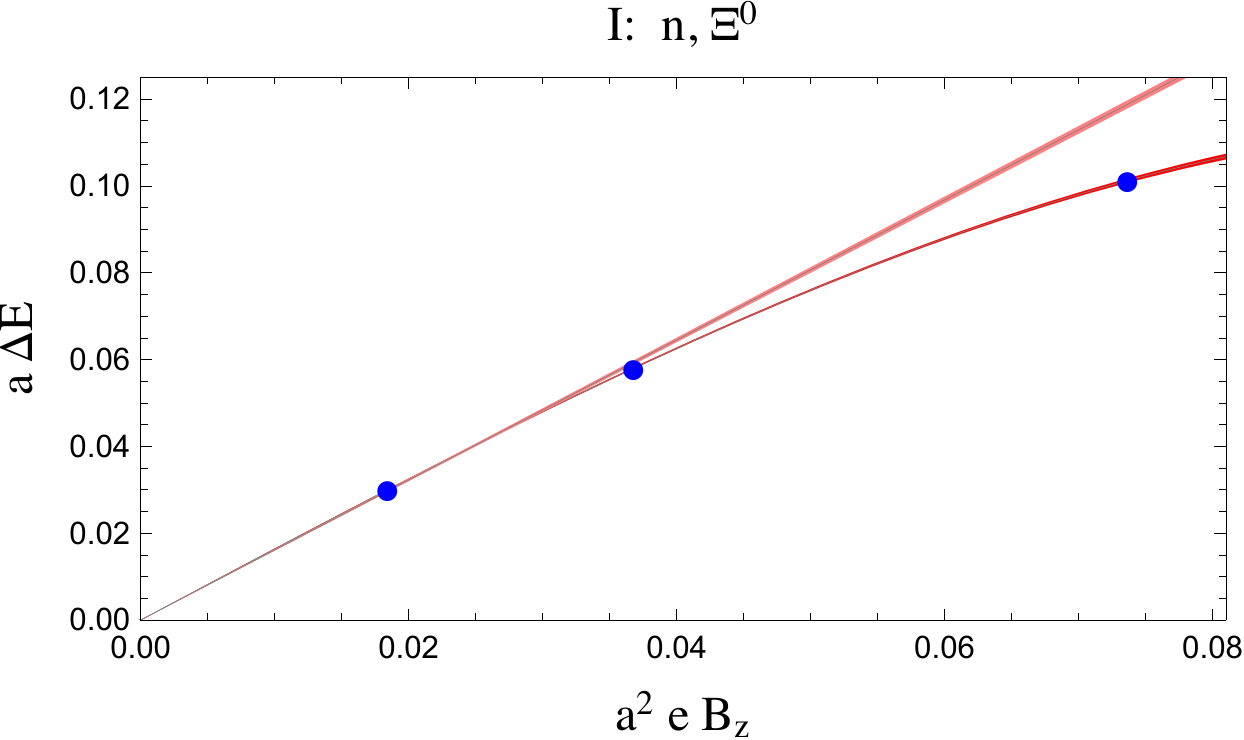}
$\quad$
\includegraphics{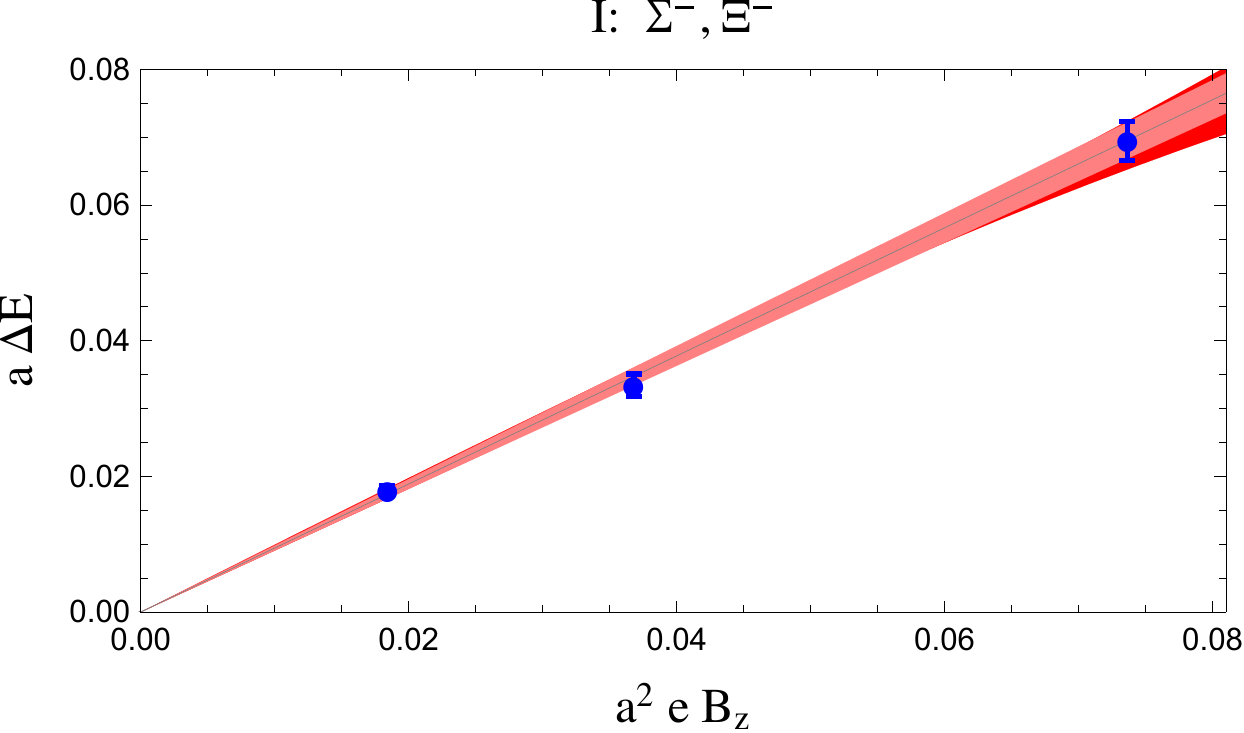}
}
\\
\medskip
\resizebox{\linewidth}{!}{
\includegraphics{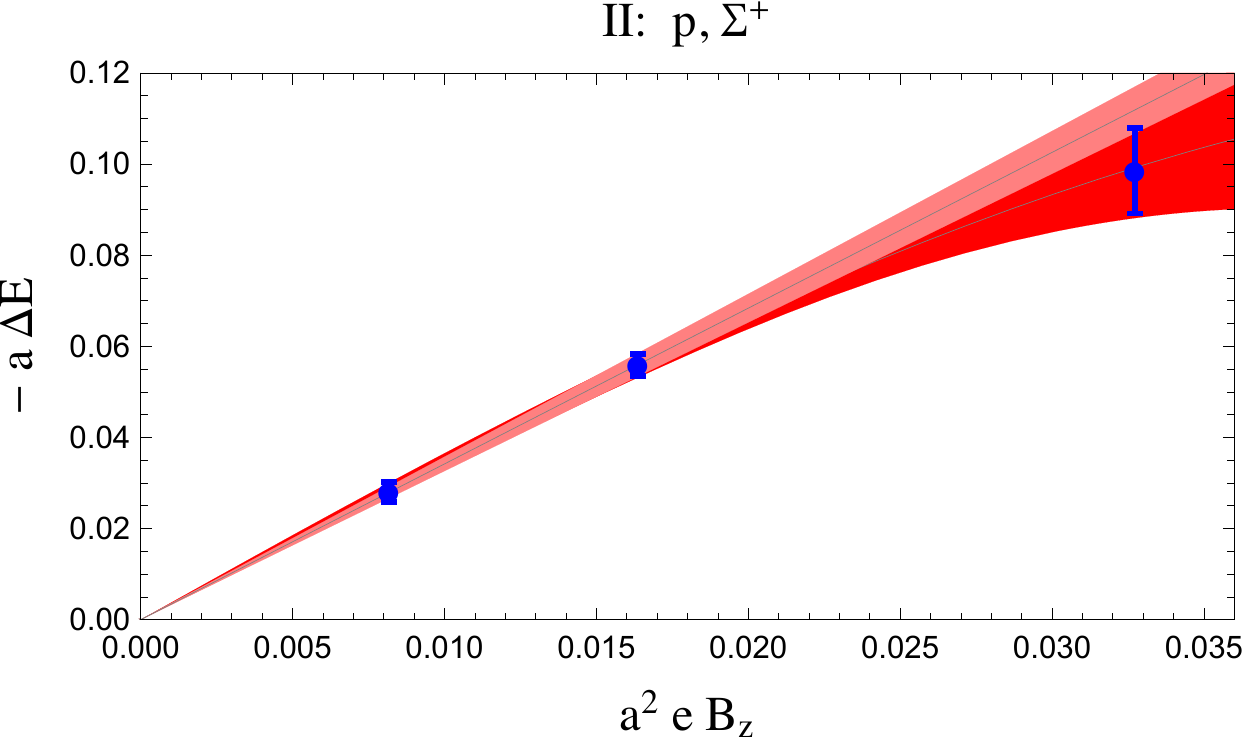}
$\quad$
\includegraphics{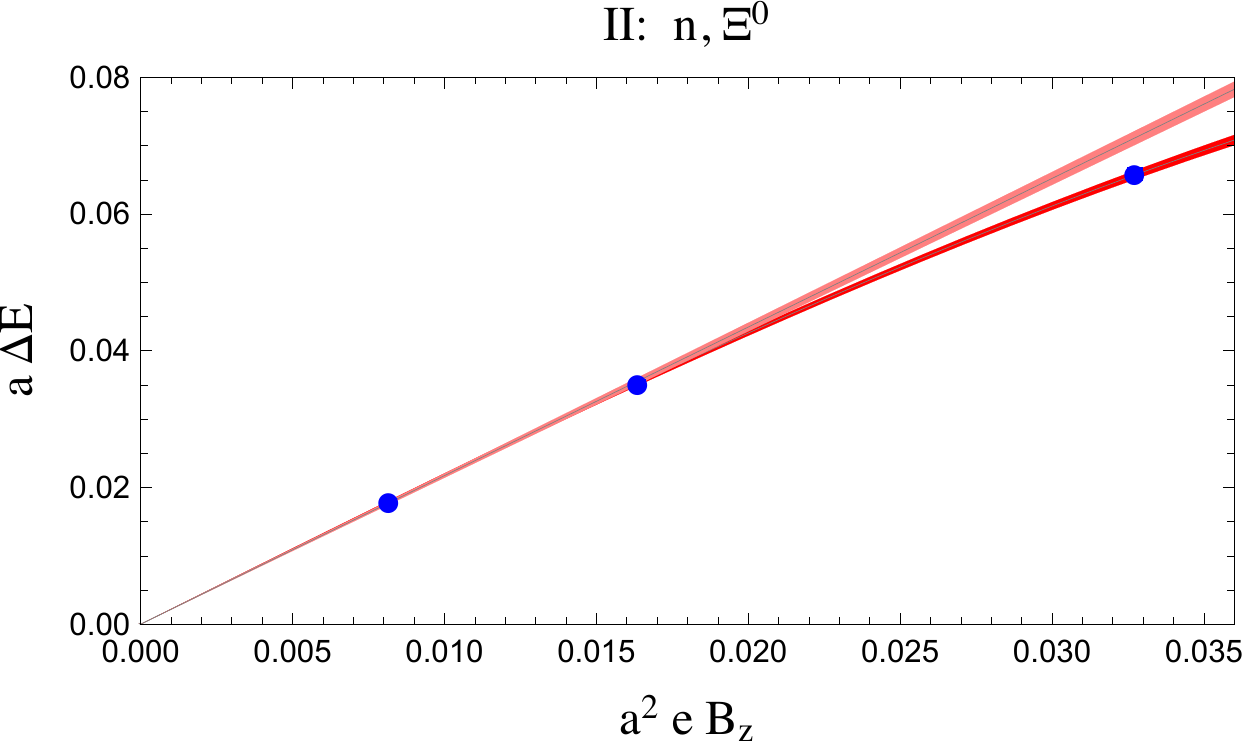}
$\quad$
\includegraphics{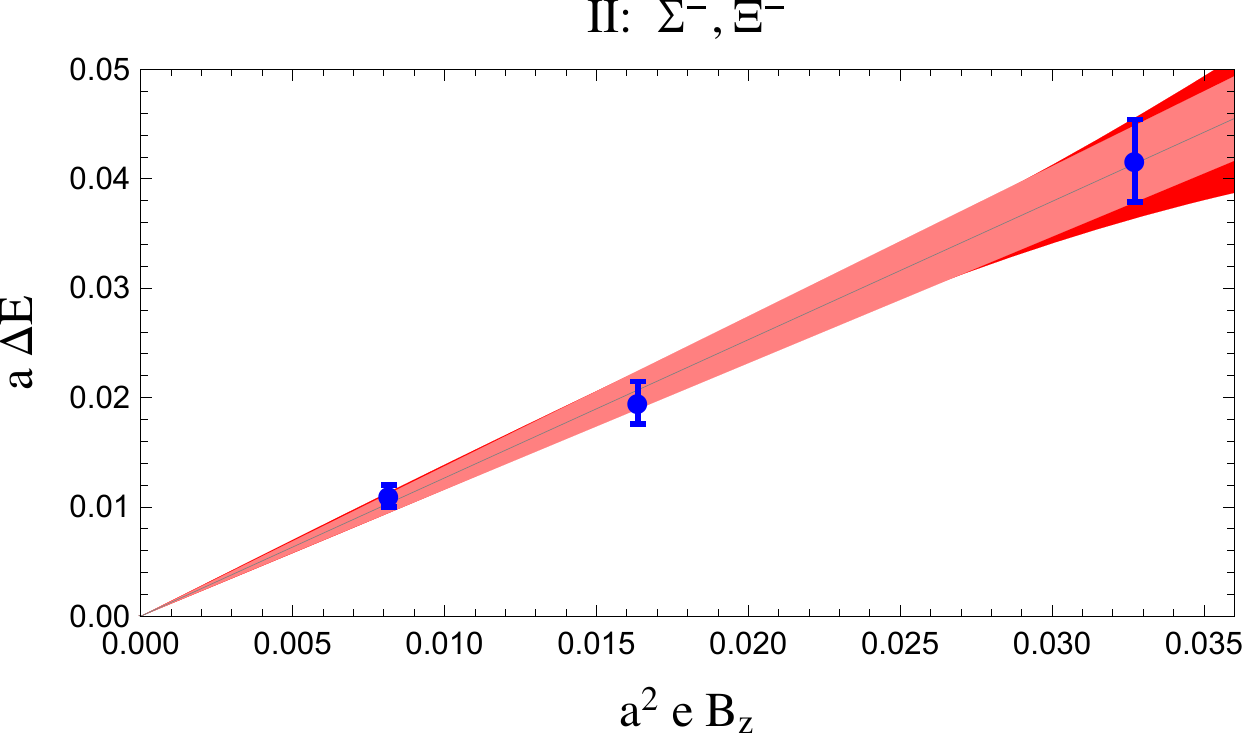}
}
\\
\medskip
\resizebox{\linewidth}{!}{
\includegraphics{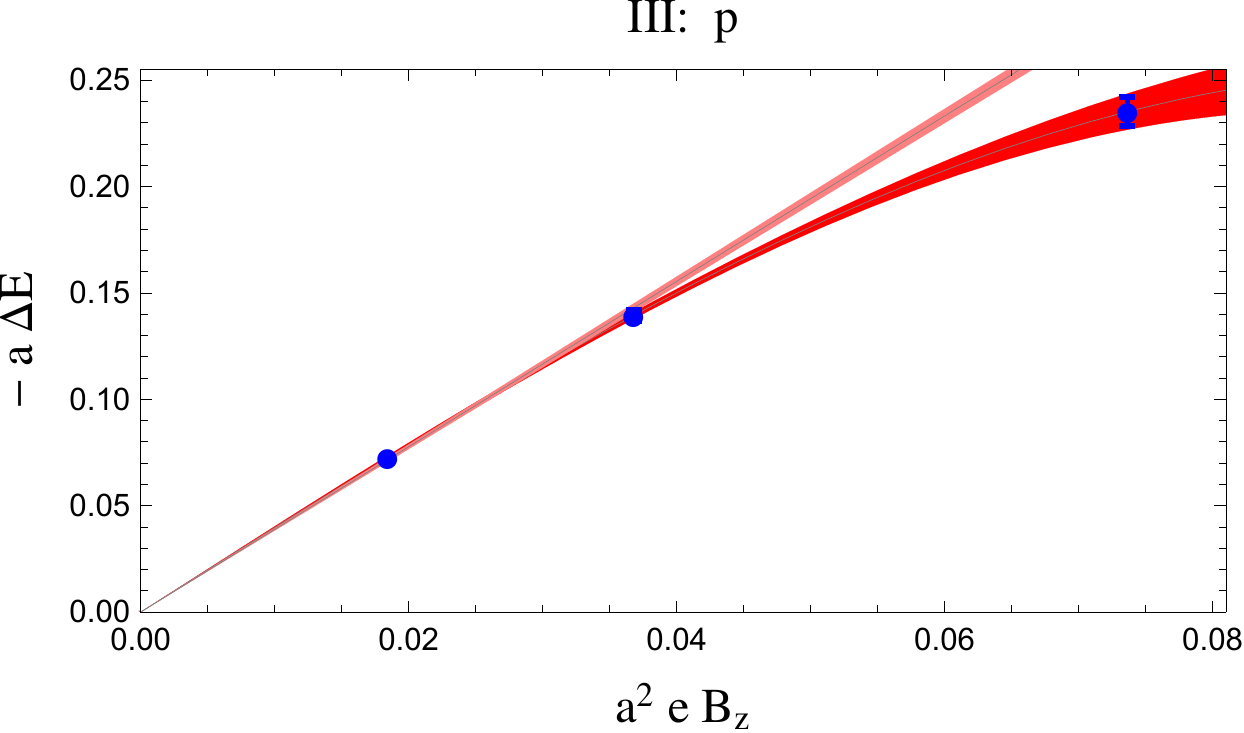}
$\quad$
\includegraphics{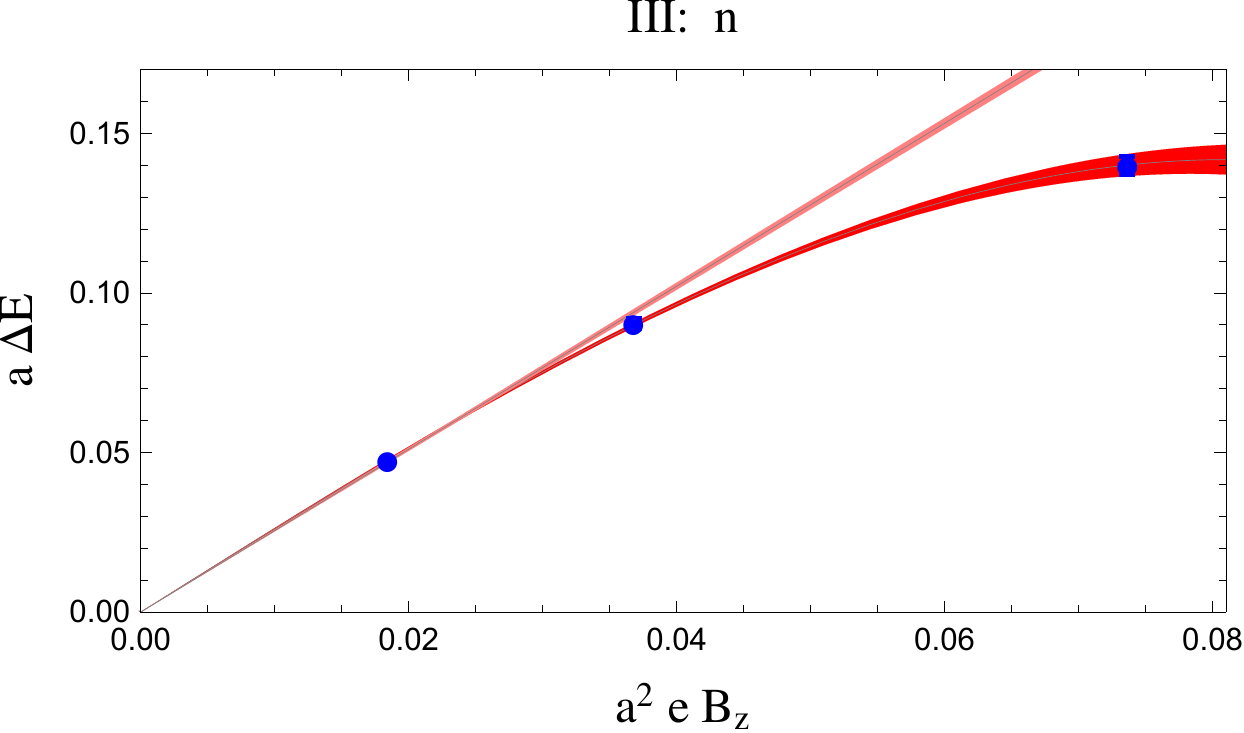}
$\quad$
\includegraphics{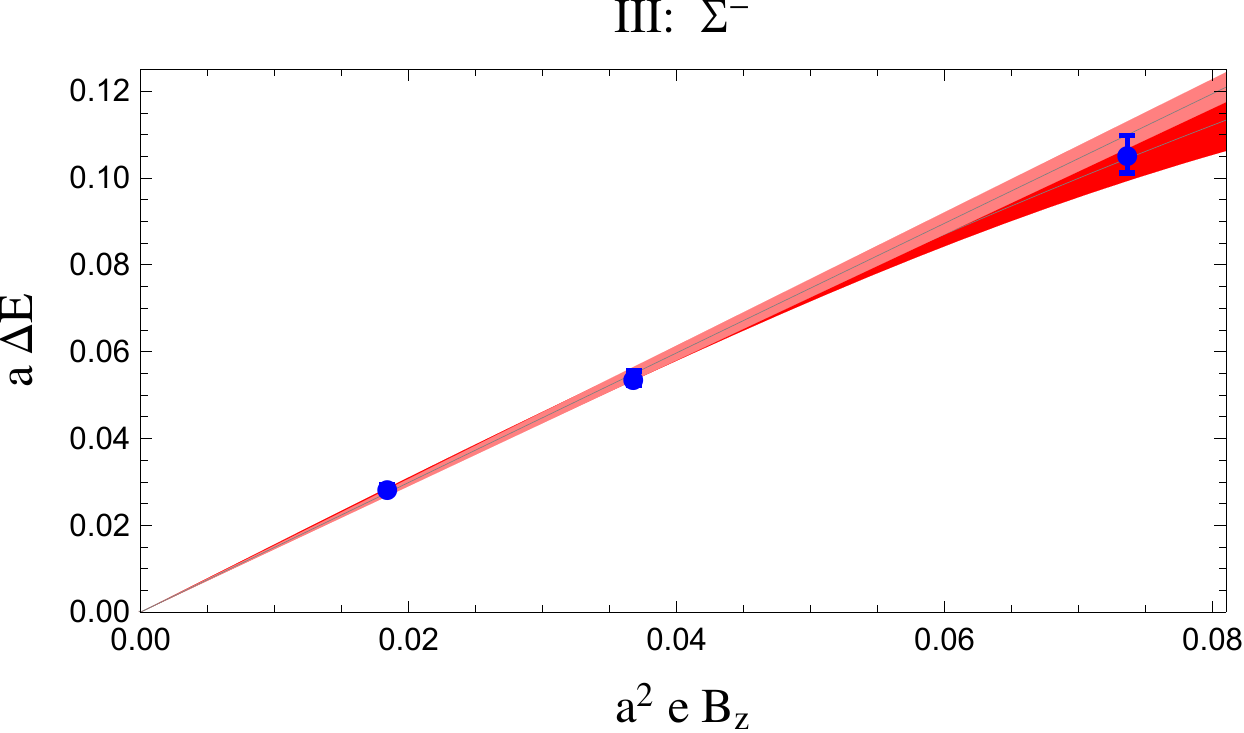}
}
\\
\medskip
\resizebox{\linewidth}{!}{
\includegraphics{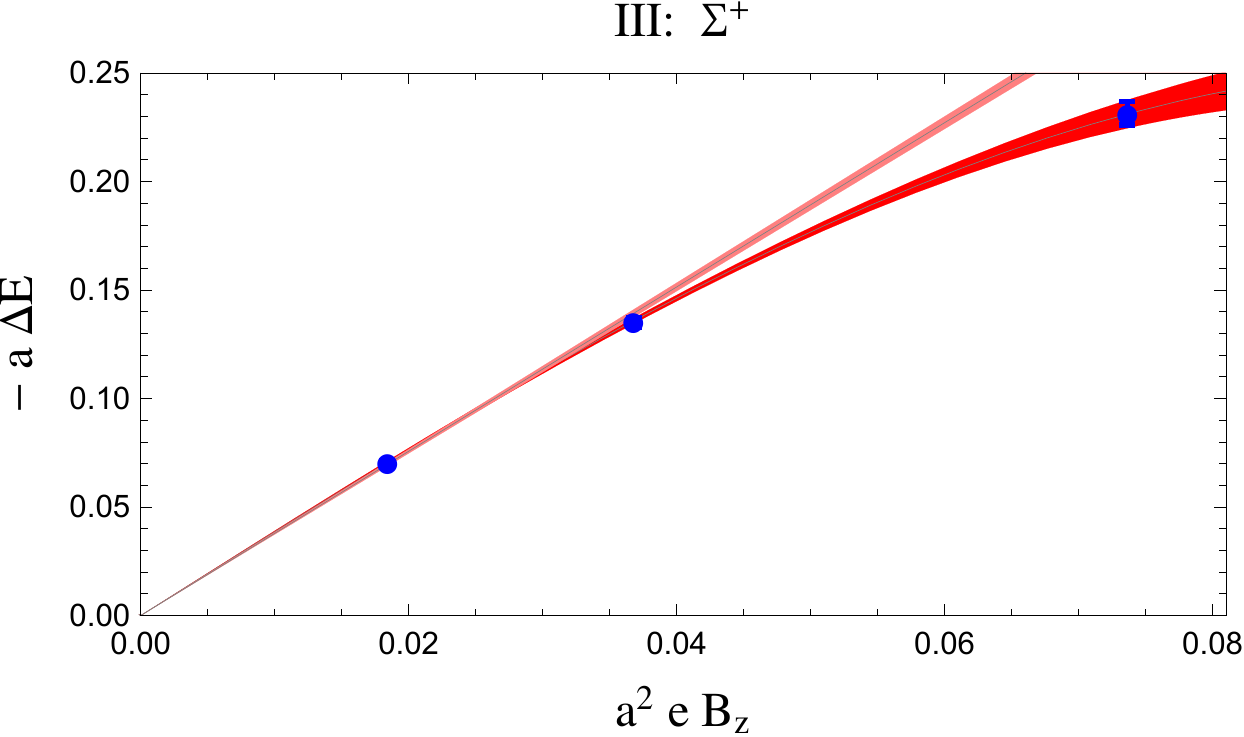}
$\quad$
\includegraphics{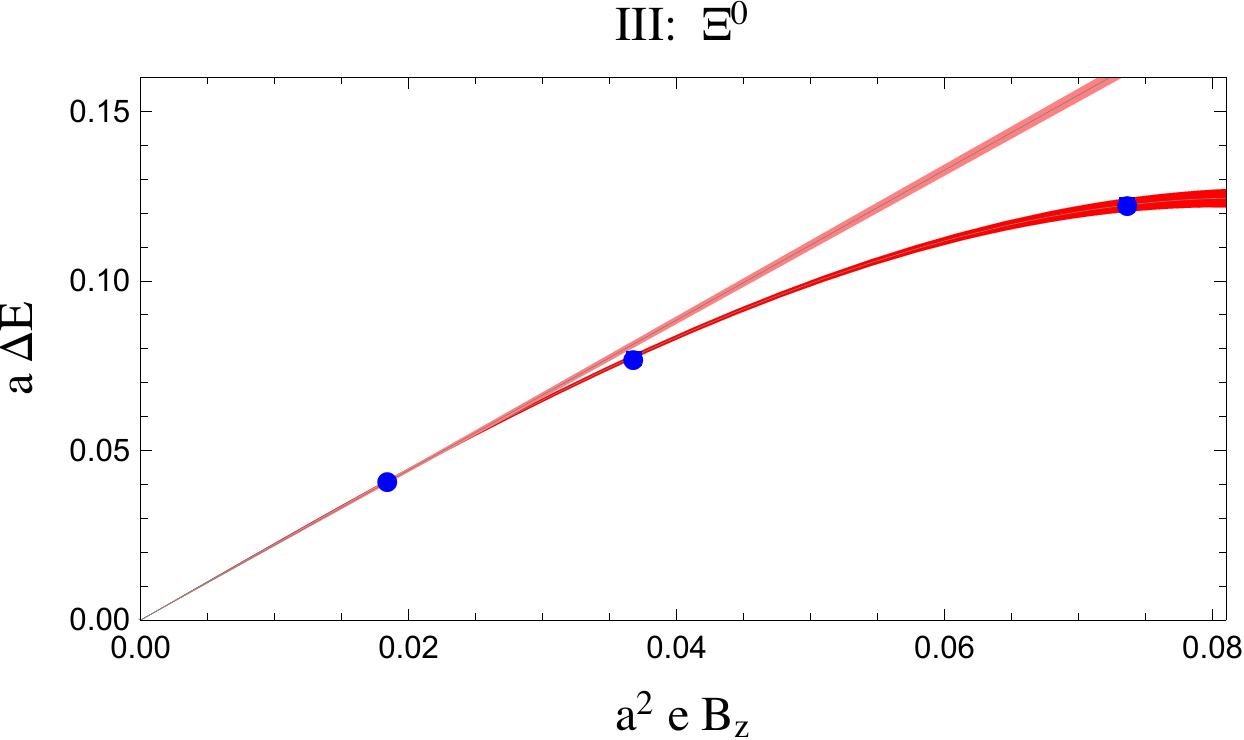}
$\quad$
\includegraphics{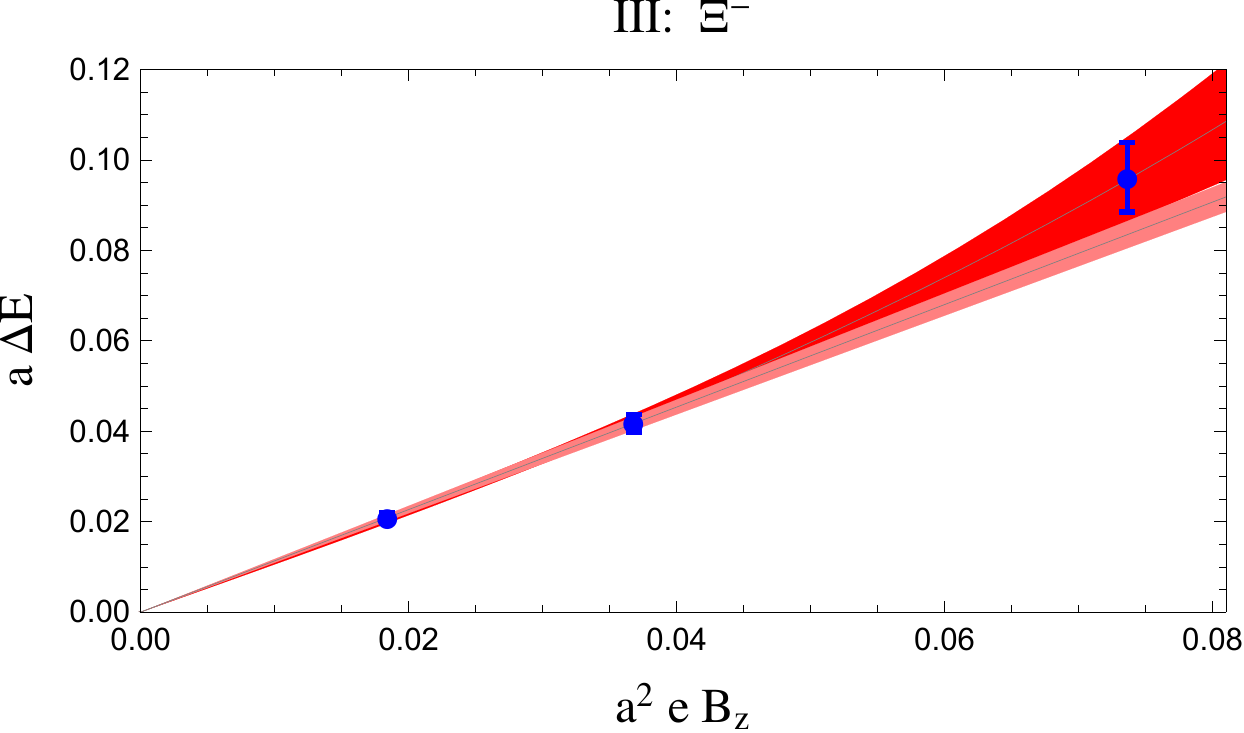}
}
\\
\medskip
\caption{
The magnetic-field dependence of baryon Zeeman splittings computed on Ensembles I--III. 
Two representative fits are shown: 
the darker bands correspond to linear plus cubic fits to all three field values,
$F_3$ in  Eq.~\eqref{eq:BFits}, 
while lighter bands correspond to linear fits that exclude the value at the largest magnetic field. 
For the positively charged baryons (appearing in the first column),
the negatives of their Zeeman splittings are shown.
} 
\label{f:BfitB}
\end{figure}
%
%
%
%
%
%
%
%

\section{Coupled $\Lambda$--$\Sigma^0$ System Analysis}
\label{s:B}

Technical details related to the coupled  $\Lambda$--$\Sigma^0$ system are contained in this Appendix. 
First, the PQ analysis of the magnetic polarizabilities of octet baryons,   necessitated
by the vanishing sea-quark electric charges in our LQCD calculations, is sketched. 
Next, the determination of the transition correlation function from diagonal baryon correlation functions is made explicit.
Finally, the analysis of the principal correlation functions obtained from solving the generalized eigenvalue problem in 
Eq.~\eqref{eq:GEVP} is detailed.

\subsection{Partially Quenched Magnetic Polarizabilities of the Octet Baryons}

The vanishing of sea-quark electric charges can be addressed using a PQ framework for baryons,
developed first in the context of baryon chiral perturbation theory, 
see~Refs.%
~\cite{Labrenz:1996jy,Savage:2001dy,Chen:2001yi}. 
The sea quarks 
$u_\text{sea}$, 
$d_\text{sea}$,
and 
$s_\text{sea}$
appear in the vector 
$\Psi_i = (u, d, s, u_\text{sea}, d_\text{sea}, s_\text{sea}, \tilde{u}, \tilde{d}, \tilde{s})_i$, 
which transforms in the fundamental representation of the $SU(6|3)$ graded group. 
Accordingly 
$u$, 
$d$, 
and
$s$
are valence quarks, 
while 
$\tilde{u}$, 
$\tilde{d}$, 
and
$\tilde{s}$
are their ghost counterparts which are not Grassman valued. 
Invariant operators often require the explicit appearance of grading factors of the form
$(-1)^{\eta_k}$, 
where $\eta_k = 1$ for fermionic indices and $\eta_k = 0$ for bosonic indices. 
The quark electric charge matrix can be  written as
\begin{equation}
\cQ
= 
\text{diag}
\left( 
Q_u, Q_d, Q_s, Q_{u_\text{sea}}, Q_{d_\text{sea}}, Q_{s_\text{sea}}, Q_u, Q_d, Q_s 
\right)
,
\end{equation}
where the ghost quarks necessarily share the electric charges of their valence counterparts, 
and the condition 
$\text{str} \, \cQ \equiv Q_{u_\text{sea}} + Q_{d_\text{sea}} +  Q_{s_\text{sea}} = 0$
ensures that no unphysical singlet operators appear. 
While all sea-quark charges vanish in our computation, 
$ Q_{u_\text{sea}} = Q_{d_\text{sea}} =  Q_{s_\text{sea}} = 0$,
it is nonetheless useful to treat quantities as functions of the valence- and sea-quark electric charges.

The lowest-lying spin-half baryons in $SU(6|3)$ are embedded in a  $\bm{240}$-dimensional supermultiplet 
$\mathcal{B}_{ijk}$~\cite{Chen:2001yi},
and 
the octet baryons $B_i \, {}^j$ formed from three valence quarks are embedded in this supermultiplet as 
$\mathcal{B}_{ijk}\big| = \frac{1}{\sqrt{6}} \left(\epsilon_{ijl} B_k \, {}^l + \epsilon_{ikl} B_j \, {}^l \right)$, 
where the 
$\big|$
notation represents the restriction of all indices to the valence sector. 
Magnetic polarizability operators are 
$SU(6|3)$
invariants constructed from 
$\overline{\mathcal{B}}$, 
$\mathcal{B}$, 
and two insertions of the PQ  charge matrix
$\cQ$. 
The effective Hamiltonian density describing  
the
magnetic polarizabilities of the 
$\bm{240}$-plet 
baryons has the form 
\begin{eqnarray}
\Delta \mathcal{H}^{(PQ)}
&=&
- \frac{1}{2} 4 \pi \bm{B}^2 
\Bigg[
\beta_1^{(PQ)}
\overline{\mathcal{B}} {}^{kji}
\mathcal{B}_{ijk}
\, \text{str} \left( \cQ^2 \right)
+
\beta_2^{(PQ)}
\overline{\mathcal{B}} {}^{kji}
\left( \cQ \cQ \right)_i \, {}^l 
\mathcal{B}_{ljk}
+
\beta_3^{(PQ)}
(-1)^{(\eta_i + \eta_j)(\eta_k + \eta_l)}
\overline{\mathcal{B}} {}^{kji}
\left( \cQ \cQ \right)_k \, {}^l 
\mathcal{B}_{ijl}
\notag \\
&& \phantom{spacing}
+
\beta_4^{(PQ)}
(-1)^{\eta_l (\eta_j + \eta_m)}
\overline{\mathcal{B}} {}^{kji}
\cQ_{i} \, {}^{l}
\cQ_{j} \, {}^{m}
\mathcal{B}_{lmk}
+
\beta_5^{(PQ)}
(-1)^{\eta_j \eta_m + 1} \,
\overline{\mathcal{B}} {}^{kji}
\cQ_{i} \, {}^{m}
\cQ_{j} \, {}^{l}
\mathcal{B}_{lmk}
\Bigg]
,
\label{eq:PQpol}
\end{eqnarray}
where we have used ${\rm str} \left(\cQ\right)=0$,
and the  
$\beta_j^{(PQ)}$ are numerical coefficients. 
Notice that the number of independent operators is one greater in the PQ theory compared to QCD,
see Eq.~\eqref{eq:BBHam}.
The relations between the five PQ coefficients and the four QCD coefficients in the QCD limit
can easily be found from matching the two expressions, 
but the full result is not required  here. 
Only the first operator in Eq.~(\ref{eq:PQpol}) depends on the electric charges of sea quarks. 
Contributions to magnetic polarizabilities from this operator are identical for all baryons, 
and zero for the transition polarizability. 
Thus 
$\beta_{\Lambda \Sigma}$
is independent of sea-quark charges in the mass-symmetric limit, 
i.e.~$\beta_{\Lambda \Sigma} = \beta_{\Lambda \Sigma}^{(c)}$, 
where the superscript  $(c)$ denotes the quark-connected part.  
In general, 
setting the sea-quark charges to zero corresponds to retaining the quark-connected parts of magnetic polarizabilities. 
For example, 
the connected part of the neutron magnetic polarizability is denoted by
$\beta_n^{(c)}$, 
and satisfies
$\beta_n^{(c)} = \beta_n - \frac{2}{3} \beta_1^{(PQ)}$, 
where
$\beta_n$
is the magnetic polarizability of the neutron in QCD.
In the $SU(3)_F$ mass-symmetric limit, 
the PQ Hamiltonian for the $I_3 = 0$  baryons at  $\cO(\bm{B}^2)$ can be written in the simple form,
\begin{equation}
\Delta H^{(PQ)}_{I_3=0}
=
- \frac{1}{2} 4 \pi \bm{B}^2 
\left[
\beta_1^{(PQ)} \text{str} \, \cQ^2
\,\, \mathbb{1}
+ 
\begin{pmatrix}
\beta_n^{(c)} + \sqrt{3} \beta_{\Lambda \Sigma}& 
\beta_{\Lambda \Sigma}
\\
\beta_{\Lambda \Sigma}
&
\beta_n^{(c)} + \frac{1}{\sqrt{3}} \beta_{\Lambda \Sigma}
\end{pmatrix}
\right]
.
\end{equation}
Upon setting the sea-quark charges to zero, 
the connected parts of 
$\Lambda$
and 
$\Sigma^0$
magnetic polarizabilities can be identified, 
$\beta_\Lambda^{(c)} = \beta_n^{(c)} + \frac{1}{\sqrt{3}} \beta_{\Lambda \Sigma}$
and
$\beta_{\Sigma^0}^{(c)} = \beta_n^{(c)} + \sqrt{3}\beta_{\Lambda \Sigma}$, 
respectively,
along with  the magnetic polarizabilities of the 
$\lambda_\pm$
eigenstates. 
These are given by
$\beta_{\lambda_+}^{(c)} = \beta_n^{(c)} + \frac{4}{\sqrt{3}} \beta_{\Lambda \Sigma}$
and
$\beta_{\lambda_-}^{(c)} = \beta_n^{(c)}$.
This implies that the only modification necessary to account for vanishing sea-quark charges in 
Eq.~\eqref{eq:LamSigSpec} is the replacement  $\beta_n \to \beta_n^{(c)}$.

\subsection{$\Lambda$--$\Sigma^0$ Transition Correlation Function and $U$-Spin Symmetry}

The transition correlation function between 
$\Lambda$
and
$\Sigma^0$ 
baryons can be obtained from the diagonal baryon two-point functions in the limit of exact 
$U$-spin symmetry. 
This result has been utilized in the analysis of 
$\Lambda$--$\Sigma^0$
mixing in 
Sec.~\ref{s:CCA}, 
and the derivation is given here. 
For simplicity, 
 the magnetic field-strength dependence is implicit below.

The neutron interpolating operator used in this work has the form 
$\chi^n_\alpha (x) = \varepsilon_{abc} \big( u^{aT}(x)  C \gamma_5 d^b(x)  \big) d^c_{\alpha}(x)$,
for which the neutron two-point function can be written in terms of the sum of two quark contractions. 
Using the spin-projection matrices 
$\mathcal{P}^{(\pm\frac{1}{2})} = \frac{1}{2} \left( 1 \pm \Sigma_3 \right)$, 
it can be expressed as
\begin{equation}
G^{(s)}_{nn} (x)
= 
\langle 0 | 
\mathcal{P}^{(s)}_{\alpha \beta}  \,
\chi^n_\beta (x)
\chi^{n \dagger}_\alpha (0)  
| 0 \rangle
=
\langle S \langle (U ,S) \rangle \rangle
+ 
\langle S (U,S) \rangle
,\end{equation}
making use of a short-hand notation for the quark-level contractions
\begin{eqnarray}
\langle 3 (1,2) \rangle
&=&
\mathcal{P}^{(s)}_{\gamma' \gamma} 
[\mathcal{G}^{(3)}(x,0)]^{cc'}_{\gamma \beta}
\,
\varepsilon_{abc} 
\,
\varepsilon_{a'b'c'}
[
\mathcal{G}^{(1)}(x,0)
C\gamma_5
]^{aa'}_{\alpha \beta}
[
C\gamma_5 \,
\mathcal{G}^{(2)}(x,0)]^{bb'}_{\alpha \gamma'}
,
\notag \\
\langle 3 \langle (1, 2) \rangle \rangle
&=&
\mathcal{P}^{(s)}_{\gamma' \gamma} 
[\mathcal{G}^{(3)}(x,0)]^{cc'}_{\gamma \gamma'}
\,
\varepsilon_{abc} 
\,
\varepsilon_{a'b'c'}
[
\mathcal{G}^{(1)}(x,0)
C\gamma_5
]^{aa'}_{\alpha \beta}
[
C\gamma_5 \,
\mathcal{G}^{(2)}(x,0)]^{bb'}_{\alpha \beta}
.\end{eqnarray}
The quantities $\mathcal{G}^{(j)}$, 
for 
$j = 1$,
$2$,
and 
$3$, 
represent the propagators for three distinguishable quark flavors, 
and the angled bracket notation denotes the traces over spinor indices. 
One can easily demonstrate the identity 
$\langle 3 \langle (1, 2) \rangle \rangle = \langle 3 \langle (2,1) \rangle \rangle$
for the second type of quark contraction. 
In the relevant $n$, $\Sigma^0$ and $\Lambda$ baryon correlation functions,  
$1 \to U$, 
and
$2,3 \to S$
on account of 
$U$-spin symmetry,
i.e.~because
$m_d = m_s$
and
$Q_d = Q_s$.

From the 
$\Sigma^0$
interpolating operator,
$\chi_\alpha^{\Sigma^0}(x) 
= 
\frac{1}{\sqrt{2}} \varepsilon_{abc}
\left[  \big(s^{aT}(x) C \gamma_5 d^b(x) \big) u^c_{\alpha}(x)  + \big(s^{aT}(x) C \gamma_5 u^b(x) \big) d^c_{\alpha}(x) \right]$, 
the two-point function has the form
\begin{equation}
G^{(s)}_{\Sigma\Sigma} (x)
= 
\langle 0 | 
\mathcal{P}^{(s)}_{\alpha \beta}  \,
\chi^{\Sigma^0}_\beta (x)
\chi^{\Sigma^0 \dagger}_\alpha (0) 
| 0 \rangle
=
\frac{1}{2}
\Big[
\langle U \langle (S,S) \rangle \rangle
+ 
\langle U (S,S) \rangle
+
\langle S \langle (S,U) \rangle \rangle
+ 
\langle S (S,U) \rangle
\Big]
,\end{equation}
making explicit use of 
$U$-spin symmetry above by writing all down-quarks propagators as strange-quark propagators. 
Similarly, from the 
$\Lambda$
interpolating operator, 
which has the form
$\chi^\Lambda_\alpha(x) 
= 
\frac{1}{\sqrt{6}} 
\varepsilon_{abc}
\left[ 
2 \big( u^{aT}(x) C \gamma_5 d^b(x) \big) s^c_{\alpha} (x)
+ 
\big(u^{aT} (x) C \gamma_5 s^b (x) \big) d^c_{\alpha} (x)
- 
\big(d^{aT} (x) C \gamma_5 s^b(x) \big) u^c_{\alpha} (x)
\right]$, 
the $\Lambda$ two-point function is given by
\begin{equation}
G^{(s)}_{\Lambda \Lambda} (x)
=
\langle 0 | 
\mathcal{P}^{(s)}_{\alpha \beta}  \,
\chi^{\Lambda}_\beta (x)
\chi^{\Lambda \dagger}_\alpha  (0)
| 0 \rangle
=
\frac{1}{4} 
\Big[
5 \, 
\langle S \langle (U,S) \rangle \rangle
+ 
4 \, 
\langle S  (U,S) \rangle
+ 
\langle U \langle (S,S) \rangle \rangle
+ 
\langle U  (S,S)  \rangle
+
\langle S ( S, U) \rangle
\Big]
.\end{equation}
Transition correlation functions between the 
$\Lambda$
and
$\Sigma^0$
baryons are defined by 
$G_{\Lambda \Sigma}^{(s)} (x)
= 
\langle 0 | 
\mathcal{P}^{(s)}_{\alpha \beta}  \,
\chi^{\Lambda}_\beta (x)
\chi^{\Sigma^0 \dagger}_\alpha (0) 
| 0 \rangle
$
and
$G_{\Sigma\Lambda}^{(s)} (x) = 
\langle 0 | 
\mathcal{P}^{(s)}_{\alpha \beta}  \,
\chi^{\Sigma^0}_\beta (x) 
\chi^{\Lambda \dagger}_\alpha (0) 
| 0 \rangle$. 
These can be written in terms of the following quark contractions 
\begin{equation}
G_{\Lambda \Sigma}^{(s)}
(x)
= 
G_{\Sigma \Lambda}^{(s)}
(x)
=
\frac{1}{2 \sqrt{3}}
\Big[
\langle U \langle (S,S) \rangle \rangle 
+ 
\langle U (S,S) \rangle
+ 
\langle S (S,U) \rangle
- 
\langle S \langle (U,S) \rangle \rangle
- 
2 \, 
\langle S (U,S) \rangle
\Big]
.\end{equation}
From this expression, 
the desired relations,
$G^{(s)}_{\Lambda \Sigma}(x) 
= G^{(s)}_{\Sigma \Lambda}(x)
= \frac{1}{\sqrt{3}} \left[ G^{(s)}_{\Sigma \Sigma} (x) - G^{(s)}_{nn} (x) \right]$, 
follow, along with
$G^{(s)}_{nn}(x)
=
\frac{1}{2} 
\left[
3 G^{(s)}_{\Lambda \Lambda}(x)
- 
G^{(s)}_{\Sigma \Sigma}(x)
\right]
$.
We have confirmed the latter relation holds configuration-by-configuration for each magnetic field in accordance with 
$U$-spin symmetry. 
It shows, 
moreover,
that the neutron correlation function can be omitted from this discussion, 
in favor of
$G^{(s)}_{\Lambda \Sigma} (x)
= 
G^{(s)}_{\Sigma \Lambda} (x)
= 
\frac{\sqrt{3}}{2} 
\left[
G^{(s)}_{\Sigma \Sigma} (x) 
-  
G^{(s)}_{\Lambda \Lambda} (x)
\right]$.

%
%
%
\begin{figure}
\resizebox{0.65\linewidth}{!}{
\includegraphics{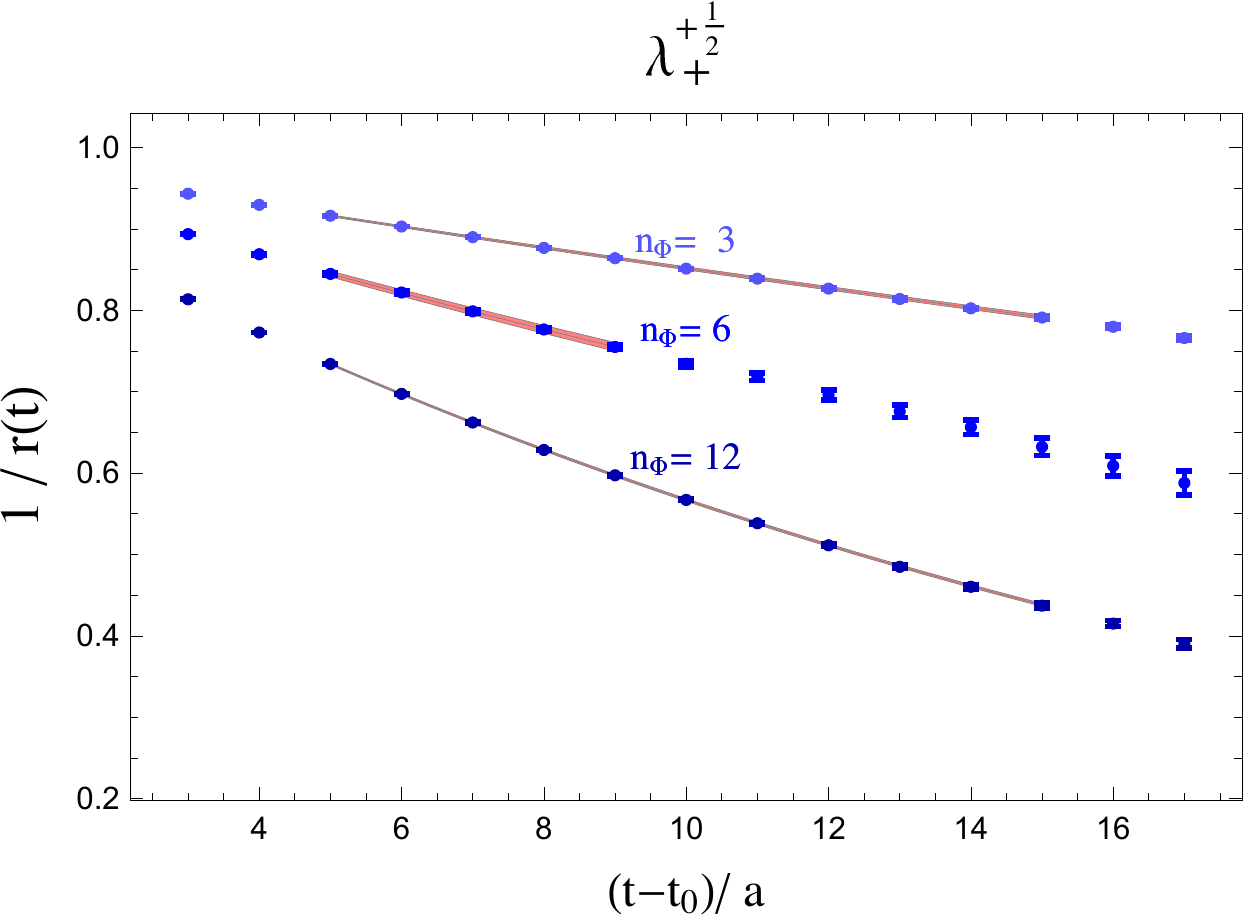}
$\quad$
\includegraphics{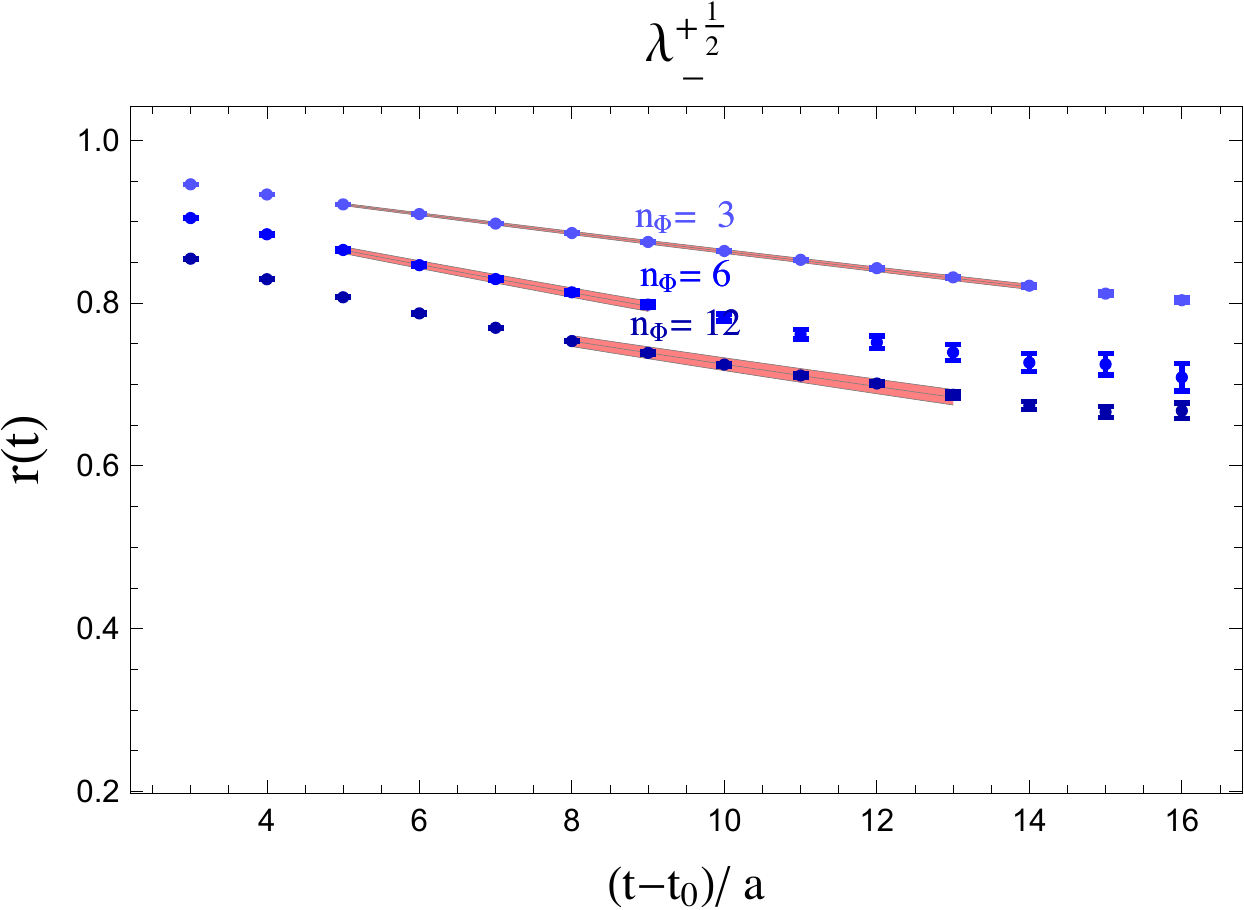}
}
\\
\medskip
\resizebox{0.65\linewidth}{!}{
\includegraphics{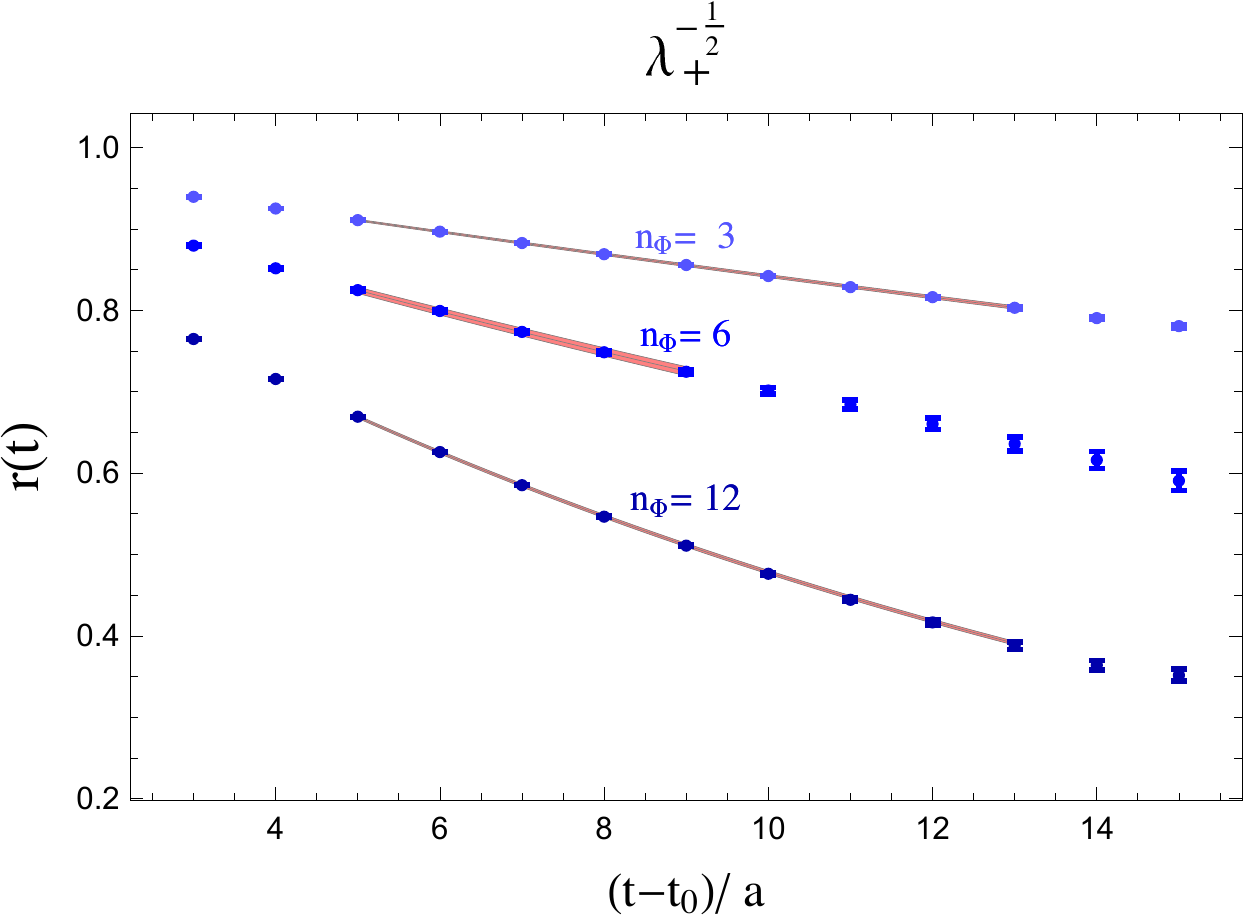}
$\quad$
\includegraphics{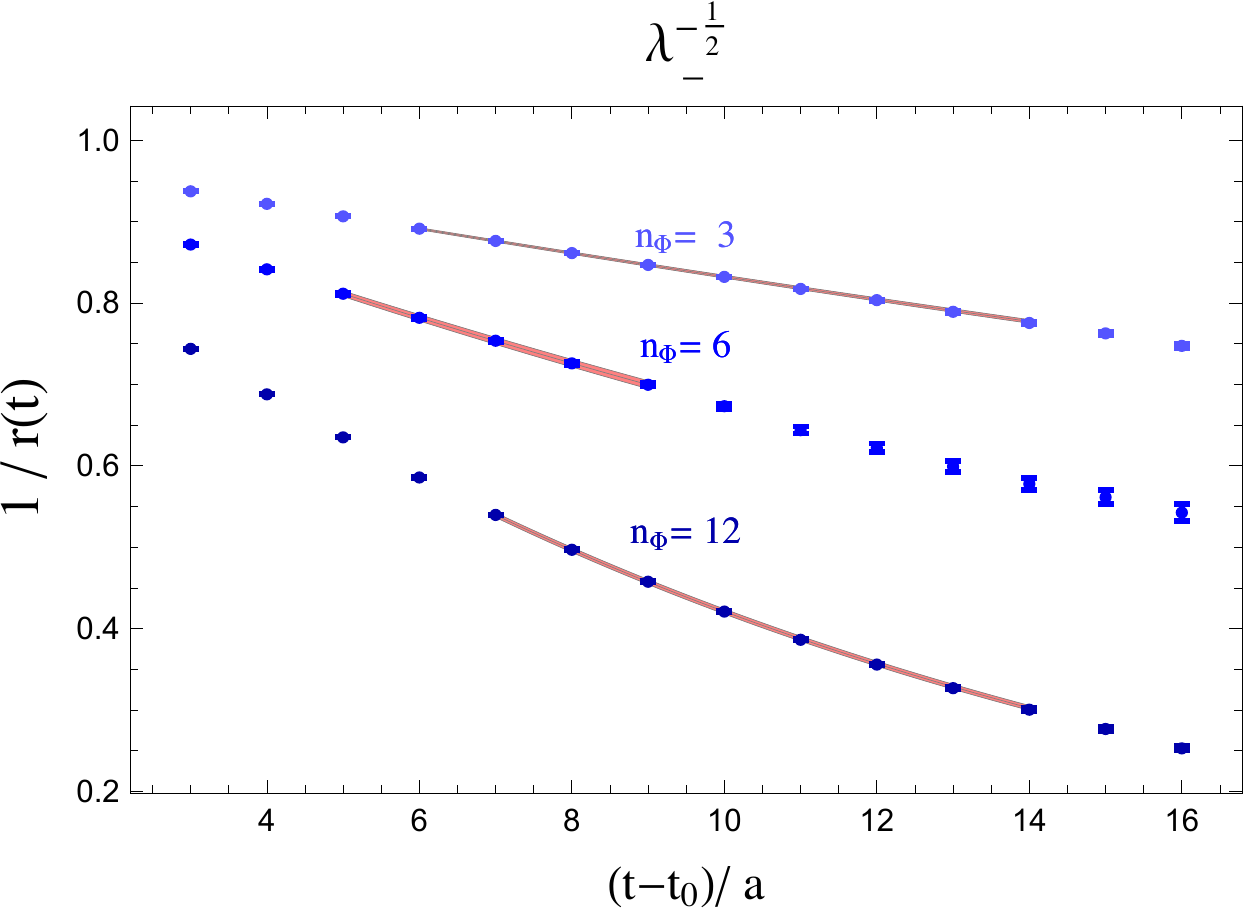}
}
\\
\medskip
\resizebox{0.65\linewidth}{!}{
\includegraphics{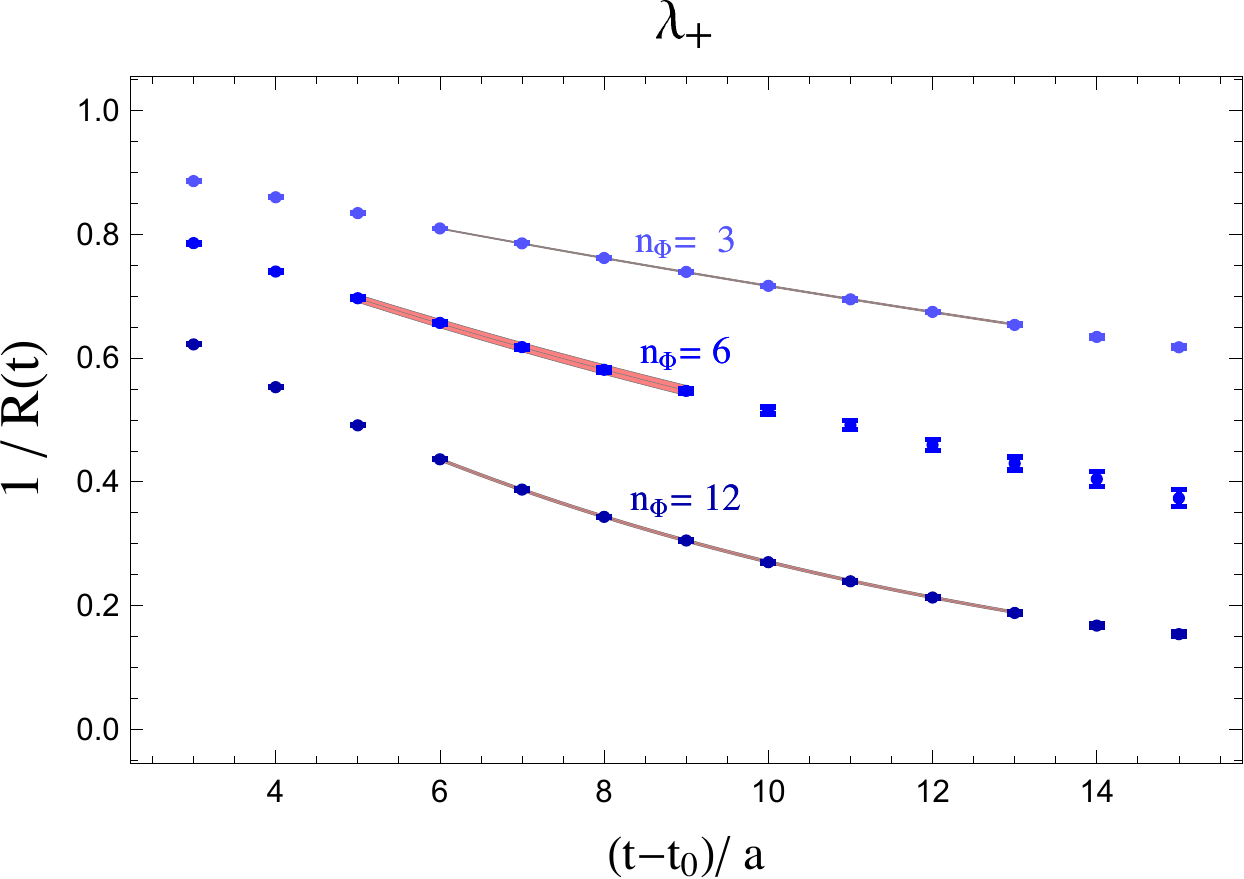}
$\quad$
\includegraphics{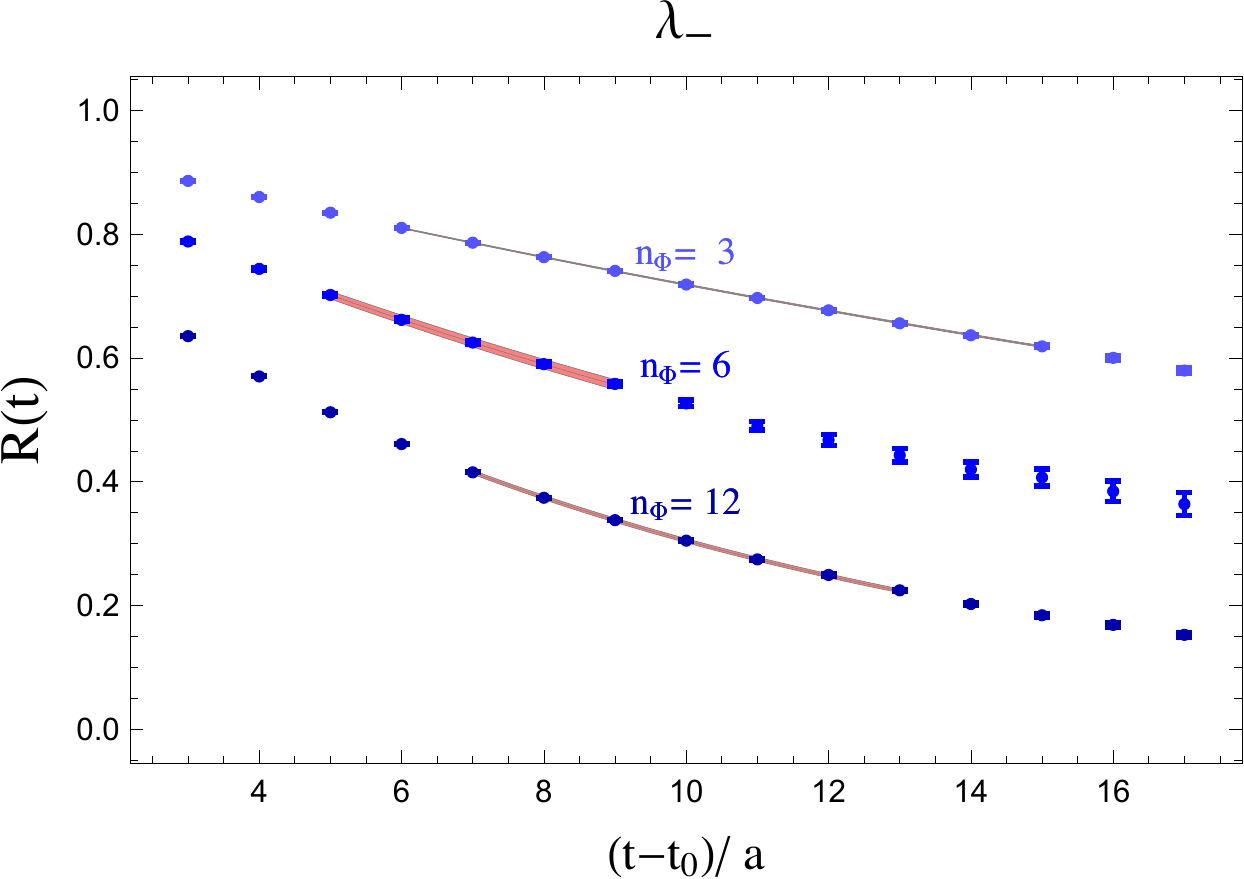}
}
\\
\medskip
\resizebox{0.65\linewidth}{!}{
\includegraphics{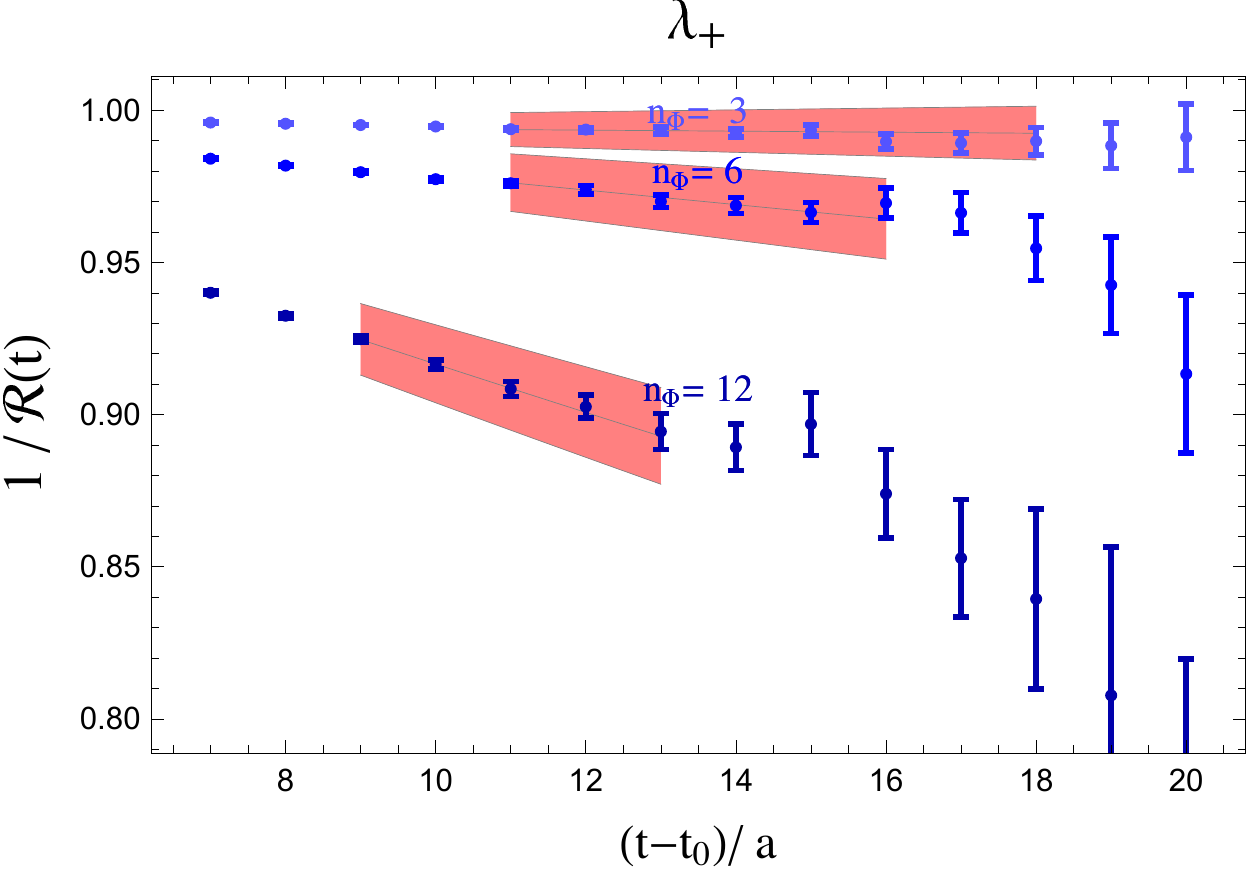}
$\quad$
\includegraphics{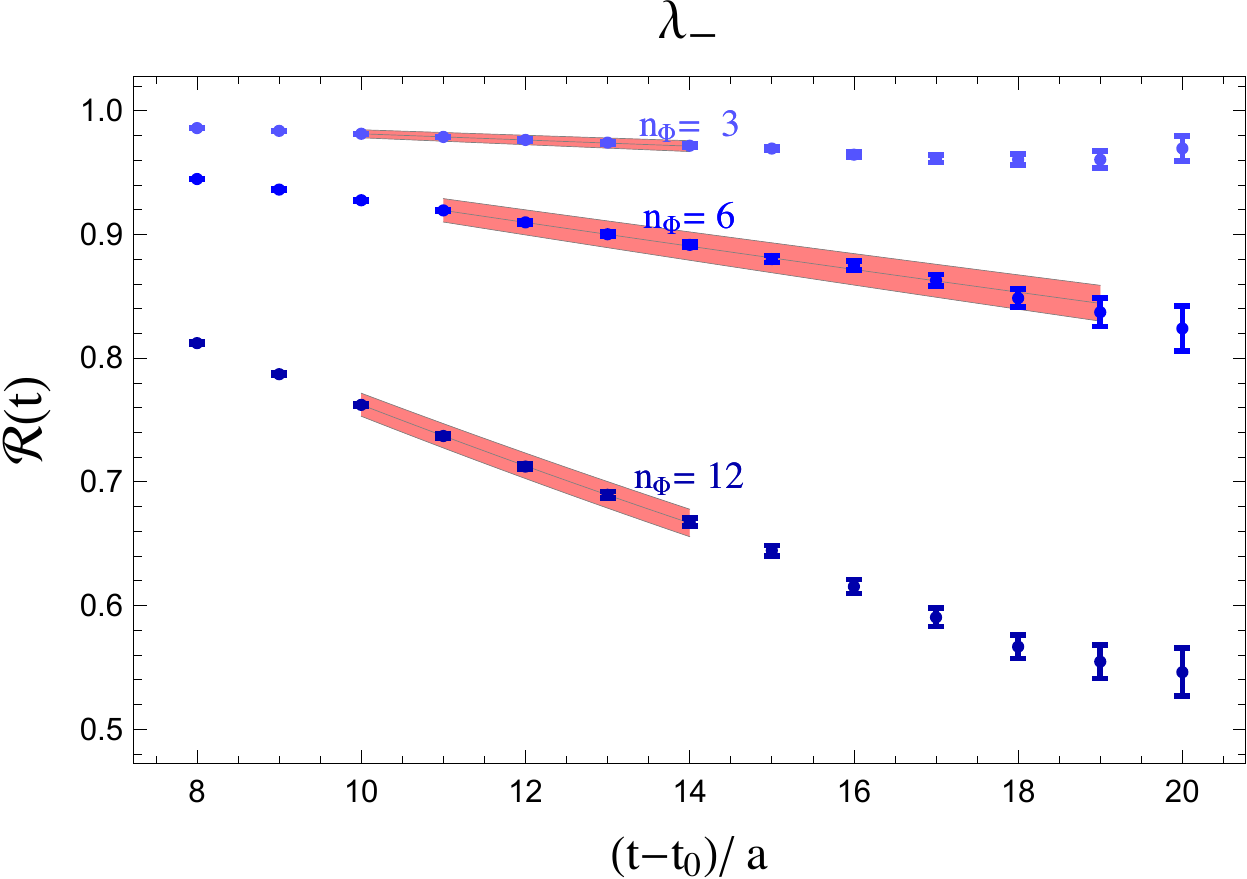}
}
\caption{
Ratios formed from spin-projected,  principal correlators in the  $\Lambda$--$\Sigma^0$ system calculated on Ensemble I. 
The ratios 
$r(t)$
defined in Eq.~\eqref{eq:PrinRatio} are used to obtain energy differences of a given spin state, 
whereas further ratios,
$R(t)$
defined in Eq.~\eqref{eq:PrinZeeman}, 
are used to obtain the Zeeman splittings. 
Ratios of spin-averaged principal correlators, 
$\mathcal{R}(t)$
in Eq.~\eqref{eq:PrinAvg}, 
are used to obtain spin-averaged energy differences.
For each ratio,  results of exponential fits to the ratios are also shown. 
Shaded bands depict the uncertainty on the extracted energy, 
and include quadrature-combined statistical and systematic uncertainties, 
with the latter arising from the fit and  choice of fit window.  
In cases where the extracted energy differences or Zeeman splittings are negative, 
 inverse ratios are presented as indicated. 
}
\label{f:SigLamEeffB}
\end{figure}
%
%
%

\end{widetext}

%
\begin{table}
\caption{%
Energy eigenvalues of the $\Lambda$--$\Sigma^0$ system from principal correlation functions  calculated on Ensemble I. 
Ratios have been normalized in 
Eq.~\eqref{eq:PrinRatio}
to produce the energy differences,
$a E(B_z) - a E(0)$. 
Zeeman splittings, 
$a \, \Delta E$, 
computed from ratios of spin-projected principal correlators, 
Eq.~\eqref{eq:PrinZeeman};
the spin-averaged energy differences,  
$\overline{E}(B_z) - \overline{E}(0)$, 
are also obtained from Eq.~\eqref{eq:PrinAvg}. 
The first uncertainty quoted is statistical, 
while the second is the systematic due to the fit and choice of fit window. 
Baryon magnetic moments,  
$\mu_B$,  
are determined in
$\texttt{[LatM]}$.
The first uncertainty quoted on magnetic moments is statistical, 
while the second is the systematic due to the fit and choice of fit function. 
Magnetic polarizabilities are determined from fits to the magnetic-field dependence of 
spin-averaged energy differences, and the associated uncertainties  are statistical, systematic, 
and additionally scale-setting for the case of standard physical units,
$\texttt{[10}^{\texttt{-4}} \,  \texttt{fm}^{\texttt{3}} \texttt{]}$. 
}
\begin{center}
\resizebox{\linewidth}{!}{
\begin{tabular}{|l|ccc|}
\hline
\hline
& 
\multicolumn{3}{c|}{$a \, E(B_z) - a E(0)$ }
\\
& 
$n_\Phi = 3$
& 
$n_\Phi = 6$
& 
$n_\Phi = 12$
\tabularnewline
\hline
\hline
$\lambda_+^{+ \frac{1}{2}}$
&
$- 0.01461(02)(20)$
&
$- 0.02801(27)(65)$
&
$- 0.05162(06)(24)$
\tabularnewline
$\lambda_+^{- \frac{1}{2}}$
&
$\phantom{-} 0.01566(03)(19)$
&
$\phantom{-} 0.03218(29)(71)$
&
$\phantom{-} 0.06721(13)(41)$
\tabularnewline
$\lambda_-^{+ \frac{1}{2}}$
&
$\phantom{-} 0.01294(02)(26)$
&
$\phantom{-} 0.02079(36)(70)$
&
$\phantom{-} 0.01924(40)(90)$
\tabularnewline
$\lambda_-^{- \frac{1}{2}}$
&
$- 0.01710(03)(19)$
&
$- 0.03706(27)(65)$
&
$- 0.08266(21)(62)$
\tabularnewline
\hline
\hline

& 
\multicolumn{3}{c|}{$a \, \Delta E$ }
\\
& 
$n_\Phi = 3$
& 
$n_\Phi = 6$
& 
$n_\Phi = 12$
\tabularnewline
\hline
\hline
$\lambda_+$
&
$- 0.03040(06)(14)$
&
$- 0.0602(06)(14)$
&
$- 0.11938(27)(78)$
\tabularnewline
$\lambda_-$
&
$\phantom{-} 0.03002(05)(13)$
&
$\phantom{-} 0.0578(06)(13)$
&
$\phantom{-} 0.10308(37)(88)$
\tabularnewline
\hline
\hline
& 
\multicolumn{3}{c|}{$a \, \Delta \overline{E}$ }
\\
& 
$n_\Phi = 3$
& 
$n_\Phi = 6$
& 
$n_\Phi = 12$
\tabularnewline
\hline
\hline
$\lambda_+$
&
$\phantom{-} 0.00017(15)(45)$
&
$\phantom{-} 0.00254(35)(72)$
&
$\phantom{-} 0.0087(05)(11)$
\tabularnewline
$\lambda_-$
&
$- 0.00256(11)(28)$
&
$- 0.01068(25)(82)$
&
$- 0.0335(04)(10)$
\tabularnewline
\hline
\hline
\multicolumn{4}{c}{}
\tabularnewline
\hline
\hline
& 
\multicolumn{3}{c|}{$\mu _B\, \texttt{[LatM]}$}
\\
& 
$s = + \frac{1}{2}$ 
& 
$s = - \frac{1}{2}$
& 
Zeeman
\tabularnewline
\hline
\hline
$\lambda_+$
&
$\quad
\phantom{-} 1.648(17)(80)
\quad$
&
$
\phantom{-} 1.661(15)(56)
$
&
$\quad
\phantom{-} 1.6532(61)(86)
\quad$
\tabularnewline
$\lambda_-$
&
$\quad
- 1.71(02)(38) \phantom{8}
\quad$
&
$- 1.73(02)(15) \phantom{8}$
&
$- 1.646(06)(18) \phantom{8}$
\tabularnewline
\hline
\hline
\multicolumn{4}{c}{ }
\tabularnewline
\hline
\hline
&
$\beta^{(c)}_B \, \texttt{[LatU]}$
&
$\hat{\beta}^{(c)}_B$
&
$\beta^{(c)}_B \, 
\texttt{[10}^{\texttt{-4}} \,  \texttt{fm}^{\texttt{3}} \texttt{]}$ 
\tabularnewline
\hline
\hline
$\lambda_+$
&
$- 1.63(12)(22)$
&
$-0.0479(36)(66)$
&
$- 0.73(05)(10)(01)$
\tabularnewline
$\lambda_-$
&
$\phantom{-} 7.77(27)(57)$
&
$\phantom{-} 0.228(08)(17) \phantom{8}$
&
$\phantom{-} 3.48(12)(26)(04)$
\tabularnewline
$\Lambda\Sigma$
&
$- 4.06(13)(27)$
&
$- 0.1196(39)(79)$
&
$- 1.82(06)(12)(02)$
\tabularnewline
\hline
\hline
\end{tabular}
}
\end{center}
\label{t:LamSigE}
\end{table}

%
%
%
\begin{figure}
\resizebox{0.725\linewidth}{!}{
\includegraphics{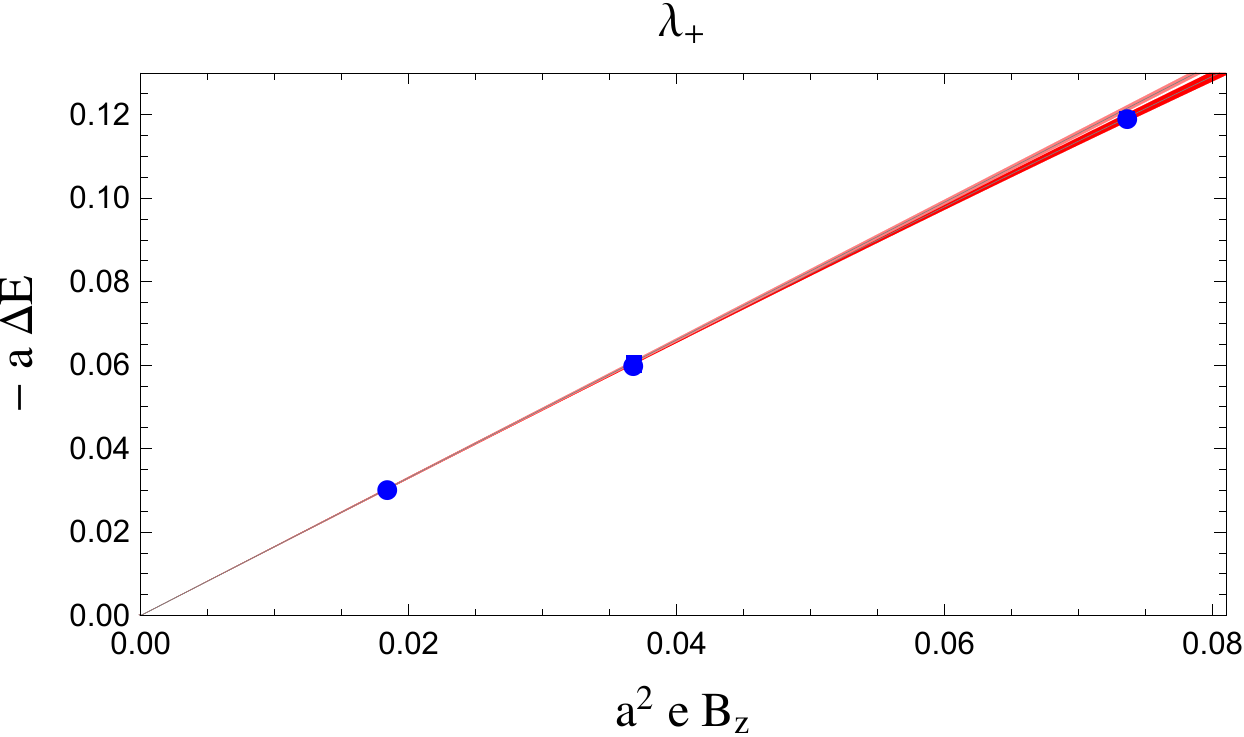}
}
\vskip 0.05in
\resizebox{0.725\linewidth}{!}{
\includegraphics{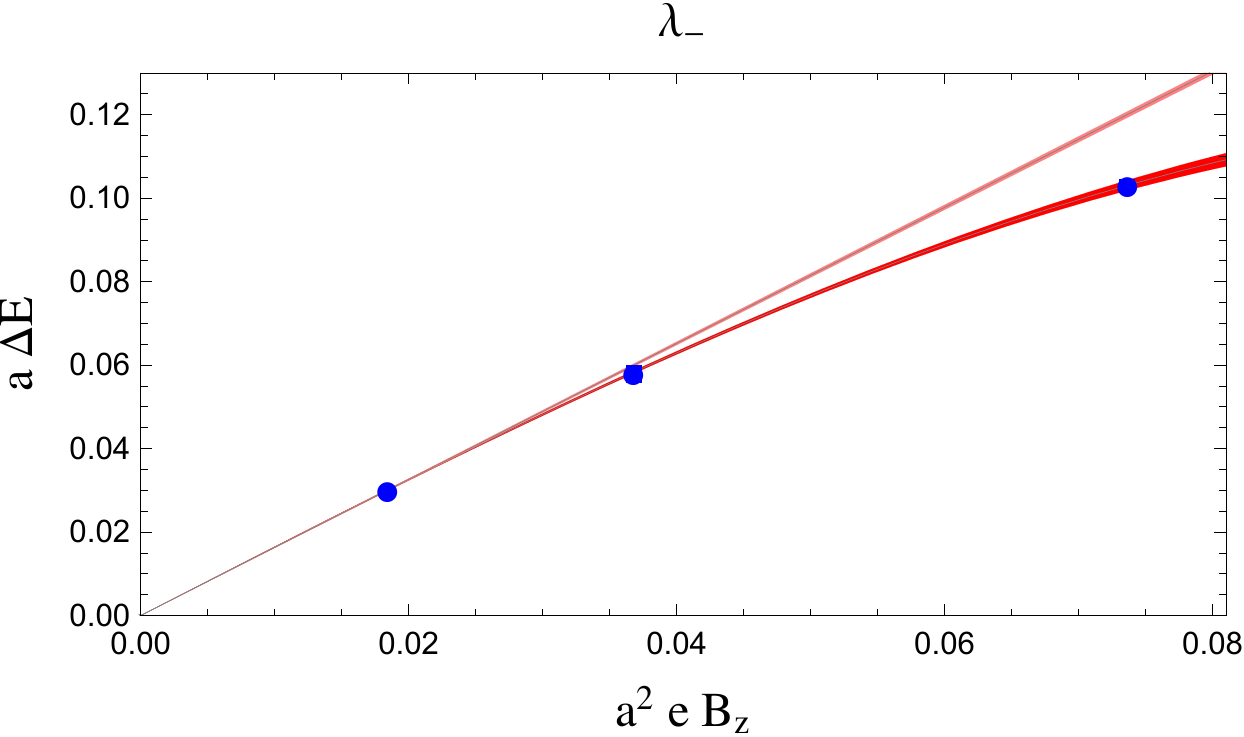}
}
\vskip 0.05in
\resizebox{0.725\linewidth}{!}{
\includegraphics{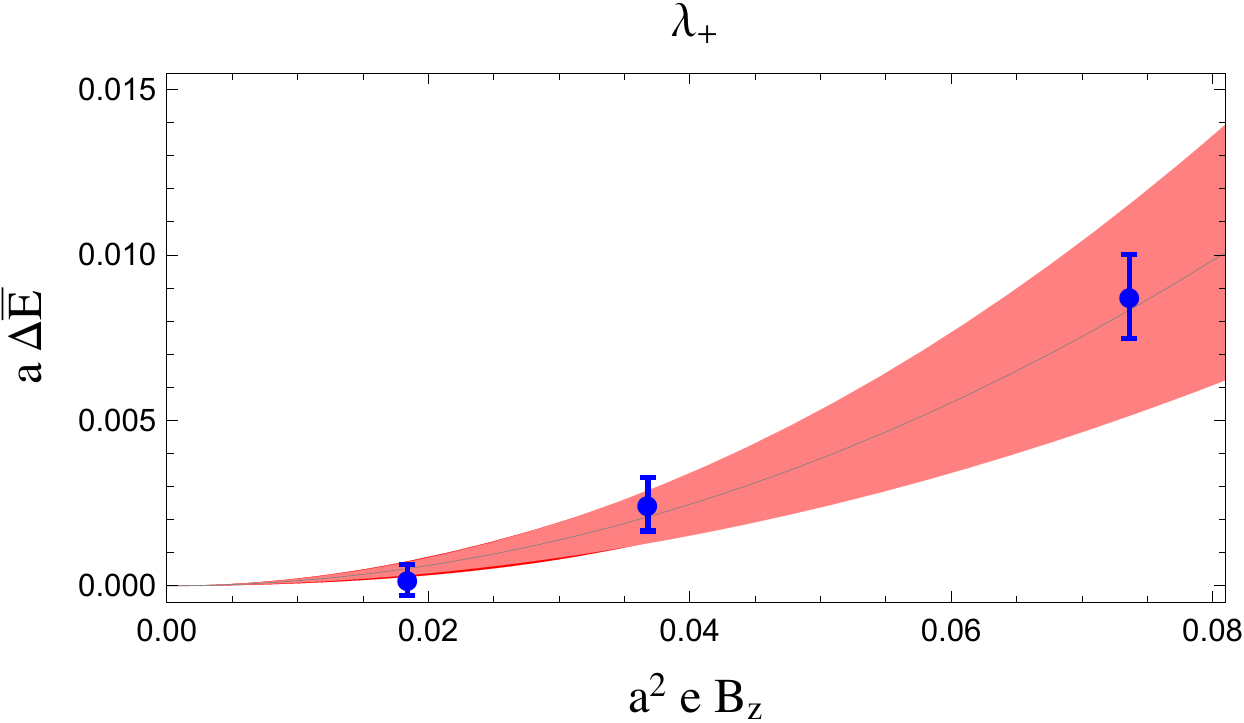}
}
\vskip 0.05in
\resizebox{0.725\linewidth}{!}{
\includegraphics{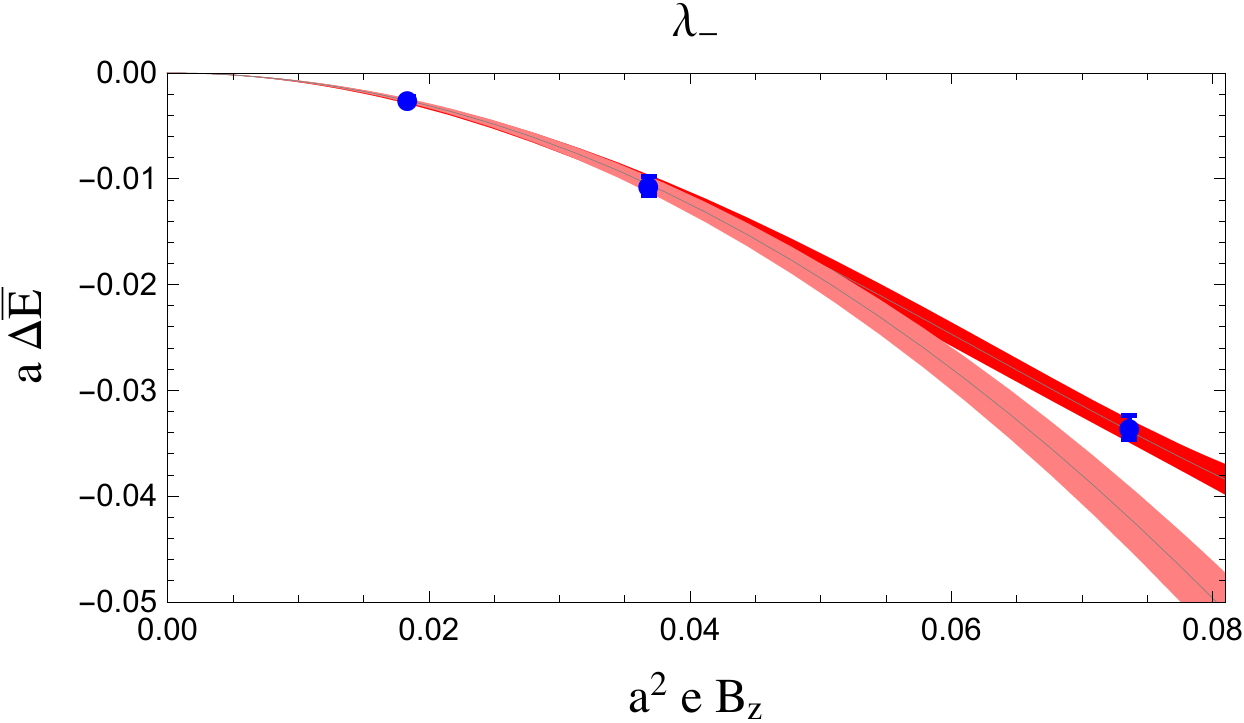}
}
\caption{
Fits to the magnetic-field dependence of Zeeman splittings and spin-averaged energy differences in the 
$\Lambda$--$\Sigma^0$  system  calculated on Ensemble I. 
For Zeeman splittings, two representative fits are shown:
the darker band corresponds to fitting all three magnetic field values  to a  linear plus cubic form, 
$F_3$ in  Eq.~\eqref{eq:BFits}, 
while the lighter band corresponds to a linear fit that excludes the largest magnetic field value. 
For the  $\lambda_+$  eigenstate,  the negative of the Zeeman splittings are shown.
For the spin-averaged energy differences,  $\Delta \overline{E}$,
the two representative fits are: a fit to all three magnetic field values using quadratic plus quartic form, 
$\mathcal{F}_4$  in  Eq.~\eqref{eq:BBFits},  shown as a darker band; 
a quadratic fit that excludes the largest magnetic field value, shown as a lighter band.  
} 
\label{f:LamSigfitB}
\end{figure}
%
%
%

\subsection{$\Lambda$--$\Sigma^0$ Correlator Analysis}

On the $U$-spin symmetric ensembles,  
the energy eigenstates of the  $\Lambda$--$\Sigma^0$
system are calculated from principal correlators  that are solutions to the generalized eigenvalue problem in 
Eq.~\eqref{eq:GEVP}, requiring the same baryon operators  for the source and sink. 
The correlation functions used in this work are constructed 
from multiple different source 
locations   on each configuration, 
making the smeared source interpolators  the same as the zero-momentum projected, smeared-sink interpolators within statistical uncertainty
(on the ensemble average, this is guaranteed by momentum conservation).
Therefore, only the SS correlation functions are utilized in this part of the analysis.

For each of the spin-projected, principal correlators, 
$G_\lambda^{(s)}(t, t_0, n_\Phi)$, 
ratios of correlation functions
\begin{equation}
r^{(s)}_\lambda
(t)
=
G_\lambda^{(s)}(t,t_0,n_\Phi)
\big/
G_{\lambda}^{(s)} (t,t_0, 0)
\label{eq:PrinRatio}
,\end{equation}
are formed,
where the magnetic-field dependence is treated as implicit.
The time offset 
$t_0$
is the same parameter employed to solve 
Eq.~\eqref{eq:GEVP}, and the value
$t_0 / a = 3$ is used.
The effect of varying 
$t_0$
is found to be numerically insignificant in this particular analysis, 
and for this reason we drop the 
$t_0$ 
label from 
$r_\lambda^{(s)}(t)$, 
which should be independent of 
$t_0$. 
For the normalization of the ratios, 
it is possible to divide the principal correlators by any linear combination of the 
diagonal 
$\Lambda$
and
$\Sigma^0$
correlation functions in zero magnetic field due to their mass degeneracy 
(including the exact linear combinations for the 
$\lambda_\pm$ 
states which are known from the analytic solution). 
For the present study, 
 the
$\Sigma^0$
correlator is used for 
$\lambda_+$, 
and the
$\Lambda$
correlator is used for 
$\lambda_-$. 
These would be the natural choices in the case of broken 
$SU(3)_F$.  
The correlation function ratios are shown in 
Fig.~\ref{f:SigLamEeffB}, 
and allow for the extraction of the energy differences, 
$E(B_z) - E(0)$. 
Results of exponential fits to these ratios are presented in 
Table~\ref{t:LamSigE}.

To compute magnetic moments, 
it is efficacious to isolate them by taking further ratios 
\begin{equation}
R(t)
=
r_\lambda^{(+\frac{1}{2})} (t) \Big/ r_\lambda^{(-\frac{1}{2})}(t)
\label{eq:PrinZeeman}
,\end{equation}
whose long-time behavior leads to the Zeeman splittings  $\Delta E$  (see also  Eq.~\eqref{eq:Ratio}). 
There is one such ratio for the  $\lambda_+$ eigenstates,  and another for the  $\lambda_-$ eigenstates. 
These double ratios and exponential fits to their time dependence are shown in 
Fig.~\ref{f:SigLamEeffB} and given  in 
Table~\ref{t:LamSigE}. 
Values of the energy differences, 
$E(B_z) - E(0)$,
and 
Zeeman splittings,
$\Delta E$,
allow determination of magnetic moments through fits to their magnetic-field dependence. 
For the spin-dependent energy differences, 
 a linear plus quadratic fit function, 
namely 
$\tilde{F}^{(s)}_2(B) = - 2 \mu \, s B + \tilde{f}^{(s)}_2 B^2$, 
for 
$s = \pm \frac{1}{2}$ is utilized. 
The fits are shown in 
Fig.~\ref{f:LamSigB}, 
and extracted values of magnetic moments are given in 
Table~\ref{t:LamSigE}. 
For Zeeman splittings, 
 the fit functions 
$F_1$
and
$F_3$
appearing in 
Eq.~\eqref{eq:BFits} are used. 
Representative fits are shown in 
Fig.~\ref{f:LamSigfitB}.

The final part of the analysis concerning the 
$\Lambda$--$\Sigma^0$
system is the determination of magnetic polarizabilities, 
which are responsible for lifting the residual degeneracy of the different eigenstates of opposite spin. 
To determine the polarizabilities, 
products of ratios of spin-projected principal correlators,
\begin{equation}
\mathcal{R}(t)
=
\sqrt{r_\lambda^{(+\frac{1}{2})} (t) \,  r_\lambda^{(-\frac{1}{2})}(t)}
\label{eq:PrinAvg}
,\end{equation}
are formed,
whose long-time exponential behavior is governed by the spin-averaged energy differences, 
$\Delta \overline{E} \equiv \overline{E}(B_z) - \overline{E}(0)$, 
where
$\overline{E}(\bm{B}) = \frac{1}{2} \left[ E^{(+\frac{1}{2})} (\bm{B}) + E^{(- \frac{1}{2})} (\bm{B}) \right]$.  
There is one such ratio for each of the two eigenstates
$\lambda_\pm$ 
and each has been plotted in 
Fig~\ref{f:SigLamEeffB}, 
along with exponential fits. 
Results of fitting the 
$\mathcal{R}(t)$
ratios are provided in 
Table~\ref{t:LamSigE}. 
Values of the spin-averaged energy differences are then fit as a function of the magnetic field to extract the magnetic polarizability, 
$\beta$, using the two fit functions
\begin{eqnarray}
\mathcal{F}_2(B) 
&=& 
- \beta B^2
, 
\notag \\
\mathcal{F}_4(B)
&=& 
- \beta B^2 + g_4 B^4
\label{eq:BBFits}
.\end{eqnarray} 
Using values of the magnetic field in lattice units, 
$a^2 e B_z$, 
leads to fit parameters 
$\beta$
in lattice polarizability units, 
$\texttt{[LatU]} = e^2 a^3$.
Table~\ref{t:LamSigE} also provides  values for 
$\hat{\beta}$, 
which are polarizabilities in units of 
$e^2 / M_B^2 (M_T - M_B)$, 
where 
$a M_T = 1.3321(10)(19)$
is the mass of the baryon decuplet on Ensemble I%
~\cite{Chang:2015qxa}, 
and values for 
$\beta$
in the conventional polarizability units of
$\texttt{[10}^{\texttt{-4}} \,  \texttt{fm}^{\texttt{3}} \texttt{]}$
using the lattice spacing given in 
Table~\ref{t:configs}.

\bibliography{bib_octetB}
\end{document}